\documentclass[11pt]{article}
\usepackage{graphicx}
\usepackage{tikz-cd}
\usepackage{grffile}
\usepackage{ctable}
\usepackage{appendix}
\usepackage{geometry}
\usepackage{subcaption}
\usepackage[export]{adjustbox}
\usepackage{amsmath,amsthm,amssymb}
\usetikzlibrary{matrix,calc}
\usepackage{epstopdf}
\usepackage [american]{babel} 
\usepackage{enumerate}
\usepackage[font=footnotesize]{caption}                     

\usepackage[skip=5pt plus1pt, indent=0pt]{parskip}

\newcommand{\Z}{\mathbb{Z}}

\newcommand{\R}{\mathbb{R}}

\newcommand{\half}{\frac{1}{2}}

\newcommand{\ep}{\varepsilon}

\newcommand{\wt}{\widetilde}

\newcommand{\eps}{\varepsilon}
\newcommand{\mnew}{{\widetilde{\mu}}}
\newcommand{\morig}{{\mu}}

\newtheorem{Lemma}{Lemma}[section]
\newtheorem{Theorem}{Theorem}
\newtheorem{Proposition}[Lemma]{Proposition}

\newtheorem{Remark}[Lemma]{Remark}

\newenvironment{Proof}[1][\unskip]%
 {\begin{trivlist} \item[]{\bf Proof #1. }}%
 {\hspace*{\fill}$\rule{.4\baselineskip}{.4\baselineskip}$\end{trivlist}}

\newenvironment{Acknowledgment}%
 {\begin{trivlist}\item[]\textbf{Acknowledgments.}}{\end{trivlist}}

 {\begin{center}\textbf{Abstract}}{\end{center}}

\makeatletter\@addtoreset{figure}{section}\makeatother

\makeatletter \@addtoreset{equation}{section} \makeatother

\makeatletter
\newsavebox{\@brx}
\newcommand{\llangle}[1][]{\savebox{\@brx}{\(\m@th{#1\langle}\)}%
  \mathopen{\copy\@brx\kern-0.5\wd\@brx\usebox{\@brx}}}
\newcommand{\rrangle}[1][]{\savebox{\@brx}{\(\m@th{#1\rangle}\)}%
  \mathclose{\copy\@brx\kern-0.5\wd\@brx\usebox{\@brx}}}
\makeatother

\definecolor{Green}{rgb}{0.,0.4,0.}

\renewcommand{\leq}{\leqslant}

\newcommand{\Rmnum}[1]{\uppercase\expandafter{\romannumeral #1\relax}}

\def\XXint#1#2#3{{\setbox0=\hbox{$#1{#2#3}{\int}$}
     \vcenter{\hbox{$#2#3$}}\kern-.5\wd0}}

\newfam\bifam
\font\tenbi=cmmib10 scaled \magstep1 \font\sevenbi=cmmib10 at 11pt
\font\fivebi=cmmib10 at 6pt \textfont\bifam = \tenbi
\scriptfont\bifam = \sevenbi \scriptscriptfont\bifam= \fivebi

\begin{document}

\begin{center}
{\fontsize{15}{15}\fontfamily{cmr}\fontseries{b}\selectfont
Vacuum bubble and fissure formation in collective motion with competing attractive and repulsive forces }\\[0.2in]
Olivia Clifton$^{1}$,  Angel Chavez$^{2}$, Antonio Madrigal$^{3}$, Annie Warren$^{2}$, Paige Yeung$^{4}$, and  Arnd Scheel$^{2}$\\[0.1in]
\textit{\footnotesize
$^1$Department of Mathematics, Univ. of Illinois at Urbana-Champaign,  Urbana, IL 61801, USA\\[0.05in]
$^2$School of Mathematics, University of Minnesota, Minneapolis, MN 55414, USA\\[0.05in]
$^3$ College of New Jersey, Ewing, NJ 08628, USA\\[0.05in]
$^4$ Massachusetts Institute of Technology, Cambridge, MA 02139, USA
}\footnote{Competing interests: The authors declare none.}
\end{center}

\begin{abstract}
\noindent 
We study the continuum limit of the motion of agents in the plane driven by competing short-range repulsion and long-range attractive forces. At a critical parameter value, we find destabilization of a trivial branch of uniformly distributed solutions and analyze bifurcating solutions. Curiously, the bifurcating branch is vertical, leading to a reversible, non-hysteretic phase transition. Near the bifurcation point, we demonstrate scaling laws for the size of vacuum regions, which can form fissures or bubbles. We also study the effect of small noise and the eventual topological transition from  vacuum bubbles to isolated particle clusters. 
\end{abstract}

\section{Introduction}

We study the continuum limit of a large number of particles moving in the plane with positions $x_j\in\R^2$, $j=1,2,\ldots,M$ driven by  an interaction potential $W_{\morig}$,
\begin{equation}\label{e:ODE}
\dot{x}_j=-\frac{1}{M} \sum_{1\leq j\leq M,j \neq \ell} \nabla W_{\morig}(x_\ell-x_{j}).
\end{equation}
We assume that the interaction potential includes a strong, very short-range repulsive and an intermediate-range attractive force, that we will assume depends on the parameter $\morig$. In a limit of large particle numbers, one approximates the distributions of particles by a density and finds formally taking the limit of particle Dirac-$\delta$ distributions as measures,  that the density evolves according to a Vlasov-type equation, 
\begin{equation}\label{e:MV_time-dep}
    u_t = \nabla\cdot\left(u\,\nabla( W_{\morig}*u)\right) ,\qquad x\in\R^2,
\end{equation}
where convolution is defined through
\[
(W_{\morig}*u)(x)=\int_{y\in\R^2} W_{\morig}(x-y)u(y) dy;
\]
see for instance \cite{burger_esposito23,jabin} and references therein.
Letting the range of the repulsive potential converge to zero while increasing the strength appropriately, one can further approximate $W_{\morig}=\delta-\morig V$, which leads to the nonlinear diffusion equation with nonlocal interaction,
\begin{equation}\label{e:MV_time-dep2}
    u_t =\nabla\cdot\left(u\,\nabla\left(u -\morig  (V*u)\right)\right), \qquad u\in\R,\  x\in\R^2,
\end{equation}
where we also made the parameter dependence explicit by varying the strength of the attractive potential $V$ as a prefactor.
We further simplify the  setting by assuming  particles are distributed periodically with respect to a lattice spanned by lattice vectors $e_{1,2}\in\R^2$. In the continuum limit, we then restrict to the fundamental torus $\Omega=\{x=\tau_1 e_1+ \tau_2 e_2|0\leq \tau_1,\tau_2<1\}=\R^2/(e_1\Z+e_2\Z)$, with periodic boundary conditions. We then have to replace the interaction potential by the periodized potential  $V_\mathrm{per}(\cdot)=\sum_{k_1,k_2\in\Z} V(\cdot + k_1 e_1+ k_2 e_2)$, thus studying 
\begin{equation}\label{e:MV_time-dep3}
    u_t =\nabla\cdot\left(u\, \nabla\left(u -\morig (V_\mathrm{per}*u)\right)\right), \qquad x\in\Omega,
\end{equation}
with periodic boundary conditions, where now,
\[
 (V_\mathrm{per}*u)(x)=\int_{y\in\Omega} V_\mathrm{per}(x-y)u(y) dy.
\]
Note the slight abuse of notation where convolution is denoted by the same symbol as in the whole plane. From now on, we shall always consider convolution in this periodic setting, thus removing any ambiguity.
We focus on the
\begin{equation}
    \begin{aligned}
    \text{(i) square lattice: } & e_1=(\pi,\pi)^T,\quad e_2=(\pi,-\pi)^T ;\qquad \text{ and the }\\  \text{(ii) hexagonal lattice: } & e_1 = \left( 2\pi,0 \right)^T, \quad e_2 =\left( - \pi, \sqrt{3}\pi  \right)^T. 
    \end{aligned}
\end{equation}
The lattices have  $D_4$ and $D_6$ dihedral symmetry generated by 
\begin{equation}\label{e:sym}
\text{reflection: } S(x,y)=( -x,y), \qquad \text{  rotation: } R(x,y)=((\cos\varphi) x+(\sin\varphi) y,-(\sin\varphi) x+(\cos \varphi )y),
\end{equation}
with $\varphi=2\pi/4 $ and $\varphi=2\pi/6$, { respectively.}

We shall also consider \eqref{e:MV_time-dep3} with an additional diffusion term 
\begin{equation}\label{e:MV_time-dep-diff}
    u_t = \eps\Delta u + \nabla\cdot\left(u\,\nabla\left(u -\morig( V_\mathrm{per}*u)\right)\right),
\end{equation}
and $0<\ep\ll 1$, reflecting the presence of noise in the particle dynamics \eqref{e:ODE}. 

We are interested in bifurcations from a constant distribution, which we choose as  $u\equiv 1$ without loss of generality by scale-invariance of the equation. Clearly, this uniform density is always an equilibrium solution to \eqref{e:MV_time-dep} or \eqref{e:MV_time-dep-diff}, and one expects this state to be stable for weak attracting forces, $0<\morig\ll 1$. Increasing $\morig$, one then indeed finds a destabilization and an ensuing phase transition, where densities will no longer be uniform but evolve into distributions with peaks, where particles cluster, and vacuum regions, where the density vanishes $u=0$. This transition curiously exhibits a vertical bifurcating branch, which leads to transitions that are reversible; that is, the bifurcation diagram does not display hysteresis; and to bifurcating patterns that exhibit vacuum regions at arbitrary small parameter distances from the bifurcation point; see for instance \cite{cky,berthelin,ccww,stevensscheel}. We analyzed this transition in quite some detail in \cite{stevensscheel}, finding universal expansions near the bifurcation point in quite general settings, as well as universal corrections for $0<\ep\ll 1$. The work there focused on one-dimensional arrangements of particles, and motivates the study of the more complex two-dimensional arrangements that we are interested in here.

\begin{figure}[h!]
\centering
    \includegraphics[width=1.7in]{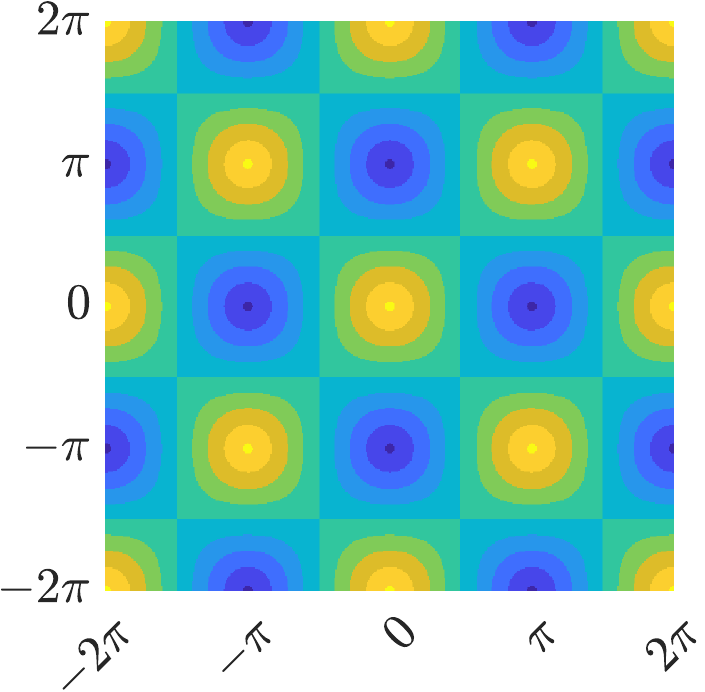}\hspace*{1in}
    \includegraphics[width=1.7in]{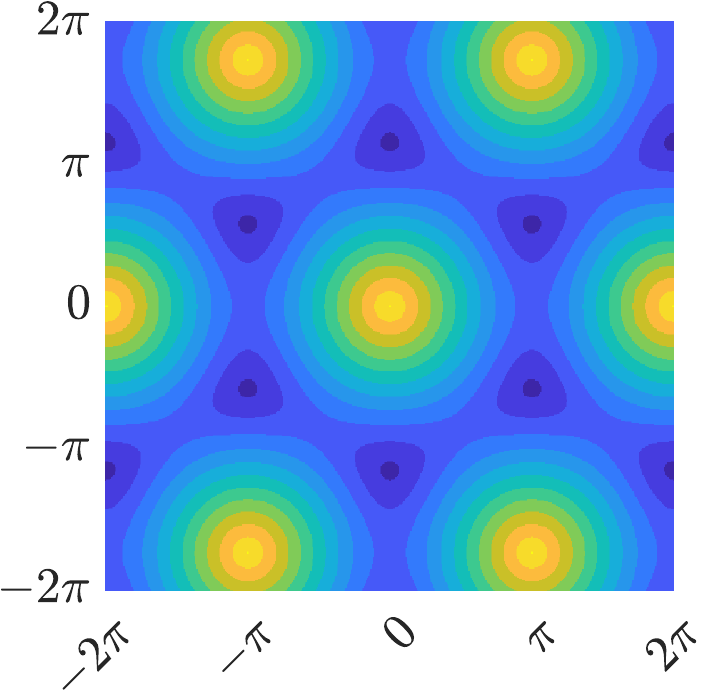}
    \caption{Contour plots of potentials $V_\mathrm{per}$ from \eqref{e:Vexpl} in the square case (left) and the hexagonal case (right) with maxima yellow, minima blue. Note that the square potential is odd, but the hexagonal potential is not: level sets near maxima and minima have square-like corrections to the leading-order round shape in the square potential; the corrections are triangular near minima and hexagonal near maxima in the hexagonal potential case.}\label{f:pot}
\end{figure}

The vertical branch of bifurcating solutions is in fact explicit, determined by the kernel of the linearized operator.  Our results focus on three aspects of this transition:
\begin{itemize}
    \item[(i)] at the endpoint of the vertical branch, solutions with vacuum regions bifurcate supercritically: we show existence of branches exhibiting vacuum regions that are roughly circular or form bands, and give expansions for the size of the vacuum region in terms of the parameter $\morig$; 
    \item[(ii)] we study the fate of the vertical branch with small diffusion $\ep\gtrsim 0$ and derive stability information in this scenario;
    \item[(iii)] we investigate the fate of solution branches further away from the bifurcation point, pointing towards, in the case of the square lattice, the presence of elliptical vacuum regions connecting roughly spherical vacuum bubbles and fissures, and a topology change from vacuum bubbles to isolated particle clusters.
\end{itemize}
In all of our results we focus on very simple potentials. In fact, due to periodization, potentials can be written as Fourier series over the fundamental domain, and we choose to keep only the first terms in these Fourier series, while retaining the lattice symmetry.  Starting with a radially symmetric potential $V$, the symmetrized potential $V_\mathrm{per}$ will have the lattice symmetries; that is, with $R,S$ from \eqref{e:sym}, $V_\mathrm{per}(R(x,y))=V_\mathrm{per}(x,y)$ and $V_\mathrm{per}(S(x,y))=V_\mathrm{per}(x,y)$, and of course $V_\mathrm{per}(x,y)=V_\mathrm{per}(x+e_{j,x},y+e_{j,y})$, $j=1,2$. To be specific, we consider the potentials
\begin{equation}\label{e:Vexpl}
    \begin{aligned}
    \text{square lattice: \qquad}  V_\mathrm{per}(x,y)=&\cos(x+y) + \cos(x-y) = 2\cos x \cos y;\\  
    \text{hexagonal lattice:\qquad  } V_\mathrm{per}(x,y)= & \cos\left(x+\frac{y}{\sqrt{3}}\right) + \cos\left(x-\frac{y}{\sqrt{3}}\right) + \cos\left(\frac{2y}{\sqrt{3}}\right)\\
    = &2\cos (x) \cos\left( \frac{y}{\sqrt{3} } \right) + \cos \left( \frac{2y}{\sqrt{3} } \right);
    \end{aligned}
\end{equation}
see Fig.~\ref{f:pot}
The functions in \eqref{e:Vexpl} are simply minimal Fourier polynomials with the dihedral symmetry and lattice periodicity. We demonstrated in the one-dimensional case \cite{stevensscheel} that results do not depend on the addition of harmonics, that is, they are universal in this sense, but we will not pursue such an analysis here. 

The maxima of $V_\mathrm{per}$ correspond to preferred distances between particles and the minima correspond to disfavored distances, when considering the long-range interaction $V_\mathrm{per}$, only. Both potentials exhibit unique maxima, up to lattice symmetries, at the origin, thus favoring clustering with zero distance, which is however a forbidden distance due to the strong short-range repulsion. In the square lattice case, $V_\mathrm{per}$ possesses a minimum at $x=0,\,y=\pi$, making this an energetically disfavored distance. Similarly, in the hexagonal case, minima and therefore energetically disfavored distances occur at $x=0,y=\pi/\sqrt{3}$ 
and, by reflection symmetry, at $x=\pi,y=2\pi/\sqrt{3}$. 
In the remainder of this paper, we drop the subscript and simply write $V=V_\mathrm{per}$.

{}\smallskip\textbf{Outline.} We derive expansions for the size of vacuum bubbles and fissures near the bifurcation point in case of both square and hexagonal lattices in \S\ref{s:gaps_proofs}. The following section,  \S\ref{s:diffusion}, contains results on diffusive corrections, $\eps\gtrsim 0$ in \eqref{e:MV_time-dep-diff}.
In this setting, we also connect our findings to a more standard center-manifold analysis as for instance carried out in \cite{carrillogvalani}, complemented there by global energetic considerations; see also \cite{balasubramanian2025structurestationarysolutionsmckeanvlasov} and \cite{shalova2025solutionsstationarymckeanvlasovequation} for generalizations to spherical geometry.
We study branches numerically and identify a secondary bifurcation mediating a change in topology and a transition between fissures and clusters in \S\ref{s:num_pde}. We conclude with numerical results in the discrete setting, $N<\infty$, in \S\ref{s:discrete} and a brief discussion, \S\ref{s:dis}.

 \begin{Acknowledgment}  The authors acknowledge support through grant NSF DMS-2205663 and NSF DMS-2506837.
 \end{Acknowledgment}
 
\section{Vacuum Formation in Continuum Model}\label{s:gaps_proofs}

We identify bifurcation points, a global vertical branch, and continue solutions from the end of this vertical branch as bubbles and fissures open up past the critical parameter value. 
\subsection{The Vertical Branch}
Solving for equilibria of \eqref{e:MV_time-dep3}, we find that $u\,\nabla\left(u -\morig (V*u)\right)=\nu\in\R^2$, a constant. If $u(x,y)=0$ at some point, then $\nu=0$. Otherwise, $\partial_x\left(u -\morig (V*u)\right)=\nu_1/u$ has a sign, which contradicts periodicity, unless $\nu_1=0$. With the same reasoning for the $y$-derivative, we conclude that $\nu=0$. For a solution without vacuum, that is, for $u > 0$, we then find
\begin{equation}
    u -  \morig(V * u) = \rho,
\end{equation}
for $\rho$ a constant. Given our simple choices of repulsive potentials,  one then has the explicit vertical branches 
\begin{align}
  \morig_* = \frac{1}{\pi^2}, \quad   u =& A_0 + A_1\cos(x)\cos(y) + B_1\sin(x)\sin(y)\nonumber \\
  &+  a_1\cos(x)\sin(y) + b_1\sin(x)\cos(y) \quad &&\textrm{(sq.)}, \label{e:vbsq}\\
  \morig_* = \frac{1}{\sqrt{3}\pi^2}, \quad   u = &A_0 + A_1\cos\left(x+\frac{y}{\sqrt{3}}\right) + B_1\cos\left(x-\frac{y}{\sqrt{3}}\right) + C_1\cos\left(\frac{2y}{\sqrt{3}}\right)\nonumber\\
  &+a_1\sin\left(x+\frac{y}{\sqrt{3}}\right) + b_1\sin\left(x-\frac{y}{\sqrt{3}}\right) + c_1\sin\left(\frac{2y}{\sqrt{3}}\right) .\quad &&\textrm{(hex.)}\label{e:vbhex}
\end{align}
We shall focus on solutions with $a_1=b_1=0$ in \eqref{e:vbsq} and $a_1=b_1=c_1=0$ in \eqref{e:vbhex} in the remainder of this paper. These subspaces consist of functions invariant under point reflections $(x,y)\mapsto (-x,-y)$. We only comment  briefly on solutions outside of this fixed point subspace. Solutions with maximal isotropy all lie within these subspaces, possibly after a spatial translation; see Remarks~\ref{r:isotropy} and~\ref{r:isotropy2}, below.

Fixing the average density to, say, 1, we may set $A_0=1$ and are left with  two- and three-dimensional subspaces of solutions, respectively. These subspaces contain one-dimensional subspaces with larger symmetry (maximal isotropy) as follows. 
In the square symmetry case,  $A_1 = B_1$ corresponds to fissures, solutions depending on $x-y$, only,  and $B_1 = 0$ (or $A_1=0$) corresponds to bubble solutions invariant under the 4-fold rotation $R$. Solutions are strictly positive, possessing no vacuum regions, as long as $\max(|A_1|,|B_1|) < A_0$.

In the hexagonal symmetry case, $A_1 = B_1 = C_1$ corresponds to bubble solutions  with 6-fold rotational symmetry $R$, while fissures correspond to solutions where only one of $A_1, B_1 ,C_1$ is nonzero. Bubble solutions are strictly positive if $-\frac{1}{3}A_0 < A_1,B_1, C_1 < \frac{2}{3}A_0$, fissure solutions for $|A_1|<A_0=1$, $B_1=C_1=0$.

\subsection{Vacuum Formation: Square Lattice Symmetry}
We analyze solutions in the vicinity of the boundary of the vertical branch, when  $\max(|A_1|,|B_1|) = A_0$.  We look for non-negative solutions with support  $\Omega_0\subset\Omega$ that have average density $1$. They therefore satisfy
\begin{align}\label{e:MV_gen}
  u - 
  \morig(V * u) - \rho &= 0, \text{ for } x\in\Omega_0&&\textrm{ (McKean-Vlasov condition),}\\
  \label{e:MV_density_gen}
  \frac{1}{\text{Area}(\Omega)} \iint_{\Omega_0} u(x,y) dydx &= 1 && \textrm{ (average density condition),}\\
  \label{e:MV_boundary_gen}
  u(x,y) &= 0 \text{ on $\partial \Omega_0$} && \textrm{ (boundary condition). }
\end{align}
The continuity condition \eqref{e:MV_boundary_gen}  follows from the fact that $u$ is a weak solution.

\begin{Theorem}[Vacuum region scaling --- square lattice]\label{t:square}
    Let $V(x,y) = 2\cos(x)\cos(y)$, and set $\morig = \frac{1}{\pi^2} + \mnew$, with $\mnew$ sufficiently small. Then solutions $u$ to \eqref{e:MV_gen}-\eqref{e:MV_boundary_gen}, invariant under point reflections $u(x,y)=u(-x,-y)$,  with vacuum regions, that is, $\Omega_0\neq \Omega$, are of the form
    \begin{equation}  \label{e:rk3}
        u(x,y) = \left( A_0 + A_1 \cos (x) \cos (y)+B_1 \sin(x)\sin(y)\right)_+,
    \end{equation}
    where $f_+=\max(f,0)$. There are then two solution branches, both supercritical, locally unique up to  translations in $x$ and $y$, parameterized by $\mnew\gtrsim 0$ as $A_0=A_0(\mnew),\, A_1 = A_1(\mnew),\, B_1 = B_1(\mnew)$:
    \begin{enumerate}
        \item \emph{fissures: } $A_1=-B_1$, 
             \begin{align*}
            A_0  =1-\frac{\pi}{2}\mnew+\mathcal{O}(\mnew^2),\qquad
            A_1  =1+\left(\frac{3}{4\sqrt{2}}\pi^3\mnew\right)^{2/3}+\mathcal{O}(\mnew).
            \end{align*}
            The half-width $\ell$  of the fissures  and area $\mathcal{A}$  of vacuum region inside $\Omega$ are 
            \[
                \ell=\frac{\pi}{2^{1/2}}\left(\frac{3}{2} \mnew\right)^{1/3}+\mathcal{O}(\mnew^{2/3}) ,\qquad \mathcal{A}=4\pi^2\left(\frac{3}{2} \mnew\right)^{1/3}+\mathcal{O}(\mnew^{2/3}).
            \]
        \item \emph{bubbles: } $B_1=0$, 
            \begin{align*}
            A_0  = 1 - \frac{\pi^2}{4} \mnew  + \mathcal{O}(\mnew^{3/2}),\qquad
            A_1  = 1 + \frac{\pi^{3/2}}{\sqrt{2}} \mnew^{1/2} - \frac{\pi^2}{4} \mnew + \mathcal{O}(\mnew^{3/2}) .
            \end{align*}
            The inner radius  $\ell$ and area $\mathcal{A}$  of the vacuum region inside $\Omega$ are
            \[
                \ell = (2\pi^3 \mnew)^{1/4} + \mathcal{O}(\mnew^{3/4} ),\qquad \mathcal{A}=(2\pi^5 \mnew)^{1/2}+\mathcal{O}(\mnew).
            \]
    \end{enumerate}
\end{Theorem}

\begin{figure}[h!]
\begin{center}
    \begin{tikzpicture}[scale=0.4]
    \def\side{2*pi}
    \def\p{3.14159}
    \def\radius{1} 

    \definecolor{lightpink}{rgb}{1.0, 0.71, 0.76}
    \definecolor{lightgreen}{rgb}{0.56, 0.93, 0.56}
    \definecolor{lightblue}{rgb}{0.4, 0.7, 1.0}
     \definecolor{lightblued}{rgb}{0, 0.65, 1.0}
      \definecolor{lightyellow}{rgb}{0.9, 1, .5}

\fontsize{10}{10}\selectfont

    \fill[lightgreen] (0, -pi) rectangle (2*pi, pi);

    \begin{scope}
        \clip (0, -pi) rectangle (2*pi, pi);
        
        \filldraw[black, fill=lightblued] (3.14, 0) circle (\radius);
        \filldraw[black, fill=lightblue] (0, 3.14) circle (\radius);
         \filldraw[black, fill=lightblue] (2*pi, 3.14) circle (\radius);
          \filldraw[black, fill=lightblue] (2*pi, -3.14) circle (\radius);
       \filldraw[black, fill=lightblued] (-3.14, 0) circle (\radius);
       \filldraw[black, fill=lightblue] (0, -3.14) circle (\radius);
    \end{scope}

    \node at (pi, 9*pi/15) { $\Omega$};


    \draw[thick] (0, -pi) rectangle (2*pi, pi);
    
    \draw[<->, thick] (-\side/2-0.5+\p,-\side/2-1.3) -- (\side/2 +0.5+\p,-\side/2-1.3) node[right] {$x$};
    \draw[<->, thick] (-\side/2-1.3+\p,-\side/2-0.5) -- (-\side/2-1.3+\p,\side/2+0.5) node[above] {$y$};

    \draw[thick] (\p,-\side/2-1.4) -- (\p, -\side/2-1.2) node[midway,below=4pt]{$\pi$};
     \draw[thick] (-\side/2+\p,-\side/2-1.4) -- (-\side/2+\p, -\side/2-1.2) node[midway,below=4pt]{$0$};
      \draw[thick] (\side/2+\p,-\side/2-1.4) -- (\side/2+\p, -\side/2-1.2) node[midway,below=4pt]{$2\pi$};
    \draw[thick] (-\side/2-1.4+\p,0) -- (-\side/2-1.2+\p,0) node[midway,left=4pt]{0};
    \draw[thick] (-\side/2-1.4+\p,-\side/2) -- (-\side/2-1.2+\p,-\side/2) node[midway,left=4pt]{$-\pi$};
    \draw[thick] (-\side/2-1.4+\p,\side/2) -- (-\side/2-1.2+\p,\side/2) node[midway,left=4pt]{$\pi$};

    \draw[thick,loosely dotted] (-\side/2+\p,0) -- (\p,-\side/2);
   \draw[thick,loosely dotted] (\p,\side/2) -- (\side/2+\p,0);
   \draw[thick,loosely dotted] (-\side/2+\p,0) -- (\p,\side/2);
   \draw[thick,loosely dotted] (\p,-\side/2) -- (\side/2+\p,0);

\end{tikzpicture}\hspace*{0.7in}
    \begin{tikzpicture}[scale=0.4]
    \def\side{2*pi}
    \def\p{3.14159}
    \def\radius{1} 

    \definecolor{lightpink}{rgb}{1.0, 0.71, 0.76}
    \definecolor{lightgreen}{rgb}{0.56, 0.93, 0.56}
    \definecolor{lightblue}{rgb}{0.4, 0.7, 1.0}
     \definecolor{lightblued}{rgb}{0, 0.65, 1.0}
      \definecolor{lightyellow}{rgb}{0.9, 1, .5}

\fontsize{10}{10}\selectfont

    \fill[lightgreen] (0, -pi) rectangle (2*pi, pi);

    \begin{scope}
        \clip (0, -pi) rectangle (2*pi, pi);

        \filldraw[black, fill=lightblue, rotate=-45] (-\side-\radius/2,-0.707*\p-\radius/2) rectangle (2*\side+\radius/2,\radius/2-0.707*\p);
        \filldraw[black, fill=lightblue, rotate=-45] (-\side-\radius/2,0.707*\p-\radius/2) rectangle (2*\side+\radius/2,0.707*\p+\radius/2);
        \filldraw[black, fill=lightblue, rotate=-45] (-\side-\radius/2,3*0.707*\p-\radius/2) rectangle (2*\side+\radius/2,3*0.707*\p+\radius/2);
    \end{scope}

     \node at (pi, 9*pi/15) { $\Omega$};
    

   \draw[thick,loosely dotted] (-\side/2+\p,0) -- (\p,-\side/2);
   \draw[thick,loosely dotted] (\p,\side/2) -- (\side/2+\p,0);
     \draw[thick,loosely dotted] (-\side/2+\p,0) -- (\p,\side/2);
   \draw[thick,loosely dotted] (\p,-\side/2) -- (\side/2+\p,0);


    \draw[thick] (0, -pi) rectangle (2*pi, pi);
    
    \draw[<->, thick] (-\side/2-0.5+\p,-\side/2-1.3) -- (\side/2 +0.5+\p,-\side/2-1.3) node[right] {$x$};
    \draw[<->, thick] (-\side/2-1.3+\p,-\side/2-0.5) -- (-\side/2-1.3+\p,\side/2+0.5) node[above] {$y$};

    \draw[thick] (\p,-\side/2-1.4) -- (\p, -\side/2-1.2) node[midway,below=4pt]{$\pi$};
     \draw[thick] (-\side/2+\p,-\side/2-1.4) -- (-\side/2+\p, -\side/2-1.2) node[midway,below=4pt]{$0$};
      \draw[thick] (\side/2+\p,-\side/2-1.4) -- (\side/2+\p, -\side/2-1.2) node[midway,below=4pt]{$2\pi$};
    \draw[thick] (-\side/2-1.4+\p,0) -- (-\side/2-1.2+\p,0) node[midway,left=4pt]{0};
    \draw[thick] (-\side/2-1.4+\p,-\side/2) -- (-\side/2-1.2+\p,-\side/2) node[midway,left=4pt]{$-\pi$};
    \draw[thick] (-\side/2-1.4+\p,\side/2) -- (-\side/2-1.2+\p,\side/2) node[midway,left=4pt]{$\pi$};
\end{tikzpicture}
\end{center}
\caption{ An illustration of vacuum bubbles (left) and fissures (right) in a square of width $2\pi$, comprising two (!) fundamental domains $\Omega$ of the square lattice. Blue regions represent areas of vacuum; $u$ is positive in the green regions $\Omega_0$. Note that there is one bubble and one fissure in each fundamental domain (boundaries of fundamental domain indicated by dashed lines); fissures have width $2\ell$ and total length $\sqrt{2}\pi$ in a fundamental domain, bubbles have radius $\ell$. }
\label{fig:2d-sq-lattice}
\end{figure}
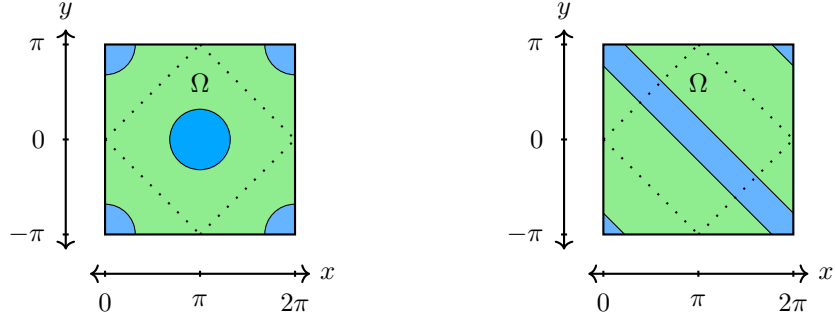

\begin{Proof} 
%
First, notice that \eqref{e:MV_gen} implies that for all $(x,y)$ where $u(x,y)>0$, we have 
\[
u(x,y)=\rho+2\morig\iint_{\Omega_0} \cos(x-\xi)\cos(y-\eta)u(\xi,\eta) d\xi d\eta,
\]
which, using trigonometric addition theorems gives the form \eqref{e:rk3}. Therefore, all solutions with (or without) vacuum are found as solutions to a finite system of equations for $A_0,A_1$, and $B_1$. We find solutions using a perturbative approach, starting with  bubbles, $B_1=0$, $A_1\sim 1$, and $A_0\sim 1$.
Substituting the ansatz $\left(A_0 + A_1 \cos (x) \cos (y)\right)_+$ into \eqref{e:MV_gen}-\eqref{e:MV_boundary_gen}, we find
\begin{align}\label{e:MV}
  A_0 + A_1 \cos (x) \cos (y) - \left(\frac{1}{\pi^2}+\mnew\right) &\iint_{\Omega_0} 2\cos(x-\xi)\cos(y-\eta) \left(A_0 + A_1 \cos (\xi) \cos (\eta)\right)d\xi d\eta - \rho = 0 \\
  \label{e:mass}\frac{1}{2\pi^2} &\iint_{\Omega_0} \left(A_0 + A_1 \cos (x) \cos (y) \right) dydx = 1 \\
   \label{e:boundary}
  &\qquad \ A_0 + A_1 \cos (x) \cos (y)  = 0 \text{ on $\partial \Omega_0$},
\end{align}
where 
\[
\Omega_0 = \left\{(x,y) \in \Omega \  \big| \ \ A_0 + A_1 \cos (x) \cos (y) > 0 \  \right\};
\]
see Fig.~\ref{fig:2d-sq-lattice}.
Assuming that there are vacuum regions, we have $A_1>A_0$, so that we may scale
\begin{align*}
  A_0 &= 1+a_0, \qquad A_1 = 1+ a_0 + z_1^2, \qquad a_0 = a_0(z_1),
\end{align*}
for a small parameter $z_1$. 
Our explicit ansatz then allows us to write the free-boundary condition \eqref{e:boundary} in polar coordinates about the center $(\pi, 0)$ of the bubble, as follows:
\begin{equation}\label{e:ell_boundary}
    \begin{split}
        0&= A_0 - A_1 \left ( 1- \frac{r^2}{2} + 
        \frac{r^4(3-\cos(4\theta))}{48}
        + \mathcal{O}(r^6) \right) \\
        &=\left( 1+a_0 \right)- \left( 1 + a_0 + z_1^2 \right) \left( 1-\frac{r^2}{2} + r^4 Q(\theta) + \mathcal{O}(r^6) \right), 
    \end{split}
\end{equation}
where $Q(\theta) := 
\frac{3-\cos(4\theta)}{48}.$
The inner radius $\ell$ of the bubble can be identified at leading order as the value of $r$ which solves \eqref{e:ell_boundary}. We may now further scale $\ell= \ell_1z_1$,
which yields, after some algebraic manipulation,
\begin{equation}
  0 =
  -\frac{1}{1+a_0+z_1^2} + \frac{\ell_1^2}{2} - \ell_1^4 z_1^2 Q(\theta) + \mathcal{O}(z_1^4) =: F(\ell_1,z_1, a_0, \theta).
\end{equation}
We find by the implicit function theorem that there exists a unique $\ell_1 = \ell_1(z_1,a_0,\theta)$ for $a_0, z_1$ sufficiently small, with $\ell_1(0,0,\theta) = \sqrt{2}$. 
Furthermore, expanding in $z_1,a_0$, we calculate 
\begin{equation}\label{e:ell_sq}
\ell_1(z_1,a_0,\theta) = \sqrt{2} - \frac{1+4Q(\theta) }{2} z_1^2 - \frac{1}{\sqrt{2} } a_0 + \mathcal{O}( z_1^4 + a_0^2).
\end{equation}
We can now use this expression for the boundary of the vacuum bubble to solve the mass constraint  \eqref{e:mass} for $a_0$. Substituting the expression from \eqref{e:ell_sq} into \eqref{e:mass}, we obtain
\begin{align*}
 0 & = 2\pi^2 - \iint_{\Omega} \left(A_0 + A_1 \cos (x) \cos (y) \right) dydx + \iint_{\Omega_0^c} \left(A_0 + A_1 \cos (x) \cos (y) \right) dydx \\
 &= 2\pi^2-2\pi^2(1 + a_0) + \iint_{\Omega_0^c} \left( 1+a_0 + \left( 1+ a_0 + z_1^2 \right)\cos (x) \cos (y) \right)dy dx \\
 & = -2\pi^2a_0 + \int_{0}^{2\pi} \int_0^{z_1\ell_1(z_1,a_0),\theta)} \left(- z_1^2 + \frac{r^2}{2} + \mathcal{O}(a_0r^2 + r^4 + r^2 z_1^2) \right) r \ drd\theta \\
 &= -2\pi^2 a_0 + \pi \left( -\half z_1^4
 \right) + \mathcal{O} (a_0z_1^4 + z_1^6).
\end{align*}
We can again solve for $a_0 = a_0(z_1)$ near $a_0 = z_1 = 0$ by the implicit function theorem, 
obtaining 
\[
a_0(z_1) = -\frac{1}{2\pi} z_1^4 +  \mathcal{O}(z_1^6).
\]
Finally, we use \eqref{e:MV} to relate these quantities to the parameter $\mnew$. Equating the coefficients of $\cos(x)\cos(y)$ in \eqref{e:MV}, we find
\begin{align*}
  0 =& A_1 - \left(\frac{1}{\pi^2} + \mnew\right)(A_1\pi^2) + \left(\frac{1}{\pi^2} + \mnew\right)\iint_{\Omega_0^c} 2\cos(\xi) \cos(\eta)\left( A_0 + A_1 \cos(\xi) \cos(\eta) \right) d \eta d \xi  \\
  = & -\mnew A_1\pi^2 \\
  &+ 2 \left(\frac{1}{\pi^2} + \mnew\right)\int_0^{2\pi} \int_0^{z_1\ell_1(a_0,z_1,\theta)}
  \left( -1+\frac{r^2}{2} + \mathcal{O}(r^4)\right)
  \left( A_0 + A_1
  \left( -1+\frac{r^2}{2} + \mathcal{O}(r^4)\right) \right) r d r d \theta  \\
  =& -\pi^2 \mnew \left( 1+ z_1^2 - \frac{1}{2\pi} z_1^4 \right) + \left( \frac{1}{\pi^2} + \mnew \right) \left( 2\pi z_1^4  \right) + \mathcal{O}(z_1^6). 
\end{align*}
We thus see that $\mnew$ scales like $z_1^4$. We solve for $z_1$ as a function of $\mnew^{1/4}$ near $z_1 = \mnew^{1/4} = 0$ by the implicit function theorem, obtaining 
\[
z_1 = \left(\frac{\pi^3}{2}\right)^{1/4}\mnew^{1/4} + \mathcal{O}(\mnew^{3/4}).
\]
Recalling that $\ell = \ell_1z_1 = \sqrt{2}z_1 + \mathcal{O}(z_1^3)$, this gives the expansion for $\ell$ as desired. 
The case of fissures is simpler and was studied in \cite{stevensscheel}.
\end{Proof}
\begin{Remark}[Maximal Isotropy --- square lattice]\label{r:isotropy}
Note that the theorem demonstrates existence of branches with maximal isotropy, but does not explicitly exclude solutions with submaximal isotropy; see \cite{GSS2,CLbook} for background. Here, the isotropy subgroup is understood as a subgroup $G$ of the group $\Gamma$ of translations and rotations that map functions with lattice periodicity to functions with lattice periodicity, in this case a semi-direct product of translations in $x$ and $y$, $S^1\times S^1$, and the lattice isotropy $D_4$. In addition to basic lattice translations, the isotropy of bubbles is $D_4$, the isotropy of fissures is generated by  $S^1$ and a reflection across fissures. Both are maximal subgroups of $\Gamma$. A rough calculation suggests that solutions with isotropy subgroups $G$ that are not maximal do not exist supercritically. We also tried to find those numerically both in a finite-rank approximation that we analyze here and in the full PDE adding some artificial viscosity and determined that solutions with $D_2$ isotropy do not exist close to $\tilde{\mu}=0$. See also Remark~\ref{r:ellb} for diffusive corrections and \S\ref{s:bs} and \S\ref{ss:2nd_vert_branch} for $D_2$-symmetric solutions at finite, non-small $\tilde{\mu}$. 
\end{Remark}

\subsection{Vacuum Formation: Hexagonal Lattice Symmetry}

We next consider the case of hexagonal symmetry. For convenience of notation, we define the scaled lattice vectors and dual lattice vectors
\begin{align}\label{e:hex_lattice_vecs}
  \mathbf{e}_1 = \left( 1,0 \right), \quad \mathbf{e}_2 = \left( -\frac{1}{2}, \frac{\sqrt{3}}{2}  \right) ,
  \quad \mathbf{e}_1^\star = \left( 1, \frac{1}{\sqrt{3} } \right), \quad \mathbf{e}_2^\star = \left( 0, \frac{2}{\sqrt{3} } \right),
\end{align}
such that for $i,j\in\{1,2\}$,
\begin{align}
  \langle \mathbf{e}_i, \mathbf{e}_j^\star \rangle = 
  \left\{\begin{array}{ll}0,& i\neq j\\1,& i=j\end{array}\right.,
\end{align}
The potential from \eqref{e:Vexpl} can then be written in the short form
\begin{align}\label{e:pote}
  V(x,y) &=
  \cos \left( \mathbf{e}_1^\star \cdot (x,y) \right) +
  \cos \left( \mathbf{e}_2^\star \cdot (x,y) \right) +
  \cos \left( (\mathbf{e}_1^\star - \mathbf{e}_2^\star ) \cdot (x,y) \right).
\end{align}

\begin{figure}
\begin{center}
 \begin{subfigure}[c]{0.09\textwidth}  \vskip 0pt
 \vspace*{-.15in}
  \includegraphics[width=\textwidth]{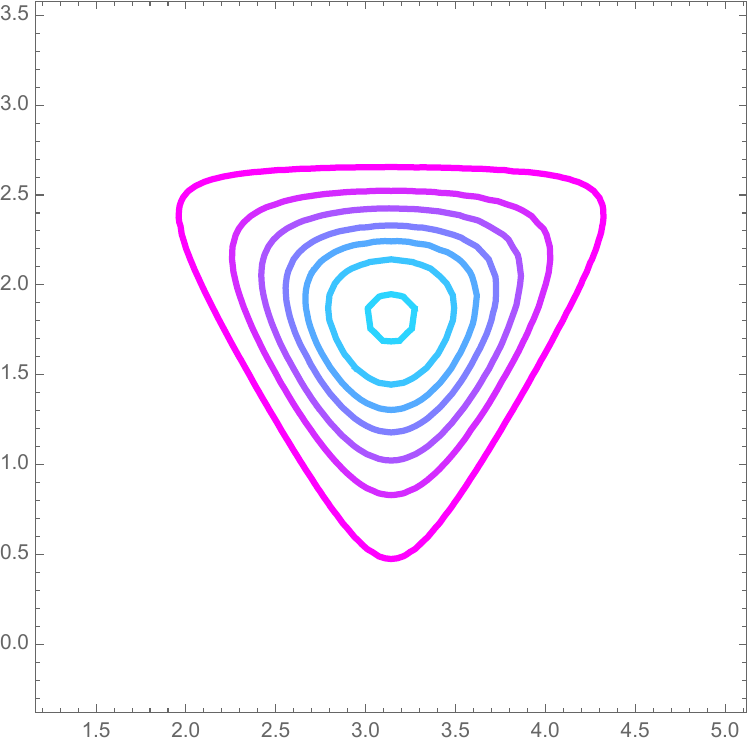}
 \end{subfigure}
 \begin{subfigure}[c]{0.69\textwidth}  \vskip 0pt
    \begin{tikzpicture}[scale=0.4]

     \def\sideLength{3.14159265359}
  \def \sqt{1.73205080757} 
  \def\circleRadius{0.5*1.189207115} 
    \definecolor{lightpink}{rgb}{1.0, 0.71, 0.76}
    \definecolor{lightpinkd}{rgb}{0.9, 0.6, 0.9}
    \definecolor{lightgreen}{rgb}{0.56, 0.93, 0.56}
    \definecolor{lightblue}{rgb}{0.4, 0.7, 1.0}
     \definecolor{lightblued}{rgb}{0, 0.65, 1.0}
      \definecolor{lightyellow}{rgb}{0.9, 1, .5}

    \def\sideLengthTwo{2*\sideLength/1.73205080757}

\fontsize{10}{10}\selectfont
\foreach \i in {0, 1, 2, 3, 4, 5} {
    \coordinate (K\i) at ({\sideLengthTwo*cos(60*\i + 30)}, {\sideLengthTwo*sin(60*\i + 30)});
}

\fill[lightyellow] (K0) -- (K1) -- (K2) -- (K3) -- (K4) -- (K5) -- cycle;

\foreach \i in {0, 1, 2, 3, 4, 5} {
    \coordinate (H\i) at ({\sideLength*cos(60*\i)}, {\sideLength*sin(60*\i)});
}

\fill[lightgreen] (H0) -- (H1) -- (H2) -- (H3) -- (H4) -- (H5) -- cycle;

\begin{scope}
\clip (K0) -- (K1) -- (K2) -- (K3) -- (K4) -- (K5) -- cycle;

\foreach \i in {0, 2, 4} {
    \filldraw[black, fill=lightblue] (K\i) circle (\circleRadius);
}
\foreach \i in { 1, 3, 5} {
    \filldraw[black, fill=lightblued] (K\i) circle (\circleRadius);
}
\end{scope}

\draw (H0) -- (H1) -- (H2) -- (H3) -- (H4) -- (H5) -- cycle;
\draw (K0) -- (K1) -- (K2) -- (K3) -- (K4) -- (K5) -- cycle;

\node at ($(K0) + (0.6,0)$) {$G_{1}$};
\node at ($(K1) + (0,0.55)$) {$G_{2}$};
\node at ($(K2) + (-0.6,0)$) {$G_{1}$}; 
\node at ($(K3) + (-0.6,0)$) {$G_{2}$}; 
\node at ($(K4) + (0,-0.55)$) {$G_{1}$}; 
\node at ($(K5) + (0.6,0)$) {$G_{2}$}; 

\draw[<->, thick] (-\sideLength/2,-\sideLength*\sqt/2+0.1) -- ++(\sideLength,0) node[midway, above] { $\pi$};
\draw[<->, thick] (\sideLengthTwo*\sqt/2 + 1.5,-\sideLengthTwo/2) -- ++(0,\sideLengthTwo) node[midway, right] {\large $\frac{2\pi}{\sqrt{3} }$};
\draw[->, thick] (0,0) -- (0,2/\sqt) node[above, above] {$\mathbf{e}_2^\star$};
\draw[->, thick] (0,0) -- (1,1/\sqt) node[right, right] { $\mathbf{e}_1^\star$};
\draw[->, opacity=0.5, thick] (0,0) -- (1,0) node[right, right] {$\mathbf{e}_1$};
\draw[->, opacity=0.5, thick] (0,0) -- (-1/2,\sqt/2) node[left, above] { $\mathbf{e}_2$};
\draw[->, thick] (0,0) -- (1,-1/\sqt) node[below,right] { $\mathbf{e}_1^\star - \mathbf{e}_2^\star$};

\draw[thick,loosely dotted] (0,0) -- (0,2*\sideLength/\sqt);
\draw[thick,loosely dotted] (0,0) -- (\sideLength,\sideLength/\sqt);
\draw[thick,loosely dotted] (0,0) -- (\sideLength, -\sideLength/\sqt);

\draw[<->, thick] (-\sideLengthTwo,-\sideLengthTwo-1) -- (\sideLengthTwo,-\sideLengthTwo-1) node[right] {$x$};
\draw[<->, thick] (-\sideLengthTwo-1,-\sideLengthTwo) -- (-\sideLengthTwo-1,\sideLengthTwo) node[above] {$y$};

\draw[thick] (0,-\sideLengthTwo-1.1) -- (0, -\sideLengthTwo-0.9) node[midway,below=4pt]{0};
\draw[thick] (-\sideLengthTwo-1.1,0) -- (-\sideLengthTwo-0.9,0) node[midway,left=4pt]{0};
\end{tikzpicture}\hspace*{0.15in}
    \begin{tikzpicture}[scale = 0.4]

 \def\sideLength{3.14159265359}
  \def \sqt{1.73205080757} 
  \def\circleRadius{0.55} 
    \definecolor{lightpink}{rgb}{.9, 0.6, 0.86}
    \definecolor{lightgreen}{rgb}{0.56, 0.93, 0.56}
    \definecolor{lightblue}{rgb}{0.4, 0.7, 1.0}
    \definecolor{lightyellow}{rgb}{0.9, 1, .5}

    \def\sideLengthTwo{2*\sideLength/1.73205080757}

\fontsize{10}{10}\selectfont
\foreach \i in {0, 1, 2, 3, 4, 5} {
    \coordinate (K\i) at ({\sideLengthTwo*cos(60*\i + 30)}, {\sideLengthTwo*sin(60*\i + 30)});
}

\fill[lightyellow] (K0) -- (K1) -- (K2) -- (K3) -- (K4) -- (K5) -- cycle;

\foreach \i in {0, 1, 2, 3, 4, 5} {
    \coordinate (H\i) at ({\sideLength*cos(60*\i)}, {\sideLength*sin(60*\i)});
}

\fill[lightgreen] (H0) -- (H1) -- (H2) -- (H3) -- (H4) -- (H5) -- cycle;

\begin{scope}
\clip (K0) -- (K1) -- (K2) -- (K3) -- (K4) -- (K5) -- cycle;

\end{scope}

\draw (H0) -- (H1) -- (H2) -- (H3) -- (H4) -- (H5) -- cycle;
\draw (K0) -- (K1) -- (K2) -- (K3) -- (K4) -- (K5) -- cycle;


\filldraw[black, fill=lightblue] (0,0) circle (\circleRadius);

\draw[<->, thick] (-\sideLength/2,-\sideLength*\sqt/2+0.1) -- ++(\sideLength,0) node[midway, above] { $\pi$};
\draw[<->, thick] (\sideLengthTwo*\sqt/2 + 1.5,-\sideLengthTwo/2) -- ++(0,\sideLengthTwo) node[midway, right] {\large $\frac{2\pi}{\sqrt{3} }$};
\draw[->, thick] (0,0) -- (0,2/\sqt) node[above, above] { $\mathbf{e}_2^\star$};
\draw[->, thick] (0,0) -- (1,1/\sqt) node[right, right] {$\mathbf{e}_1^\star$};
\draw[->, opacity=0.5, thick] (0,0) -- (1,0) node[right, right] { $\mathbf{e}_1$};
\draw[->, opacity=0.5, thick] (0,0) -- (-1/2,\sqt/2) node[left, above] {$\mathbf{e}_2$};
\draw[->, thick] (0,0) -- (1,-1/\sqt) node[below,right] { $\mathbf{e}_1^\star - \mathbf{e}_2^\star$};

\draw[thick,loosely dotted] (0,0) -- (0,2*\sideLength/\sqt);
\draw[thick,loosely dotted] (0,0) -- (\sideLength,\sideLength/\sqt);
\draw[thick,loosely dotted] (0,0) -- (\sideLength, -\sideLength/\sqt);

\draw[<->, thick] (-\sideLengthTwo,-\sideLengthTwo-1) -- (\sideLengthTwo,-\sideLengthTwo-1) node[right] {$x$};
\draw[<->, thick] (-\sideLengthTwo-1,-\sideLengthTwo) -- (-\sideLengthTwo-1,\sideLengthTwo) node[above] {$y$};

\draw[thick] (0,-\sideLengthTwo-1.1) -- (0, -\sideLengthTwo-0.9) node[midway,below=4pt]{0};
\draw[thick] (-\sideLengthTwo-1.1,0) -- (-\sideLengthTwo-0.9,0) node[midway,left=4pt]{0};
\end{tikzpicture}
    \end{subfigure} \hspace*{-.1in}
    \begin{subfigure}[c]{0.14\textwidth}  \vskip 0pt  \vspace*{-.2in}
  \includegraphics[width=\textwidth]{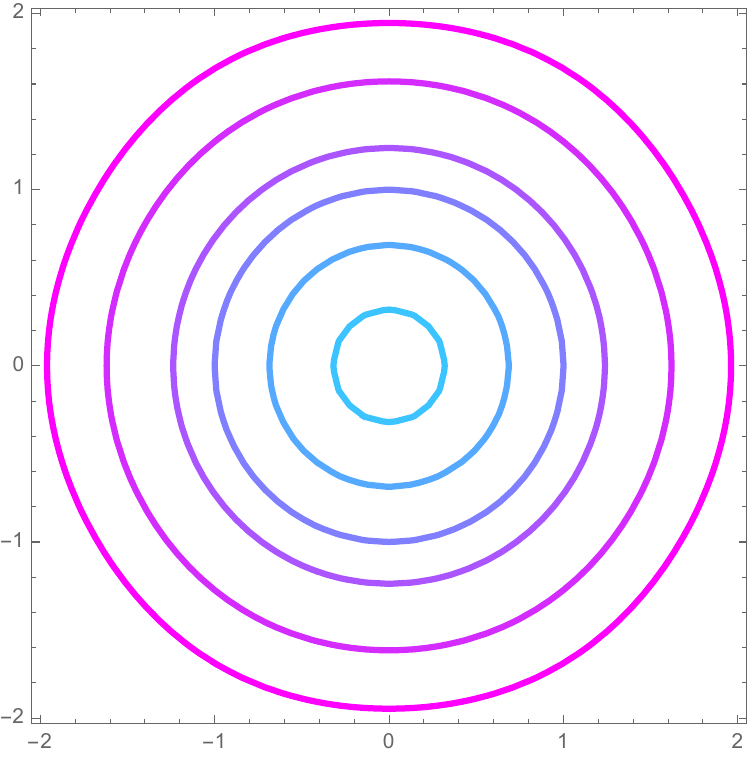}
 \end{subfigure} 

\end{center}
\caption{Inner figures (blue and green): An illustration of the vacuum bubbles on the hexagonal lattice. Left, triangles; right, hexagons. Blue regions represent areas of vacuum. In the case of triangles, bubbles are initially approximately circular, as pictured here; the triangular corrections become more apparent as $\mnew$ increases. All bubbles have (inner) radius $\ell$. Outer figures (pink and blue): example boundaries of triangle and hexagon vacuum regions from numerics, for various $\wt{\mu}$ (size not to scale with inner diagram). Triangles: $\wt{\mu} =$ 1e-5, 5e-5, 1e-4, 2.5e-4, 5e-4, 1e-3, 2e-3.  Hexagons: $\wt{\mu} =$ 5e-5, .001, .005, .01, .03, .07. 
}
  \label{fig:2d-hex-lattice}
\end{figure}
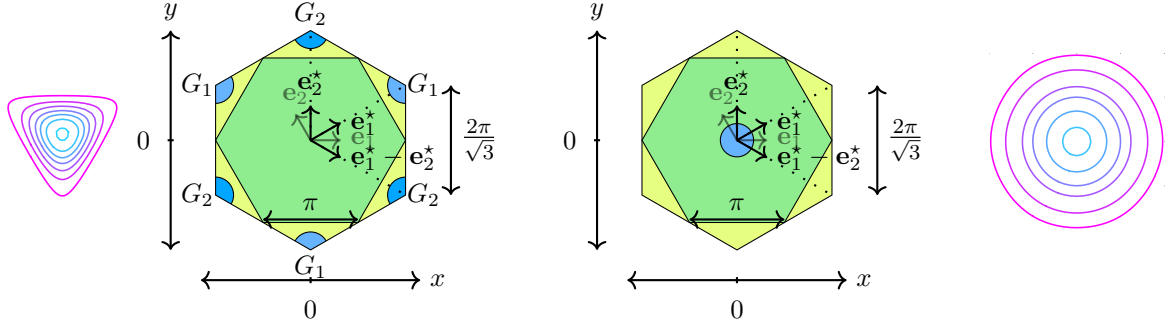

Reasoning equivalent to the case of squares will show that solutions can be found in a 3-dimensional reduced equation for the amplitudes of Fourier modes. Among those, we shall again find fissures and, similar to the bubbles in the square case, bubbles with dihedral isotropy $D_6$. Those bubbles arise in the form $u(x,y)=(A_0+A_1V(x,y))_+$. Different from the case of square lattices, however, the cases of $A_1>0$ and $A_1<0$ are different, leading to bubbles located in the corners of hexagons, taking an asymptotic shape of triangular corrections to circles, and bubbles with  hexagonal corrections to circular shapes, located at the centers of hexagons; see Fig.~\ref{fig:2d-hex-lattice}. We refer to the former patterns as triangles and to the latter ones as hexagons.

\begin{Theorem}[Vacuum region scaling --- hexagonal lattice]\label{t:hexlat}
    Let $V(x,y)$ as in \eqref{e:pote}, and $\morig = \frac{1}{\sqrt{3}\pi^2}+ \mnew$, with $\mnew$ sufficiently small. Then solutions $u$ to \eqref{e:MV_gen}-\eqref{e:MV_boundary_gen}, invariant under point reflections $u(x,y)=u(-x,-y)$ and with vacuum regions, are of the form
    \begin{equation}  \label{e:rk32}
        u(x,y) =\left( A_0 +  A_1  \cos \left( \mathbf{e}_1^\star \cdot (x,y) \right) +
        B_1 \cos \left( \mathbf{e}_2^\star \cdot (x,y) \right) +C_1
        \cos \left( (\mathbf{e}_1^\star - \mathbf{e}_2^\star ) \cdot (x,y) \right)\right)_+,
    \end{equation}
    where $f_+=\max(f,0)$. There are three solution branches, all supercritical, and parameterized by $\mnew\gtrsim 0$ as $A_0=A_0(\mnew),\, A_1 = A_1(\mnew),\, B_1 = B_1(\mnew),\, C_1 = C_1(\mnew)$, locally unique up to translations in $x$ and $y$:
    \begin{enumerate}
        \item \emph{fissures: } $B_1=C_1=0$, 
             \begin{align*}
            A_0  =1-\frac{\pi}{2}\mnew+\mathcal{O}(\mnew^2),\qquad
            A_1  =1+\left(\frac{3}{4\sqrt{2}}\pi^3\mnew\right)^{2/3}+\mathcal{O}(\mnew).
            \end{align*}
            The half-width of the fissures $\ell$ and area $\mathcal{A}$  of vacuum region are 
            \[
                \ell= \frac{3^{5/6}}{2^{4/3}}\pi\mnew^{1/3}+\mathcal{O}(\mnew^{2/3}),\qquad \mathcal{A}= \frac{3^{5/6}}{2^{1/3}}\pi^2\mnew^{1/3}+\mathcal{O}(\mnew^{2/3}).
            \]
        \item \emph{triangles: } $A_1=B_1=C_1>0$, 
            \begin{align*}
            A_0  = 1  -\frac{2\pi^2}{\sqrt{3}} \mnew   + \mathcal{O}(\mnew^{{3}/{2}}),\qquad
            A_1  =  \frac{2}{3} + \frac{4\pi^{3/2}}{3\sqrt{3}}\mnew^{{1}/{2}}  -\frac{4\pi^2 }{3\sqrt{3}}\mnew   + \mathcal{O}(\mnew^{{3}/{2}}).
            \end{align*}
            The inner radius  of bubbles $\ell$ and area $\mathcal{A}$  of vacuum region are 
            \[
                \ell= (12\pi^3 \mnew)^{1/4} + \mathcal{O}(\mnew^{{1}/{2}}),\qquad \mathcal{A}=2\sqrt{3}\pi^{5/2}\mnew^{1/2}+\mathcal{O}(\mnew).
            \] 
           \item \emph{hexagons: } $A_1=B_1=C_1<0$, 
            \begin{align*}
                A_0  = 1 -\frac{\pi^2}{18\sqrt{3}}\mnew+ \mathcal{O}(\mnew^{{3}/{2}}),\qquad
                A_1  =-\frac{1}{3} - \frac{\sqrt{2\pi^3}}{9}\mnew^{1/2} +\frac{\pi^2}{18\sqrt{3}}\mnew + \mathcal{O}(\mnew^{{3}/{2}}).
            \end{align*}
            The inner radius  of bubbles $\ell$ and area $\mathcal{A}$  of vacuum region are 
            \[
                \ell= (6\pi^3 \mnew)^{1/4} + \mathcal{O}(\mnew^{{3}/{4}}),\qquad \mathcal{A}=\sqrt{6}\pi^{5/2}\mnew^{1/2}+\mathcal{O}(\mnew).
            \] 
    \end{enumerate}
\end{Theorem}

\begin{Remark}[Vacuum area, hexagon versus triangle]
Accounting for the number of vacuum bubbles per unit hexagon (see Fig.~\ref{fig:2d-hex-lattice}), one can compare the areas of vacuum in the triangles and hexagon cases, to find
 \begin{align*}
    \textrm{(hexagons)}&   &&A_\textrm{hex}   = \sqrt{6}\pi^{{5}/{2}}\mnew^{1/2} , \\
  \textrm{(triangles)}&    &&A_\textrm{tri}  = 4\sqrt{3}\pi^{{5}/{2}}\mnew^{1/2}  =2\sqrt{2}A_\textrm{hex} .&&
 \end{align*}
 The triangle solutions thus possess not only a larger maximum density than the hexagons, but a larger area of vacuum. 
  \end{Remark}
We shall prove existence and establish expansions for triangles and hexagons in the next two sections. Existence and expansion for fissures again follow from the one-dimensional case \cite{stevensscheel}.
Similar to the square case, Remark~\ref{r:isotropy}, the branches found here have maximal isotropy.

\begin{Remark}[Polygonal shapes]
    One can easily plot the sublevel sets of the potential $V$ and its negative to identify the shape of vacuum regions; compare Fig.~\ref{f:pot}. In the case of square lattices, the vacuum bubbles are asymptotically discs with small square-symmetric corrections. This is also apparent from the formula \eqref{e:ell_boundary}. In the case of the hexagonal lattice, the corrections to the shape are triangular and hexagonal, respectively; see again  Fig.~\ref{f:pot} and~\ref{fig:2d-hex-lattice}. 
\end{Remark}

\begin{Remark}[Maximal Isotropy --- hexagonal lattice]\label{r:isotropy2}
Similar to the case of the square lattice, we also investigated the possibility of solutions with submaximal isotropy. We did not find such solutions either in the finite-rank approximation or numerically including small viscosity. We particularly investigated the possibility of triangles, which notably do not possess the point reflection symmetry $(x,y)\mapsto (-x,-y)$, with isotropy $D_3$, submaximal and did not find such solutions in this finite-rank approximation. We also did not find those numerically, nor does there appear to exist a secondary branch of triangles as sometimes observed in bifurcations with this symmetry group of the hexagonal lattice; see Fig.~\ref{fig:rank-3-sq} 
\end{Remark}
\subsubsection{Proof of Theorem~\ref{t:hexlat} (ii) --- triangles}
We begin with the case $A_1 > 0$ and proceed analogously to the proof of Theorem~\ref{t:square}. We scale 
    \begin{align*}
  A_0 &= 1+a_0, \qquad A_1 = \frac{2}{3}(1+ a_0 + z_1^2), \qquad a_0 = a_0(z_1),
\end{align*}
for a small parameter $z_1$. 
The free-boundary condition \eqref{e:MV_boundary_gen}, expanded in polar coordinates about the centers of the bubbles, gives 
\begin{equation}\label{e:bdy_hex1}
  0 = 1+a_0 + \frac{2}{3}\left(1+a_0+z_1^2\right)\left(
    -\frac{3}{2} + \frac{r^2}{2} \pm \ \frac{1}{6} \sin(3\theta) r^3 - \frac{r^4}{24} + \mathcal{O}(r^5)\right),
\end{equation}
where the minus sign is taken for bubbles labeled $G_1$ and the plus sign is used for bubbles labeled $G_2$ (see Fig.~\ref{fig:2d-hex-lattice}). We identify the bubble radius $\ell$ as the solution $r$ to \eqref{e:bdy_hex1}, and further scale $\ell = \ell_1z_1$. In this scaling, after some algebraic manipulation, \eqref{e:bdy_hex1} becomes
 \begin{equation}
    \label{e:ell_hex1}
    0 = - \frac{3}{2(1+a_0+z_1^2)} + \frac{\ell_1^2}{2}+ z_1\ell_1^3Q(\theta) 
  - \frac{z_1^2\ell_1^4}{24} + \mathcal{O}(z_1^5),
  \end{equation}                                                                         with $Q(\theta) = \pm\frac{1}{6} \sin(3\theta).$  The implicit function theorem then yields a unique $\ell_1 = \ell_1(z_1,a_0,\theta)$, for $z_1, a_0$ sufficiently small and for arbitrary $\theta$. Expanding in $z_1, a_0$, we find 
     \begin{align*}
    \ell_1(z_1,a_0,\theta)&= \sqrt{3} - 3Q(\theta)z_1 +\left(\frac{45Q(\theta)^2}{2\sqrt{3} }-\frac{9}{8\sqrt{3} }\right) z_1^2 - \frac{\sqrt{3} }{2} a_0 + \mathcal{O}\left( |z_1^3| + |a_0z_1| + |a_0^2| \right).
  \end{align*}
  Again, we can use this expression for $\ell_1(z_1,a_0,\theta)$ to solve for $a_0$ using the average-density condition \eqref{e:MV_density_gen}. Substituting, we obtain
  \begin{align*}
    0 &= 2\sqrt{3}\pi^2 - \left(\iint_{\Omega}\left(A_0 + A_1V(x,y)\right)dydx - \iint_{
    \Omega_0^c}\left(A_0 + A_1V(x,y)\right)dydx\right)\\
       &= -2\sqrt{3}\pi^2\cdot a_0 + \iint_{\Omega_0^c}\left(\left(1 + a_0\right) + \frac{2}{3}(1 + a_0 + z_1^2)\left(2\cos (x) \cos\left( \frac{y}{\sqrt{3} } \right) + \cos \left( \frac{2y}{\sqrt{3} }\right) \right) \right)dydx \\
    &=-2\sqrt{3}\pi^2\cdot a_0 + 2\int_0^{2\pi}\int_0^{z_1\ell_1(z_1,a_0,\theta)} \left[(1+a_0)\frac{r^2}{3}
   + z_1^2\left(
    -1 + \frac{r^2}{3} 
    \right)+ A_1Q(\theta)r^3 + \mathcal{O}(r^4)\right]r \ drd\theta \\
    &= -2\sqrt{3}\pi^2\cdot a_0 - 3\pi(1-a_0) z_1^4 + \mathcal{O} \left( z_1^6 \right),
  \end{align*}
  and solve by the implicit function theorem to find, for $z_0$ sufficiently small,
  \begin{equation}
       a_0 = -\frac{\sqrt{3} }{2\pi} z_1^4 + \mathcal{O} \left( z_1^6 \right).
  \end{equation}
  Lastly, we use \eqref{e:MV_gen} to relate these quantities to $\mnew$:
    \begin{align*} 
  0 =& V(x,y)A_1- \left(\frac{1}{\sqrt{3}\pi^2} + \mnew\right)\iint_{\Omega -\Omega_0^c} V(x-\xi,y-\eta) \left( A_0 + A_1 \cdot V(\xi,\eta) \right) d\xi d\eta  \\
   =& V(x,y)\left(A_1- \left(\frac{1}{\sqrt{3}\pi^2} + \mnew\right)(A_1\sqrt{3}\pi^2)\right) \\
  &+ \left(\frac{1}{\sqrt{3}\pi^2} + \mnew\right)\iint_{\Omega_0^c} V(x-\xi,y-\eta) \left( A_0 + A_1 \cdot V(\xi,\eta) \right) d\xi d\eta.
   \end{align*} 
Equating coefficients of $V(x,y)$, using polar coordinates for the integral over the bubbles, we find
    \begin{align*} 
0 =& -\sqrt{3}\mnew\pi^2A_1 + \left(\frac{1}{\sqrt{3}\pi^2} + \mnew\right)\iint_{\Omega_0^c}\frac{1}{3}V(\xi,\eta)(A_0+A_1 V(\xi,\eta))d\xi d\eta\\
 =& -\sqrt{3}\mnew\pi^2A_1 + 2\left(\frac{1}{\sqrt{3}\pi^2} + \mnew\right)\int_0^{2\pi}\int_0^{z_1\ell_1}\left( \frac{1}{2}z_1^2 - \frac{1}{6}r^2 -\frac{Q(\theta)}{3}r^3 + \mathcal{O}(r^4,r^2z_1^2) \right)r \ dr d\theta \\
  =& -\frac{2}{\sqrt{3}} \mnew \pi^2 \left(1 + z_1^2 -\frac{\sqrt{3} }{2\pi} z_1^4 \right) + \left(\frac{1}{\sqrt{3}\pi^2} + \mnew\right)\left(\frac{3\pi}{2}z_1^4 + \mathcal{O}(z_1^6)\right). 
  \end{align*}
  As in the square symmetry case for bubbles, we find that $z_1$ scales like $\mnew^{{1}/{4}}$, and we obtain a unique $z_1 = z_1(\mnew^{{1}/{4}})$, for $\mnew^{1/4}$ sufficiently small, by the implicit function theorem, with
  \begin{equation}
      z_1 = \left(\frac{4\pi^3}{3}\mnew\right)^{1/4} + \mathcal{O}(\mnew^{1/2}).
  \end{equation}
  Using the relation $\ell = \sqrt{3}z_1 +\mathcal{O}(z_1^2)$, this gives us the expression in Theorem~\ref{t:hexlat} as desired. 

\subsubsection{Proof of Theorem~\ref{t:hexlat} (iii) --- hexagons}

We now turn to the case $A_1<0$. 
and proceed analogously. We scale
\begin{align*}
    A_0 &= 1+a_0, \qquad A_1 =- \frac{1}{3}\left(1 + a_0 + z_1^2\right), \qquad a_0 = a_0(z_1),
\end{align*}
for a small parameter $z_1$. 
Setting the solution equal to 0 in the free-boundary equation \eqref{e:MV_boundary_gen} and expanding in polar coordinates yields 
\begin{equation}\label{e:bdy_hex2}
0= 1+a_0 - \frac{1}{3}\left(1+a_0 +z_1^2\right) \left( 3 - r^2 + \frac{r^4}{12} + \mathcal{O}(r^6) \right),
\end{equation}
where $r$ in \eqref{e:bdy_hex2} will be the radius $\ell$ of the bubble. Scaling $\ell = \ell_1z_1$, and simplifying, we find
\begin{equation}\label{e:ell_hex2}
    0 = \frac{3}{1+a_0 + z_1^2} - \ell_1^2 + \frac{\ell_1^4 z_1^2}{12} + \mathcal{O}(\ell_1^6 z_1^4),
\end{equation}
which we can solve by the implicit function theorem to obtain $\ell_1 = \ell_1(z_1,a_0,\theta)$, for $z_1, a_0$ small, and for arbitrary $\theta$. Expanding in $z_1, a_0$, we find
\begin{equation}
\ell_1 =\sqrt{3} -\frac{3\sqrt{3}}{8}z_0^2 -\frac{\sqrt{3}}{2}a_0 + \mathcal{O}(a_0^2 + z_1^4).
\end{equation}
 We substitute this expression for $\ell(z_1,a_0,\theta)$ to solve for $a_0$ in the average-density condition \eqref{e:MV_density_gen},
    \begin{align*} 
    0 &= 2\sqrt{3}\pi^2 - \iint_{\Omega}\left(A_0 + A_1V(x,y)\right)dydx + \iint_{
    \Omega_0^c}\left(A_0 + A_1V(x,y)\right)dydx\\
      &= -2\sqrt{3}\pi^2\cdot a_0 + \iint_{\Omega_0^c}\left(\left(1 + a_0\right) -\frac{1}{3}(1 + a_0 + z_1^2)\left(2\cos (x) \cos\left( \frac{y}{\sqrt{3} } \right) + \cos \left( \frac{2y}{\sqrt{3} }\right) \right) \right)dydx \\
    &=-2\sqrt{3}\pi^2\cdot a_0 + \int_0^{2\pi}\int_0^{z_1\ell_1(z_1,a_0,\theta)} \left[(1+a_0)\frac{r^2}{3}  
   + z_1^2\left(
    -1 + \frac{r^2}{3}  
    \right)+ \mathcal{O}(r^4)\right]r \ drd\theta \\
    &= -2\sqrt{3}\pi^2\cdot a_0 - \frac{3\pi}{2}(1-a_0) z_1^4 + \mathcal{O} \left( z_1^6 \right),
  \end{align*}
  and solve by the implicit function theorem to find, for $z_0$ sufficiently small,
  \begin{equation}
       a_0(z_1) = -\frac{\sqrt{3} }{4\pi} z_1^4 + \mathcal{O} \left( z_1^6 \right).
  \end{equation}
    Lastly, we use \eqref{e:MV_gen} to relate previous quantities to $\mnew$:
    \begin{align*} 
      0 =& V(x,y)A_1- \left(\frac{1}{\sqrt{3}\pi^2} + \mnew\right)\iint_{\Omega -\Omega_0^c} V(x-\xi,y-\eta) \left( A_0 + A_1 \cdot V(\xi,\eta) \right) d\xi d\eta  \\
     =& V(x,y)\left(A_1 - \left(\frac{1}{\sqrt{3}\pi^2} + \mnew\right)(A_1\sqrt{3}\pi^2) \right)\\
    &+ \left(\frac{1}{\sqrt{3}\pi^2} + \mnew\right)\iint_{\Omega_0^c} V(x-\xi,y-\eta) \left( A_0 - A_1 V(\xi,\eta) \right) dxdy. 
    \end{align*}
    Equating coefficients of $V(x,y)$, using polar coordinates for the integral over the bubble, we calculate
    \begin{align*}
      0 =& -\sqrt{3}\mnew\pi^2A_1 + \left(\frac{1}{\sqrt{3}\pi^2} + \mnew\right)\iint_{\Omega_0^c}\frac{1}{3}V(\xi,\eta)(A_0+A_1 V(\xi,\eta))d\xi d\eta\\
  =& \frac{\mnew\pi^2}{\sqrt{3}}\left(1 + z_1^2 -\frac{\sqrt{3} }{4\pi} z_1^4 \right) + \left(\frac{1}{\sqrt{3}\pi^2} + \mnew\right)\int_0^{2\pi}\int_0^{z_1\ell_1}\left( -z_1^2 + \frac{r^3}{3} + \mathcal{O}(r^4,  r^2z_1^2) \right)r \ dr d\theta \\
  =&\frac{\mnew\pi^2}{\sqrt{3}}\left(1 + z_1^2 -\frac{\sqrt{3} }{4\pi} z_1^4 \right) + \left(\frac{1}{\sqrt{3}\pi^2} + \mnew\right)\left(-\frac{3\pi}{2}z_1^4 + \mathcal{O}(z_1^6)\right). 
  \end{align*}

 Again, we find that $z_1$ scales like $\mnew^{1/4}$, and we obtain a unique $z_1 = z_1(\mnew^{1/4})$, for $\mnew^{1/4}$ sufficiently small, by the implicit function theorem, with
  \begin{equation}
      z_1 = \left(\frac{2\pi^3}{3}\mnew\right)^{1/4} + \mathcal{O}(\mnew^{3/4}).
  \end{equation}
  Using the relation $\ell = \sqrt{3}z_1 + \mathcal{O}(z_1^3)$, this gives us the expression in Theorem~\ref{t:hexlat} as desired.

\section{Diffusive Corrections: Almost-Vertical Branches}\label{s:diffusion}

In this section, we consider the system with small diffusion,
\begin{equation}\label{e:V_diffusion}
    u_t = \ep \Delta u + \nabla \cdot \left(u  \,\nabla \left(u - \left(\mu_*+\mnew\right)(V* u)\right)\right),
\end{equation}
representing small noise in the particle dynamics. We study the fate of the vertical solution branches, 
and find that in the presence of noise, as in \cite{stevensscheel}, the vertical bifurcation branches become almost-vertical branches, with deviation from the vertical branch depending on the strength of $\ep$. Different from the one-dimensional case, we find that the direction of bending from the vertical branch depends on the solution. 
The argument is perturbative and allows for the calculation of expansions of solutions. It also  predicts $\mathcal{O}(\eps)$ noise-induced hysteresis. In addition, the perturbed bifurcation diagram allows for a direct stability analysis, which then lets us analyze the competition between bubbles and fissures for small $\ep$ through a center-manifold expansion. 

\subsection{Square lattice symmetry}\label{ss:diffusion_sq}
We begin with the case of the square lattice.
\begin{Proposition}[Diffusive corrections --- square bubbles]\label{e:diffusion_gaps_B=0}
    For $\eps\gtrsim 0$, the McKean-Vlasov equation 
    \begin{equation}\label{e:V_diffusion_sq}
    u_t = \ep \Delta u + \nabla \cdot \left(u \, \nabla\left( u - \left(\frac{1}{\pi^2}+\mnew\right)(V*u)\right)\right),
\end{equation}
possesses an almost-vertical branch of stationary solutions $u_*(x,y;A,\ep), 0 < A < 1$,
$D_4$-symmetric and periodic, for $\mnew = \mnew_*(A,\ep)$. Normalizing average density to 1, it has the expansion
    \begin{equation}
        \mnew_*(A,\ep) = \ep \mnew_1(A)  + \mathcal{O}(\ep^2), \qquad u_*(x,y;A,\ep) = 1 + A\cos(x)\cos(y) + \mathcal{O}(\ep),
    \end{equation}
    where 
    \begin{equation}  
        A=\frac{1}{\pi^2}\int_{-\pi}^\pi \int_{-\pi}^\pi u_*(x,y)\cos(x)\cos(y)dxdy, 
    \end{equation}
    and, explicitly,  
    \begin{equation}\label{e:mu_pred_sq}
        \mnew_1(A) =\frac{2}{\pi^3A^2}  \left(2  \pi - 
     2  \left(\textrm{E}(A^2) + \sqrt{1 - A^2}  \textrm{E}\left(A^2/(-1 + A^2)\right)\right)\right),  \end{equation} 
     where $\textrm{E}(k)$ is the complete elliptic integral, $E(k) = \int_{0}^{\frac{\pi}{2}}\sqrt{1-k^2 \sin^2\theta}d\theta$. The limits of $\mnew_1(A)$ are $\mnew_1(0) = \frac{1}{\pi^2}, \mnew_1(1) = \frac{4}{\pi^2}-\frac{8}{\pi^3}.$
\end{Proposition}

\begin{Remark}[Diffusive corrections --- square fissures]
    For fissures, $\mnew = \mnew_1\ep + \mathcal{O}(\ep^2)$, 
    \[\mnew_1(A) = \frac{4\pi^2\left(\frac{1-\sqrt{1-A^2}}{A^2}\right)}{2\pi^4} = \frac{2}{\pi^2}\left(\frac{1-\sqrt{1-A^2}}{A^2}\right), \text{ with limits } \mnew_1(0) = \frac{1}{\pi^2}, \mnew_1(1)=\frac{2}{\pi^2}.
    \]
    The proof in this case is analogous to the proof in \cite{stevensscheel}.
\end{Remark}

\begin{Proof}
    Noting that $\Delta u = \nabla \cdot \left(u\, \nabla (\log u)\right)$, we can write the fixed point equation in the form
    \begin{equation}
        0 = \nabla \cdot \left(u  \,\nabla \left(\ep \log u +  u - \left(\frac{1}{\pi^2}+\mnew\right)(V*u)\right)\right).
    \end{equation}
    Restricting to strictly positive solutions $u$, this is equivalent to 
    \begin{equation}
        \ep \log u +  u - \left(\frac{1}{\pi^2}+\mnew\right)(V*u) = \rho,
    \end{equation}
    for $\rho$ a constant. 
    We append normalization conditions, solve for $\rho$, and define $F,F_0,F_1$ through
    \begin{equation} \label{e:Fsys1} 
        \begin{split}
            F(u,\mnew,\ep) &:= \tilde{F}(u,\mnew,\ep)-\rho,\quad   \tilde{F}(u,\mnew,\ep)=\ep \log u +  u - \left(\frac{1}{\pi^2}+\mnew\right)(V*u),  \quad  \rho=\frac{1}{4\pi^2}\iint_\Omega  \tilde{F}  \\
            F_0(u) &:= \int_{-\pi}^\pi \int_{-\pi}^\pi u(x,y)dxdy - 4\pi^2 \\
              F_1(u) &:= \int_{-\pi}^\pi \int_{-\pi}^\pi u(x,y)\cos(x)\cos(y)dxdy - A\pi^2,
        \end{split}
    \end{equation}
    where the parameter $A$ is fixed, $|A| < 1$. The system \eqref{e:Fsys1} defines a map 
    \[
        G(u,\mnew,\ep)=(F,F_0,F_1)(u,\mnew,\ep): \left( H^2_{\mathrm{e,p}} \right) \times \R^2 \to \overset{\circ}{L}{}^2_{\mathrm{e,p}} \times \R^2,
    \]  
    where the subscripts $\{\mathrm{e},\mathrm{p}\}$ indicate the restriction of function spaces to $D_4$-symmetric, periodic functions, and $\overset{\circ}{L}{}^2$ indicates the restriction to zero average. 
    One verifies that $G$ is well-defined and smooth, and 
    \[
    G(u_*^0(\cdot,\cdot;A),0,0) = 0, \quad \textrm{ for } \ \  u_*^0(x,y;A) = 1 + A\cos(x)\cos(y).
    \]
    The derivative at $(u_*^0,0,0)$ is given by 
    \[
    \mathcal{A} = \begin{pmatrix} \mathcal{L} & \partial_\mnew F_* & \partial_\ep F_* \\
    \langle 1, \cdot \rangle & 0 & 0 \\
    \langle \cos(x)\cos(y), \cdot\rangle & 0 & 0
    \end{pmatrix},
    \]
    where 
    \begin{align*}
        \mathcal{L}u = u - \left(\frac{1}{\pi^2}+\mnew\right)(V*u),\qquad
        \partial_\mnew F_* =  -\pi^2A\cos(x)\cos(y),\qquad
        \partial_\ep F_*  = \log(1 + A\cos(x)\cos(y)).
    \end{align*}
    The operator $\mathcal{L}$ is strictly elliptic since $|A| < 1$, and self-adjoint with domain $H^2_{\textrm{e,p}}$. It is therefore Fredholm index 1 due to the restriction of the codomain to average 0 functions.
    Bordering lemmas for Fredholm operators then imply that $\mathcal{A}$ is also Fredholm of index 1, so that we expect a 1-dimensional family of solutions. 

    We turn ourselves to showing that the map $D_{u,\mnew}G(u_*^0(\cdot,\cdot;A),0,0)$, that is, the first two columns of $\mathcal{A}$, is invertible. 
    Clearly, the kernel of $\mathcal{L}$ is spanned by
    $\{1, \cos(x)\cos(y) \}$, and the cokernel, by restriction to average zero functions, is spanned by $\{ \cos(x)\cos(y) \}$. 

    As a consequence, we find that $\partial_\mnew F_* \notin \textrm{Rg}(\mathcal{L})$, since 
    \[
    \langle\pi^2A \cos(x)\cos(y), \cos(x)\cos(y) \rangle = \pi^4A \neq 0. 
    \]
    We can thus see that $\mathcal{A}(u,\mnew,0) = 0$ if and only if $ u = 0, \mnew = 0$, since the second and third components of $\mathcal{A}$ imply $u \notin \ker(\mathcal{L})$, and therefore $\mathcal{L}u \neq \partial_\mnew F_* \mnew$, unless $u = 0, \mnew = 0$. We therefore have that $D_{u,\mnew}G(u_*^0,0,0)$ is invertible, and we can solve for $u, \mnew$ as functions of $\ep$. We then obtain the leading order expansion of $\mnew$ by projecting the first equation onto the cokernel. 
    We evaluate 
    \begin{align*}
    \langle \partial_\ep F_*, \cos(x)\cos(y)\rangle &= \int_{-\pi}^\pi \int_{-\pi}^\pi \cos(x)\cos(y)\cdot \log\left(1 + A\cos(x)\cos(y)\right)dxdy \\
    &= \frac{2\pi}{A}  \left(2  \pi - 
     2  \left(\textrm{E}(A^2) + \sqrt{1 - A^2}  \textrm{E}(A^2/(-1 + A^2))\right)\right) \neq 0,
    \end{align*}
    and thus find 
    \begin{align*}
        \mnew &= \mnew_1\ep + \mathcal{O}(\ep^2), \\
        \mnew_1 &= -\frac{\left\langle \partial_\ep F_*,\cos(x)\cos(y) \right\rangle }{\left\langle \partial_\mnew F_*,\cos(x)\cos(y) \right\rangle } = \frac{2}{\pi^3A^2}  \left(2  \pi - 
     2  \left(\textrm{E}\left(A^2\right) + \sqrt{1 - A^2}  \textrm{E}\left(A^2/(-1 + A^2)\right)\right)\right).
    \end{align*}
\end{Proof}

\begin{Remark}[Diffusive corrections --- no elliptical bubbles]\label{r:ellb}
    If one were to attempt to repeat the argument with $u_*^0 = 1 + A\cos (x) \cos (y) + B\sin(x)\sin(y) , $ for $B \neq A, \ \  A,B \neq 0$, one would not find nontrivial solutions for $\ep > 0$. 
    The corresponding nonlinear function
    \begin{equation}
     \begin{split}
            \widetilde{F}(u,\mnew,\ep) &:= \ep \log u +  u - \left(\frac{1}{\pi^2}+\mnew\right)(V*u) - \rho \\
            \widetilde{F}_0(u) &:= \int_{-\pi}^\pi \int_{-\pi}^\pi u(x,y)dxdy - 4\pi^2 \\
              \widetilde{F}_1(u) &:= \int_{-\pi}^\pi \int_{-\pi}^\pi u(x,y)\cos(x)\cos(y)dxdy - A\pi^2 \\
              \widetilde{F}_2(u) &:=\int_{-\pi}^\pi \int_{-\pi}^\pi u(x,y)\sin(x)\sin(y)dxdy - B\pi^2 ,
        \end{split}
        \end{equation}
where $\rho$ is the average of the other terms, defines a map
    $\widetilde{G}(u,\mnew,\ep): H^2_\mathrm{s,p}\times \R^2 \to \overset{\circ}{L}{}^2_{\mathrm{s,p}}  \times \R^2$ ,
    where we now can restrict function spaces to $D_2$-symmetric, periodic functions, only, generated by reflections $x\leftrightarrow y$ and rotations by $\pi$, denoted by subscripts $\mathrm{s,p}$. It 
    has linearization
      \[
    \mathcal{\widetilde{A}} = \begin{pmatrix} \mathcal{L} & \partial_\mnew F_* & \partial_\ep F_* \\
    \langle 1, \cdot \rangle & 0 & 0 \\
    \langle \cos(x)\cos(y), \cdot\rangle & 0 & 0 \\
     \langle \sin(x)\sin(y), \cdot\rangle & 0 & 0 
    \end{pmatrix},
    \]
    which is no longer Fredholm index 1, but Fredholm index 0 due to the additional normalization condition.  

    In order to determine whether $\mathcal{\widetilde{A}} $ possesses a kernel, we need to check whether the two vectors $\partial_\mnew F_* = A\cos(x)\cos(y)+B\sin(x)\sin(y)$ and $\partial_\ep F_* = \log(1 + A\cos(x)\cos(y)+B\sin(x)\sin(y))$ span a complement of the range, that is, if their orthogonal projections on the basis $\{\cos(x)\cos(y), \sin(x)\sin(y)\}$ are linearly independent.

    
    Evaluating the scalar products and changing integration variables leads to the condition
    \[
        \int_{\eta=0}^{2\pi}  \int_{\xi=0}^{2\pi} 
        (A\cos\xi - B \cos\eta)\log(1+A\cos\eta+B\cos\xi)d\xi d\eta\neq 0,
    \]
    which after partial integration simplifies to 
        \[
        AB\int_{\eta=0}^{2\pi}  \int_{\xi=0}^{2\pi}  \frac{\cos^2\xi-\cos^2\eta}{1+A\cos\eta+B\cos\xi} d\xi d\eta\neq 0.
    \]
    This integral clearly vanishes when $A=B$. We evaluated the integral numerically and found that it is strictly monotone on lines $A+B=const$, hence  nonzero for $A\neq B$.     
    As a consequence,  $\mathcal{\widetilde{A}} $  is invertible and the solution branch at $\eps=0$ is unique, that is, there are no solutions with submaximal isotropy for $\eps\gtrsim 0$. This is a notable difference from the $\ep = 0$ case, where the vertical branch
    does contain mixed-mode solutions. 
\end{Remark}
\subsection{Hexagonal Lattice Symmetry}\label{ss:diffusion_hex}
We now turn to the case of the hexagonal lattice.
\begin{Proposition}[Diffusive corrections --- triangles and hexagons]\label{e:diffusion_bubbles_hex}
    For $\eps\gtrsim 0$, the McKean-Vlasov equation 
    \begin{equation}\label{e:V_diffusion_hex}
        u_t = \ep \Delta u + \nabla \cdot \left(u\,\nabla \left( u - \left(\frac{1}{\sqrt{3}\pi^2}+\mnew\right)(V*u)\right)\right)
    \end{equation}
    possesses an almost-vertical 
    branch of stationary solutions $u_*(x,y;A,\ep)$ with $D_6$ isotropy and average 1, parameterized by  $-\frac{1}{3}  < A < \frac{2}{3}$, $A \neq 0$, for $\mnew = \mnew_*(A,\ep)$ with expansion
    \begin{equation}
        \mnew_*(A,\ep) = \ep \mnew_1(A)  + \mathcal{O}(\ep^2), \qquad u_*(x,y;A,\ep) = 1 + AV(x,y) + \mathcal{O}(\ep),
    \end{equation}
    where 
    \begin{equation}
       A=  \frac{1}{3\sqrt{3}\pi^2} \iint_{\Omega} u_*(x,y)V(x,y)dxdy , \\
    \end{equation}
   and, explicitly, 
   \begin{equation}\label{e:mu1hex}
       \mnew_1(A)=\frac{1}{3\pi^3 A^2}\int_{-\frac{\sqrt{3}\pi}{2}}^{\frac{\sqrt{3}\pi}{2}}\left(1 - \sqrt{\left(1+A\cos\left(\frac{2y}{\sqrt{3}}\right)\right)^2 -4A^2\cos^2\left(\frac{y}{\sqrt{3}}\right) }\right)dy.
   \end{equation}
   The limits of $\mnew_1(A)$ are 
          $\mnew_1(0) = \frac{1}{\sqrt{3}\pi^2}, \mnew_1(\frac{2}{3}) = \frac{2}{\sqrt{3}\pi^2}-\frac{3}{2\pi^3}$, $\mnew_1(-\frac{1}{3}) = \frac{5}{\sqrt{3}\pi^2}-\frac{6}{\pi^3}$.
\end{Proposition}

\begin{Remark}[Diffusive corrections --- hexagonal fissures]
For fissures in this setting, we find 
\[\mnew_1 = \frac{2   (1 - \sqrt{1 - A^2})}{{\sqrt{3}A^2\pi^2}}, \quad \text{ with limits } \mnew_1(0) = \frac{1}{\sqrt{3}\pi^2},\  \mnew_1(1) = \mnew_1(-1) = \frac{2}{\sqrt{3}\pi^2},
\]
simply adapting the results in \cite{stevensscheel}.
\end{Remark}
\begin{figure}
    \centering
    \includegraphics[width=0.49\textwidth]{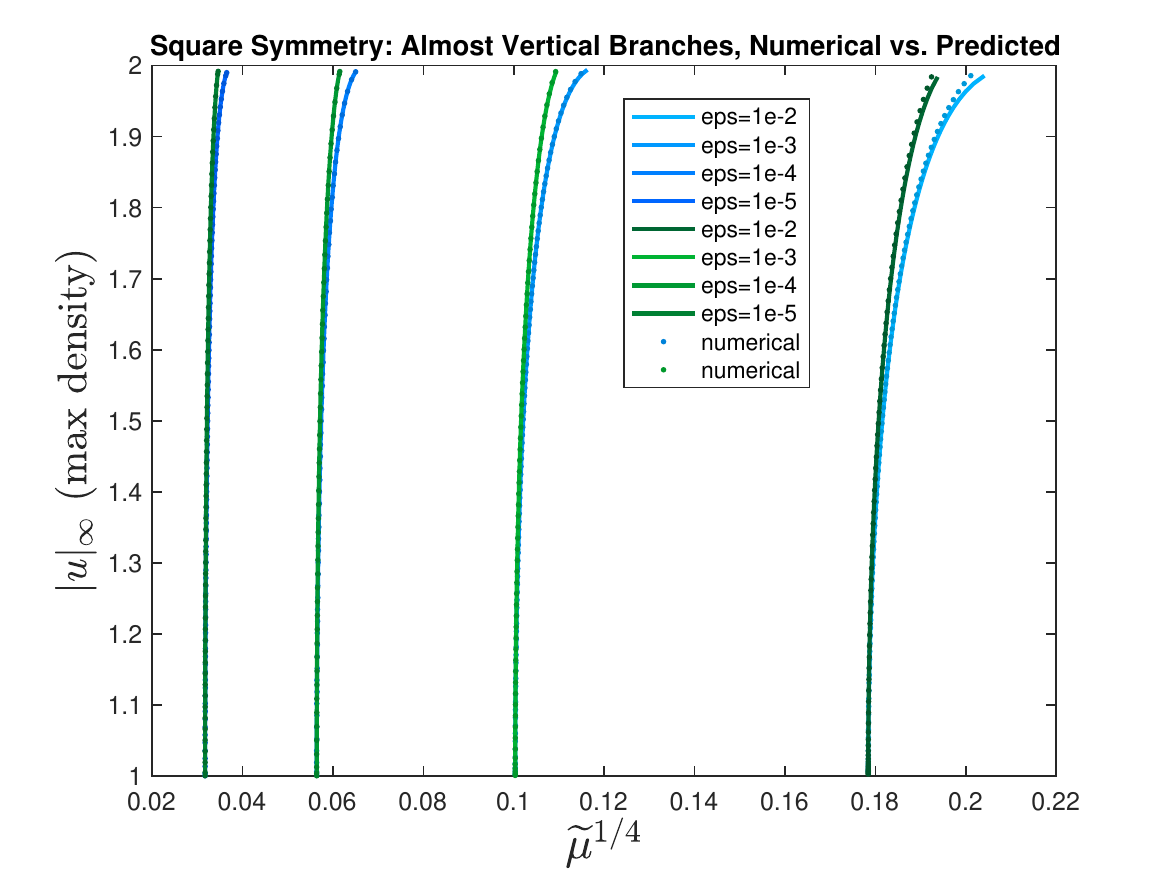}
    \includegraphics[width=0.49\textwidth]{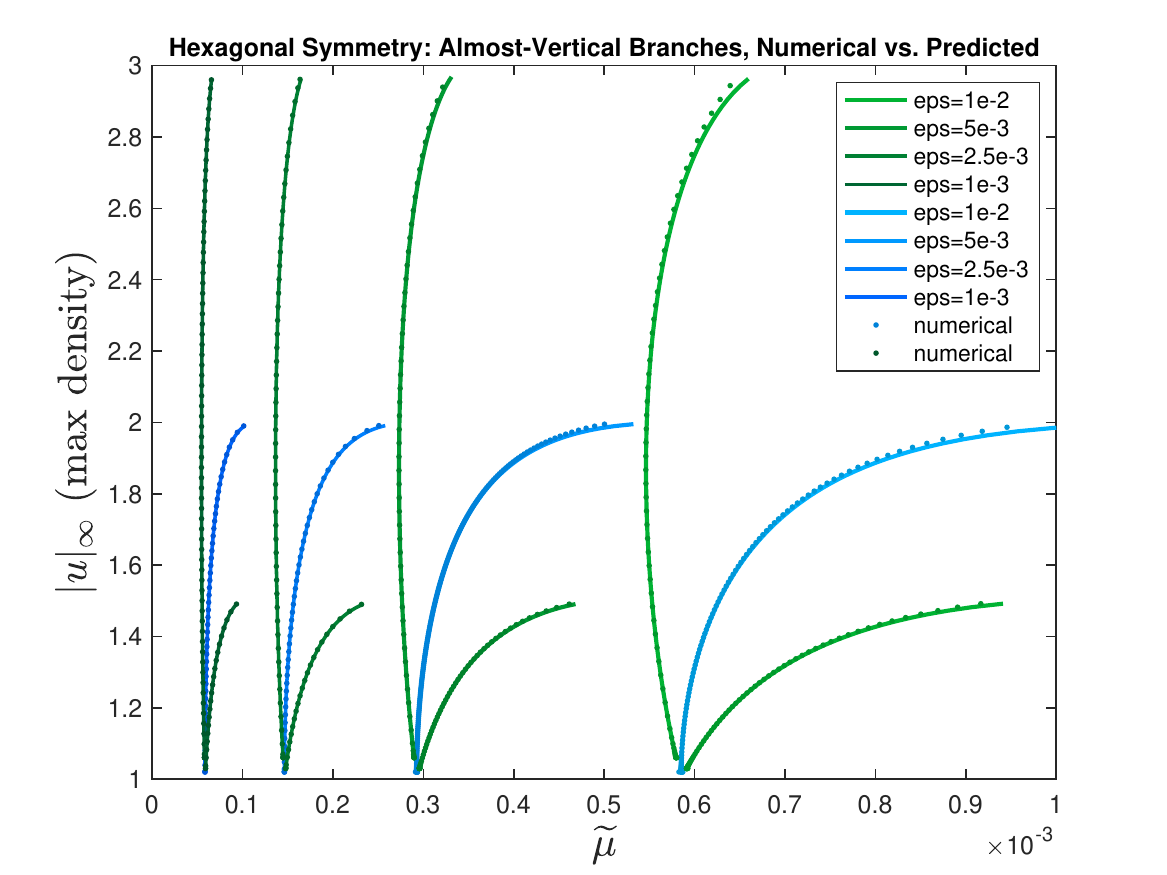}
    \caption{ Numerical agreement and comparison for the almost-vertical branches. Dots are numerical; curves are the $\mathcal{O}(\ep)$ predictions from \eqref{e:mu_pred_sq} and \eqref{e:mu1hex};  square symmetry (left) and  hexagonal symmetry (right) shown. Bubble branches are in green; fissure branches are in blue.}
    \label{fig:hex_stripes}
\end{figure}

\begin{Proof} 
    Using $\Delta u = \nabla \cdot \left(u \,\nabla (\log u)\right)$, we can again write the fixed point equation in the form
    \begin{equation}
        0 = \nabla \cdot \left(u \, \nabla \left(\ep \log u +  u - \left(\frac{1}{\sqrt{3}\pi^2}+\mnew\right)(V*u)\right)\right).
    \end{equation}
    Restricting to strictly positive solutions $u$, this is equivalent to 
    \begin{equation}
        \ep \log u +  u - \left(\frac{1}{\sqrt{3}\pi^2}+\mnew\right)(V*u) = \rho,
    \end{equation}
    for $\rho$ a constant, which we necessarily choose as the average of the left-hand side. 
    We consider the nonlinear functions
    \begin{equation}\label{e:Fsys}
        \begin{split}
            F(u,\mnew,\ep) &:= \ep \log u +  u - \left(\frac{1}{\sqrt{3}\pi^2}+\mnew\right)(V*u) - \rho \\
            F_0(u) &:= \iint_{\Omega} u(x,y)dxdy - 2\sqrt{3}\pi^2 \\
              F_1(u) &:= \iint_{\Omega} u(x,y)V(x,y) - 3\sqrt{3}A\pi^2,
        \end{split}
    \end{equation}
where  $A$ is fixed, $A \neq 0, -\frac{1}{3} < A < \frac{2}{3}$. These functions \eqref{e:Fsys} define a map 
    \[G(u,\mnew,\ep):  H^2_{\textrm{e,p}}\times \R^2 \to \overset{\circ}{L}{}^2_{\textrm{e,p}}  \times \R^2 ,\]  
    where  the subscript $\{\textrm{e,p}\}$ refers to the fact that functions are  lattice-periodic and invariant under $D_6$, the symmetry of the fundamental hexagon and $\overset{\circ}{L}{}^2_{\textrm{e,p}} $ refers to the space of symmetric, mean-zero functions.
    One verifies that $G$ is well-defined and smooth, and 
    \[
    G(u_*^0(\cdot,\cdot;A),0,0) = 0, \quad \textrm{ for } \ \  u_*^0(x,y;A) = 1 + AV(x,y).
    \]
    The derivative at $(u_*^0,0,0)$ is given by 
    \[
    \mathcal{A} = \begin{pmatrix} \mathcal{L} & \partial_\mnew F_* & \partial_\ep F_* \\
    \langle 1, \cdot \rangle & 0 & 0 \\
    \langle V(x,y), \cdot\rangle & 0 & 0
    \end{pmatrix},
    \]
    where 
    \begin{align*}
        \mathcal{L}u = u - \left(\frac{1}{\sqrt{3}\pi^2}+\mnew\right)(V*u),\qquad
        \partial_\mnew F_* = -\sqrt{3} \pi^2AV(x),\qquad
        \partial_\ep F_*  = \log(1 + AV(x)).
    \end{align*}
    The operator $\mathcal{L}$ is strictly elliptic since $\frac{1}{3} < A < \frac{2}{3}$, and self-adjoint with domain $H^2_{\textrm{e,p}}$. It is therefore Fredholm index 1 due to the restriction  to average 0 functions in the codomain.
    Bordering lemmas for Fredholm operators then imply that $\mathcal{A}$ is Fredholm of index 1, so that we expect a 1-dimensional family of solutions. 

    We now turn to showing that the map $D_{u,\mnew}G(u_*^0(\cdot,\cdot;A),0,0)$ consisting of the first two columns of $\mathcal{A}$, is invertible. We first notice that the kernel of $\mathcal{L}$ is explicitly the span of $\{1, V(x,y) \}$, and the cokernel is  the span of $\{ V(x,y) \}$. 
    As a consequence, we find that $\partial_\mnew F_* \notin \textrm{Rg}(\mathcal{L})$, through 
    \[
      - \langle \sqrt{3}\pi^2A V(x,y), V(x,y) \rangle = -9\pi^4A \neq 0. 
    \]
    We can thus see that $\mathcal{A}(u,\mnew) = 0$ if and only if $ u = 0, \mnew = 0$, since the second and third components of $\mathcal{A}$ imply $u \notin \ker(\mathcal{L})$, and therefore $\mathcal{L}u \neq \partial_\mnew F_* \mnew$, unless $u = 0, \mnew = 0$. We therefore have that $D_{u,\mnew}G(u_*^0,0,0)$ is invertible, and we can solve for $u, \mnew$ as functions of $\ep$. We then obtain the leading order expansion of $\mnew$ by projecting the first equation onto the cokernel. 
    We evaluate 
    \begin{align*}
    \langle \partial_\ep F_*, V(x,y)\rangle &=  \int_{-\frac{\sqrt{3}\pi}{2}}^{\frac{\sqrt{3}\pi}{2}}\int_{-\pi}^\pi V(x,y)\cdot \log\left(1 + AV(x,y)\right)dxdy \\
    &= \int_{-\frac{\sqrt{3}\pi}{2}}^{\frac{\sqrt{3}\pi}{2}}\frac{3\pi}{A}\left(1 - \sqrt{\left(1+A\cos\left(\frac{2y}{\sqrt{3}}\right)\right)^2 -4A^2\cos^2\left(\frac{y}{\sqrt{3}}\right) }\right)dy   \\
    &= \int_{-\frac{\sqrt{3}\pi}{2}}^{\frac{\sqrt{3}\pi}{2}}\frac{3\pi}{A}\left(1 - \sqrt{\left(1+A\cos\left(\frac{2y}{\sqrt{3}}\right)\right)^2 -4A^2\cos^2\left(\frac{y}{\sqrt{3}}\right) }+A\cos\left(\frac{2y}{\sqrt{3}}\right)\right)dy   \\
    &>0,
    \end{align*}
    since the integrand is non-negative and not identically equal to 0. 
    We thus find 
    \begin{align*}
        \mnew &= \mnew_1\ep + \mathcal{O}(\ep^2), \\
          \mnew_1 &= -\frac{\left\langle \partial_\ep F_*,\cos(x)\cos(y) \right\rangle }{\left\langle \partial_\mnew F_*,\cos(x)\cos(y) \right\rangle } \\
        \ &= \int_{-\frac{\sqrt{3}\pi}{2}}^{\frac{\sqrt{3}\pi}{2}}\frac{1}{3\pi^3 A^2}\left(1 - \sqrt{\left(1+A\cos\left(\frac{2y}{\sqrt{3}}\right)\right)^2 -4A^2\cos^2\left(\frac{y}{\sqrt{3}}\right) }\right)dy,
    \end{align*}
    with explicit limits 
    $\mnew_1(0) = \frac{1}{\sqrt{3}\pi^2}, \mnew_1(\frac{2}{3}) = \frac{2}{\sqrt{3}\pi^2}-\frac{3}{2\pi^3}$, $\mnew_1(-\frac{1}{3}) = \frac{5}{\sqrt{3}\pi^2}-\frac{6}{\pi^3}$.
\end{Proof}

  \subsection{Stability near onset: Center manifold expansions}
The characterization of the almost-vertical branches also encodes information about the stability of the solutions. The explicit dependence of the solutions on $\mnew$, along with the symmetries of the system, allow one to compute the coefficients in the center manifold expansion for small $\mnew$. 

{}\smallskip\textbf{Square lattice.}
We take $\ep > 0$ fixed, small, and let $\mnew_c = \mnew_*(0,\ep)$ defined in Prop.~\ref{e:diffusion_gaps_B=0}. We can then perform a parameter-dependent center manifold reduction in the space of functions that are invariant under the reflections $x\leftrightarrow y$ and $x\leftrightarrow -y$.  We decompose  $u = u_c + u_h$,  $u_c = A\cos(x-y)+B\cos(x+y)$ parameterizing the center eigenspace, and  $u_h\perp u_c$. The center manifold is then given as a graph $u_h=h(u_c)$, with $ h(0) = 0$ and $h'(0)=0$, where we suppress parameter dependence. Within the space of even functions, we have the action of the dihedral group $D_4$ generated by the rotation of order 4 in the plane and the reflection $x\leftrightarrow -x$, as well as translations by $\pi$ in $x$ and $y$. On the center eigenspace, and therefore on the reduced vector field, we find they induce a standard action of $D_2$, generated by $A\leftrightarrow B$ and $A\rightarrow -A$; see for instance \cite{GSS2}. This enforces a Taylor expansion of the reduced vector field of the form,
\begin{equation}\label{e:cmfdamp}
    \begin{split}
        A' &= (\mnew-\mnew_c)A + A(\alpha A^2 + \beta B^2) + \mathcal{O}((\mnew-\mnew_c)^2(|A|+|B|) +|A|^5+|B|^5)\\
        B' &= (\mnew-\mnew_c)B +B(\alpha B^2 + \beta A^2) + \mathcal{O}((\mnew-\mnew_c)^2 (|A|+|B|)+|A|^5+|B|^5).
    \end{split}
\end{equation}
Fissure solutions correspond to $B=0,$ and bubbles correspond to $A=B$, both one-dimensional invariant lines in this reduced phase plane description, enforced by symmetry. Thus steady-state solutions have, for small $\mnew-\mnew_c$, leading-order amplitudes 
\[A = \pm\sqrt{\frac{-(\mnew-\mnew_c)}{\alpha}} \textrm{ (fissures); \qquad and  } \qquad B = \pm\sqrt{\frac{-(\mnew-\mnew_c)}{\alpha+\beta}} \textrm{ (bubbles) }.\]
We can compare these expressions to the asymptotics of $\mnew_1$ derived above in \S\ref{ss:diffusion_sq}. Starting with fissure solutions, we expand $\mnew_1(A) = \frac{2}{\pi^2}\left(\frac{1-\sqrt{1-A^2}}{A^2}\right)$ about $A=0$, and find
\[
\mnew_1 - \mnew_1(0) = \frac{1}{4\pi^2}A^2 + \mathcal{O}(A^4).
\]
This implies that  $\alpha = \frac{-1}{4\pi^2}\ep + \mathcal{O}(\ep^2)$. 

Similarly, we expand the expression for $\mnew_1(A)$ from \S\ref{ss:diffusion_sq} for the bubble solutions, noting that in the coordinates used there, $A$ corresponds to $2B$ here. We then obtain
\[
\mnew_1 - \mnew_1(0) = \frac{3}{4\pi^2}B^2 + \mathcal{O}(B^4),
\]
from which we deduce $\alpha + \beta = \frac{-3}{4\pi^2}\ep + \mathcal{O}(\ep^2)$, so $\beta = \frac{-1}{2\pi^2}\ep + \mathcal{O}(\ep^2)$. 
Summarizing, we have 
\begin{equation}\label{e:ccoeff}
    \alpha= \frac{-4}{2\pi^2}\ep + \mathcal{O}(\ep^2),\qquad \beta = \frac{-1}{2\pi^2}\ep + \mathcal{O}(\ep^2),
\end{equation}
in \eqref{e:cmfdamp}. 
Notably, these dynamics for $A$ and $B$ near onset correspond to strong competition favoring the fissure branch. In other words, near onset, the fissure equilibria with $B=0$ (and the ones with $A=0$) are stable while the $A=B$ bubble equilibria are unstable. 

{}\smallskip\textbf{Hexagonal lattice.}
We proceed in an analogous fashion. We restrict to a function space with reflection symmetry $y\mapsto -y$ and sixfold rotation in the plane. 
The center subspace is 3-dimensional, parameterized via  $u_c = A\cos(x + \frac{y}{\sqrt{3}}) + B\cos(x - \frac{y}{\sqrt{3}}) + C\cos(\frac{2y}{\sqrt{3}})$. On the center-subspace, symmetries are generated by rotations of $\pi/3$, $A\rightarrow B\rightarrow C\rightarrow A$, reflections in $y$, $A\leftrightarrow B$, and translations in $x$ by $\pi$, $(A,B)\mapsto -(A,B)$ and give the reduced equation, at leading order, 
\begin{equation}\label{e:CMhex}
    \begin{split}
        A' &= (\mnew-\mnew_c)A + \alpha BC + A(\beta A^2 + \gamma( B^2+ C^2)) \\
        B' &= (\mnew-\mnew_c)B + \alpha AC +B(\beta B^2 + \gamma ( A^2+C^2)) \\
        C' &= (\mnew-\mnew_c)B+ \alpha AB +C(\beta C^2 + \gamma ( A^2+B^2)).
    \end{split}
\end{equation}
Fissure solutions correspond to $B = C = 0$, and bubbles to $|A| = |B| = |C|$, with amplitudes solving
\begin{align*}
  \ \mnew-\mnew_c &= -\beta A^2 + \mathcal{O}(A^4) \ \ \textrm{ (fissures), } \\
 \ \mnew-\mnew_c &= -\alpha A - (\beta + 2\gamma)A^2 + \mathcal{O}(A^3) \ \ \textrm{ (bubbles). }
 \end{align*}

In order to determine $\beta$, we expand our analytical expression for the fissure coefficient
$\mnew_1(A) = \frac{2}{\pi^2}\left(\frac{1-\sqrt{1-A^2}}{A^2}\right)$ at $A=0$, and find
\[
\mnew_1 - \mnew_1(0) = \frac{1}{4\sqrt{3}\pi^2}A^2 + \mathcal{O}(A^4),
\]
thus $\beta = \frac{-1}{4\sqrt{3}\pi^2}\ep + \mathcal{O}(\ep^2)$. 
Next, to find $\alpha, \gamma$, we expand the analytical expression from \eqref{e:mu1hex} for the bubble coefficient $\mnew_1(A)$ at $A = 0$ to find 
\[
\mnew_1 - \mnew_1(0) = -\frac{1}{2\sqrt{3}\pi^2}A + \frac{5}{4\sqrt{3}\pi^2}A^2 + \mathcal{O}(A^3),
\]
which, when comparing, gives $\alpha = \frac{1}{2\sqrt{3}\pi^2}\ep + \mathcal{O}(\ep^2)$ and $\gamma = \frac{-1}{2\sqrt{3}\pi^2}\ep + \mathcal{O}(\ep^2)$. 

\begin{figure}
    \centering
    \includegraphics[width=0.49\linewidth]{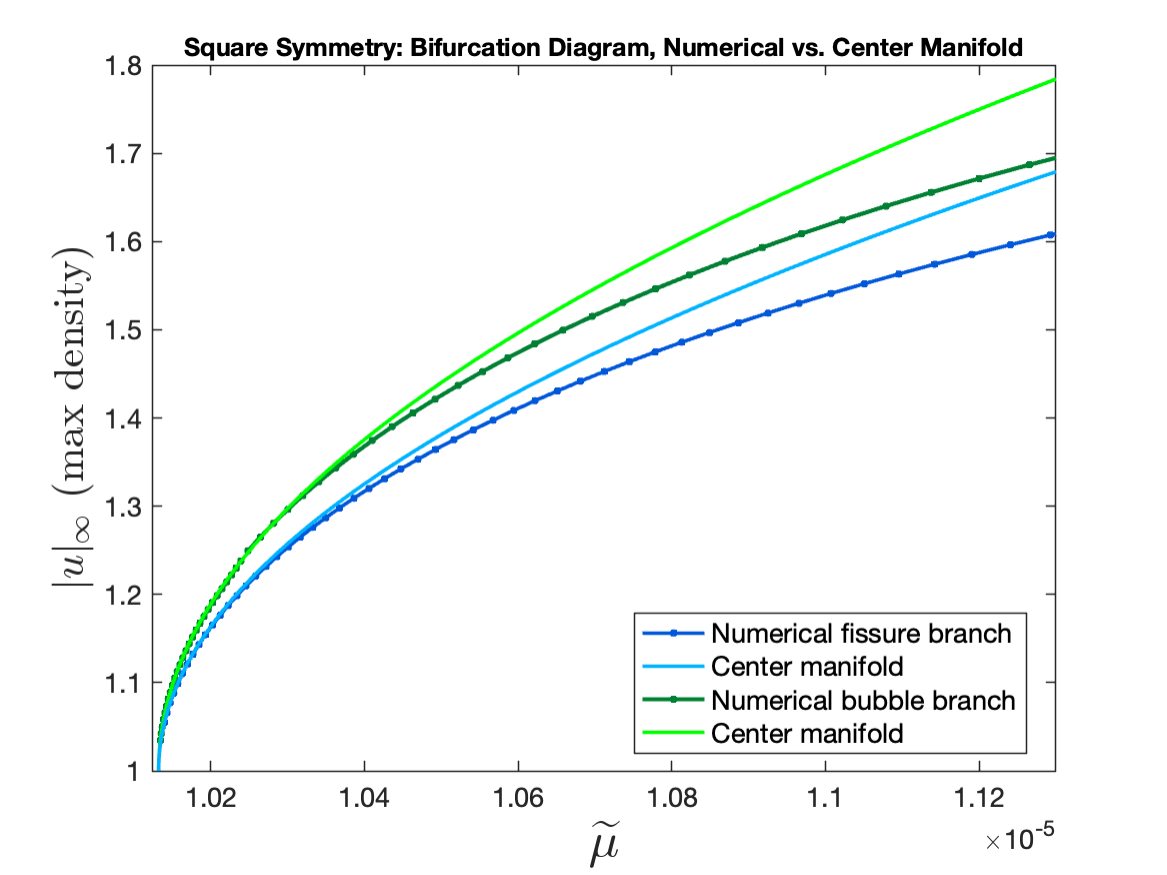}
    \includegraphics[width=0.49\linewidth]{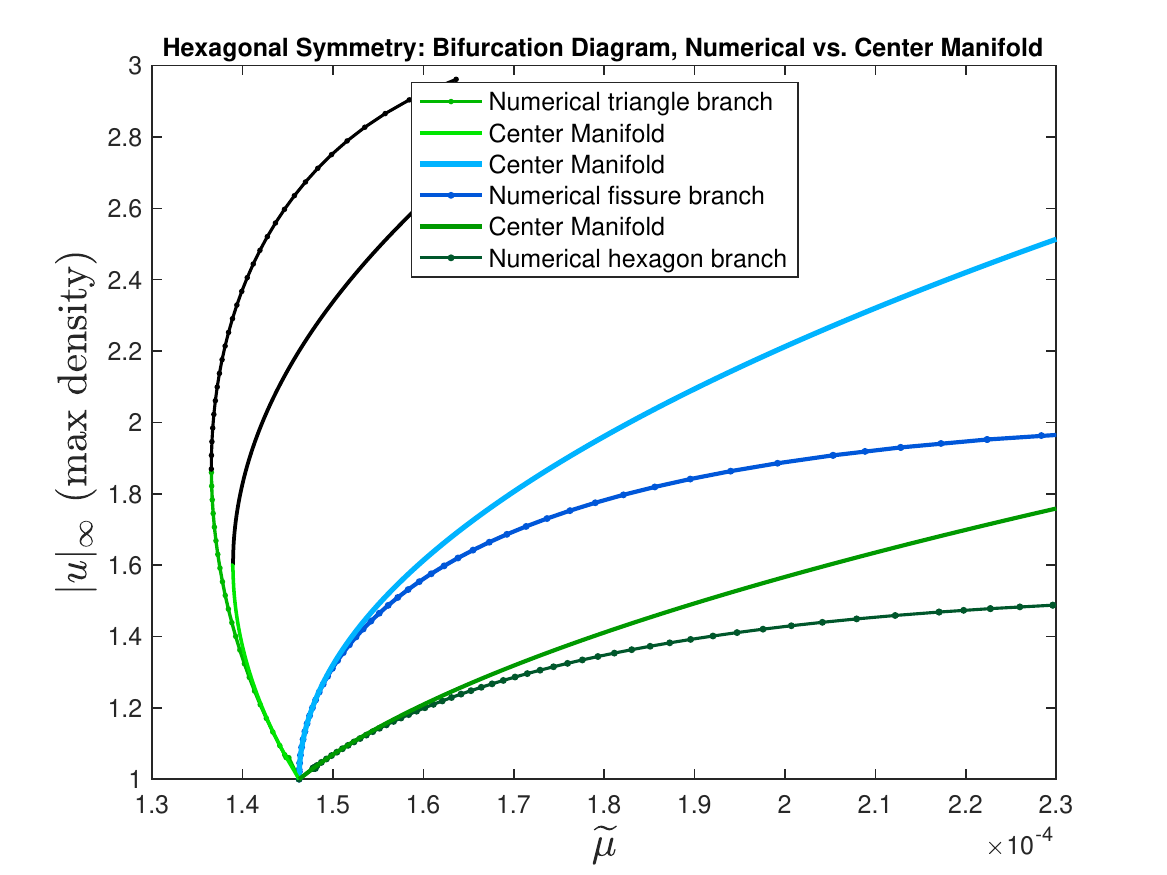}
    \caption{Center manifold predictions vs. numerical values,
    for $\ep = 10^{-4}$ (square symmetry, left) and $\ep = 2.5 10^{-3}$ (hexagonal symmetry, right). The parameter is shifted relative to the bifurcation point at $\ep=0$. Center manifold predictions are in lighter colors. Fissures are in blue; bubbles are in green and black. Black represents branches found to be stable in center manifold predictions and in numerical continuation, respectively.}
   \label{fig:hex_bifdia_CM}
   \end{figure}

Investigating stability in this reduced ODE, we find that at small amplitudes all solutions are unstable, as is common in the generic bifurcation scenario on hexagonal lattices \cite{buzgol}. The triangles stabilize at finite amplitude and parameter value $\mnew_- = \frac{-1}{20\sqrt{3}\pi^2}\ep + \mathcal{O}(\ep^2)$ in a saddle-node bifurcation. 
We compared  predictions from center-manifold with the perturbative analysis, also indicating stability of solutions with good agreement away from vacuum formation; see Fig.~\ref{fig:hex_bifdia_CM}.

\section{Numerical continuation and a second vertical branch}\label{s:num_pde}

We explore branches away from $\mnew=0$, including in particular the growth of vacuum regions and their eventual change of topology. We use two complementary numerical approaches, both relying on numerical secant continuation of stationary solutions to the deterministic limit
\begin{equation}
  \label{e:MV_timedep_nodiff}  u_t =  \nabla \cdot \left( u \, \nabla\left( u - \morig (V * u)\right) \right),
\end{equation}
and the diffusive regularization
\begin{equation}
  \label{e:MV_timedep_diff}  u_t = \ep \Delta u + \nabla \cdot \left( u \,\nabla \left( u - \morig (V*u)\right) \right).
\end{equation}
We solve the regularized problem \eqref{e:MV_timedep_diff} using a spectral discretization, \S\ref{s:spectral}, and the limiting \eqref{e:MV_timedep_nodiff} problem  via the finite-dimensional reduction permitted by the specific potential supported on finitely many Fourier modes, \S\ref{s:fr}. 
Both methods allow us to explore solution branches and their stability far from the bifurcation point $\mnew=0$; see \S\ref{s:bs}. Curiously, we find a change of stability in the case of square symmetry due to a secondary bifurcation with a  vertical branch (almost vertical for $\ep>0$)  which coincides with a change of topology in solutions; see \S\ref{ss:2nd_vert_branch}.
\subsection{Numerical Continuation: Full Equation with Small Diffusion}\label{s:spectral}
We implement secant continuation for the potentials in  \eqref{e:Vexpl} on both square and hexagonal lattices. In the case of the hexagonal lattice, we transform coordinates mapping the basis lattice vectors $e_1$ and $e_2$ from the hexagonal lattice onto the canonical basis and then computing on a square domain, with suitably transformed differential operators and potential. We impose phase conditions in $x$ and $y$ with reference solution $u^{(0)}$ to exclude the family of translations of the solution (see for instance \cite{Roose_Szalai_2007}), and a mass condition to select a branch with fixed average density. Mass and positional constraints are compensated for by dummy variables, adding a mass loss term $\alpha u$ and drift terms $s^xu_x$ and $s^yu_y$ to the equation. Numerically, we found $\alpha=s^x=s^y=0$ to numerical precision as expected.  When solving for fissure branches, only one of the phase conditions is used and the drift in the $y$-direction is omitted. Together, we solve  
\begin{align}
   \ep \Delta u +  \nabla \cdot \left( u \,\nabla \left( u - \mnew (V*u)\right) \right) -\alpha u + s^x u_x + s^yu_y = 0 && \text{(evolution equation),} \\
    \frac{1}{4\pi^2}\int_{-\pi}^{\pi}\int_{-\pi}^{\pi} u(x,y)dydx = 1 && \text{(mass condition),} \\
    \begin{cases}
        \int_{-\pi}^\pi u_t^{(0)}(x,y)(u^{(0)}(x,y)-u(x,y))dx = 0 \\
    \int_{-\pi}^\pi u_t^{(0)}(x,y)(u^{(0)}(x,y)-u(x,y))dy = 0
    \end{cases}
    && \text{(phase conditions),} \label{eq:num-secant-sys-phase}
\end{align}
in the case of square lattice periodicity, with suitably transformed differential operators in the case of hexagonal lattice periodicity. For continuation, we add the usual secant condition. 
The solution $u$ is discretized to $\underline{u}$ using $N^2$ Fourier modes in the two-dimensional domain.
A Newton method is implemented in  \textsc{Matlab} using its \texttt{gmres} linear solver for the  $N^2 + 4$ variables $(\underline{u},s^x,s^y,\alpha,\mnew)$ (resp. $N^2 +3$ variables for fissures). 
The diffusion term $\varepsilon$ serves as a regularizer and preconditioner for the linear solver. 

\begin{figure}
        \includegraphics[width=.49\linewidth]{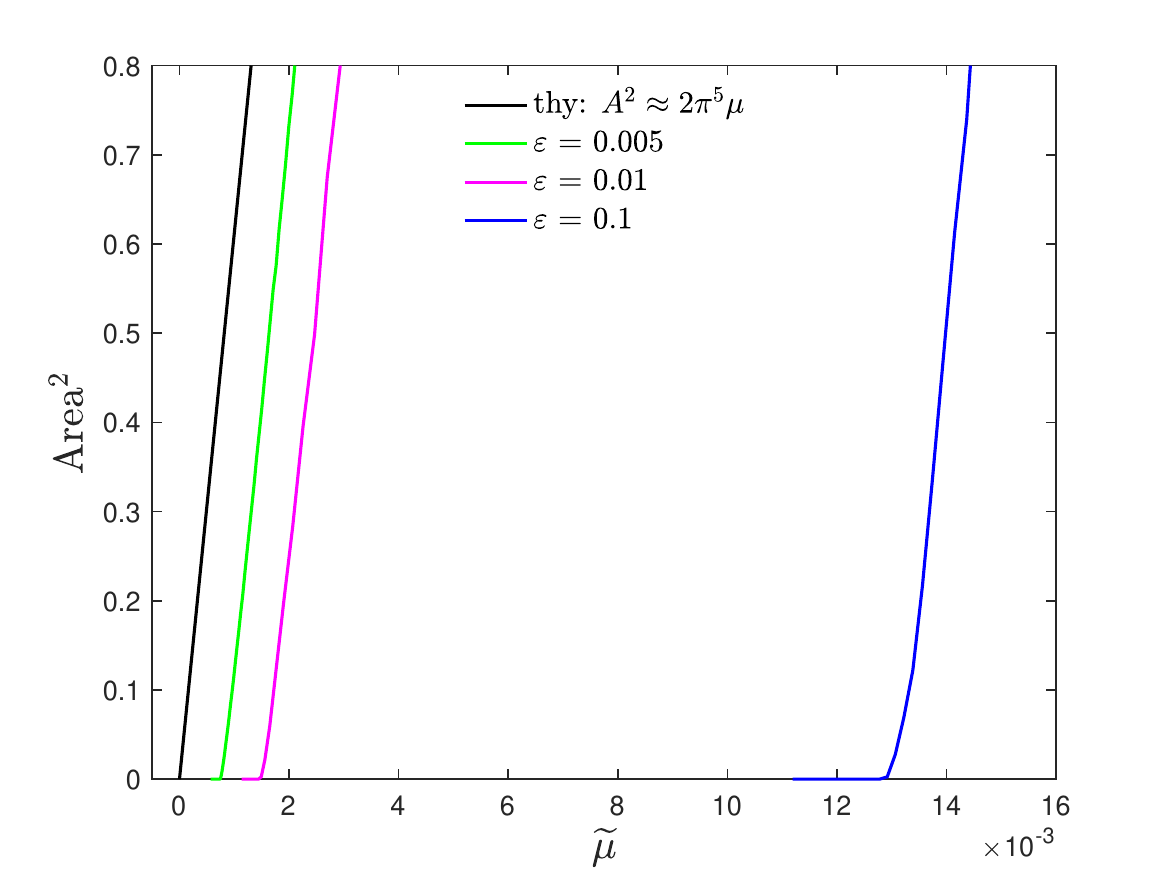}
        \label{fig:num-onset-precoeff-sq}
        \includegraphics[width=0.49\linewidth]{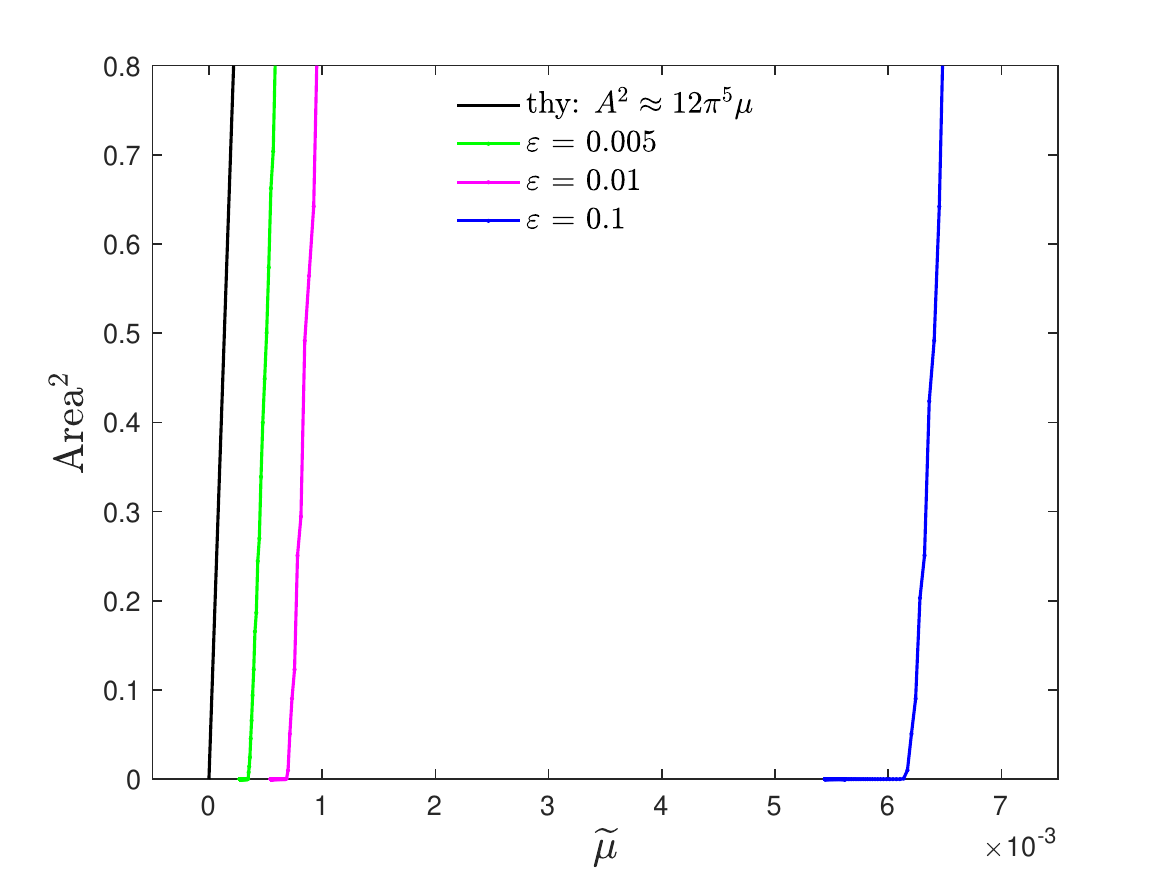}
        \label{fig:num-onset-precoeff-hex}
    \centering
    \caption{ Numerical continuation of branches forming vacuum bubbles in case of square lattice periodicity (left) and hexagonal lattice periodicity (right, triangle branch), for different values of $\ep$. Plotted is the square of the size of the vacuum area, computed as the area where $u<\ep$, together with the theoretical predictions from Theorems~\ref{t:square} and~\ref{t:hexlat}, respectively.  
    }
    \label{fig:num-onset-precoeff}
\end{figure}

Results are displayed in Fig.~\ref{fig:num-onset-precoeff} and Fig.~\ref{fig:bifdia_pde_cont_stab}. Fig.~\ref{fig:num-onset-precoeff} gives a quantitative comparison to the theory, displaying convergence of numerically computed branches to the explicit prediction. As predicted, the square of the vacuum area for bubbles grows linearly in the parameter, with slope matching our explicit calculation for the growth of vacuum bubbles, in both square and hexagonal periodicity (here for the triangle branch). Convergence in $\ep$ is linear as predicted. 
%
Different branches, for both the square lattice and the hexagonal lattice periodicity are shown in  Fig.~\ref{fig:bifdia_pde_cont_stab}. 
Calculations include stability information which agrees with the information obtained from the center-manifold analysis.

\subsection{Numerical Continuation: Finite Rank System, No Diffusion}\label{s:fr}

In order to continue solutions of \eqref{e:MV_timedep_nodiff}, we use the somewhat explicit form of solutions as in \S\ref{s:gaps_proofs}, and numerically solve equations \eqref{e:MV_gen}-\eqref{e:MV_density_gen} for the coefficients of Fourier modes. 

{}\smallskip\textbf{Square Symmetry.} In order to find bubble and fissure solutions, we set
\[
u =\big(A_0 + A_1\cos(x)\cos(y) + B_1\sin( x)\sin(y)\big)_+,
\]
where $f_+$ denotes the positive part of $f$. 
Bubble branches correspond to $B_1 = 0$, while fissure branches will have $A_1 = B_1 \neq 0$. We then numerically solve the system
\begin{equation}\label{e:finite_rank_sq}
\begin{split}
 0& = 2\pi^2 -  \iint_{\Omega_0} \left( A_0 + A_1 \cos(x) \cos(y)+ B_1\sin(x)\sin(y) \right) d x d y \\
      0& = A_1 -   \left(\frac{1}{\pi^2} + \mnew\right)\iint_{\Omega_0} 2\cos(x) \cos(y)\left( A_0 + A_1 \cos(x) \cos(y) \right) d x d y\\
        0& = B_1 -   \left(\frac{1}{\pi^2} + \mnew\right)\iint_{\Omega_0} 2B_1\sin^2(x) \sin^2(y) d x dy,
\end{split}
\end{equation}
where the domain $\Omega_0 $ is the subset of $[-\pi,\pi]^2$ on which $u$ is positive. 

{}\smallskip\textbf{Hexagonal Symmetry.} 
For the case of hexagonal symmetry, we make the ansatz 
\[
u = \big(A_0 + A_1\cos(x-\frac{y}{\sqrt{3}}) + B_1\cos(x+\frac{y}{\sqrt{3}})\sin(y)+C_1\cos(\frac{2y}{\sqrt{3}})\big)_+.
\]
Here, a bubble branch satisfies $A_1 = B_1 = C_1 \neq 0$, while a fissure branch branch has $A_1 \neq 0, B_1 = C_1 = 0$. We solve the system 
\begin{equation}\label{e:finite_rank_hex}
\begin{split}
 0& = 2\sqrt{3}\pi^2 -  \iint_{\Omega_0} \left( A_0 + A_1\cos(x-\frac{y}{\sqrt{3}}) + B_1\cos(x+\frac{y}{\sqrt{3}})\sin(y)+C_1\cos(\frac{2y}{\sqrt{3}}) \right) dxdy \\
      0& = A_1 -   \left(\frac{1}{\sqrt{3}\pi^2} + \mnew\right)\iint_{\Omega_0} \cos(x-\frac{y}{\sqrt{3}})\left( A_0 + A_1\cos(x-\frac{y}{\sqrt{3}})\right) dxdy\\
        0& = B_1 -   \left(\frac{1}{\sqrt{3}\pi^2} + \mnew\right)\iint_{\Omega_0}\cos(x+\frac{y}{\sqrt{3}})\left( A_0 + B_1\cos(x+\frac{y}{\sqrt{3}})\sin(y)\right)dxdy\\
               0& = C_1 -   \left(\frac{1}{\sqrt{3}\pi^2} + \mnew\right)\iint_{\Omega_0}\cos(\frac{2y}{\sqrt{3}})\left( A_0 + C_1\cos(\frac{2y}{\sqrt{3}})\sin(y)\right)dxdy,
\end{split}
\end{equation}
numerically, where $\Omega_0 $ is the subset of the domain on which $u$ is positive. Integrals are evaluated numerically in case of both square and hexagonal lattice periodicity, using a square grid of $N^2$ grid points, $N = 3000$. 

\subsection{Branches and Stability}\label{s:bs}

In the case of square lattice symmetry, numerical continuation, both with and without diffusion, reveals a secondary bifurcation where the solution branches exchange stability. Such a bifurcation is accompanied by a branch of mixed-mode solutions connecting the bubble and fissure branches. In the case of no diffusion, this occurs at exactly $\morig = 2\mnew_*$. In this case, the branch is again vertical, and its existence is explicit; see \S\ref{ss:2nd_vert_branch}. With diffusion, similarly to the primary vertical branch, this branch is almost-vertical and slightly shifted in parameter space. The mixed-mode branch consists of solutions with elliptical bubbles  which interpolate between bubbles and fissures. Modulations of fissures or more generally striped patterns have been observed in many other pattern-forming scenario, for instance in the Taylor-Couette problem, where they are known as wavy rolls \cite{chossatiooss}. Again, the situation here is particular as the exchange of stability between fissures and bubbles does not display hysteresis at $\eps=0$.

\begin{figure}
    \centering
    \includegraphics[width=0.49\textwidth]{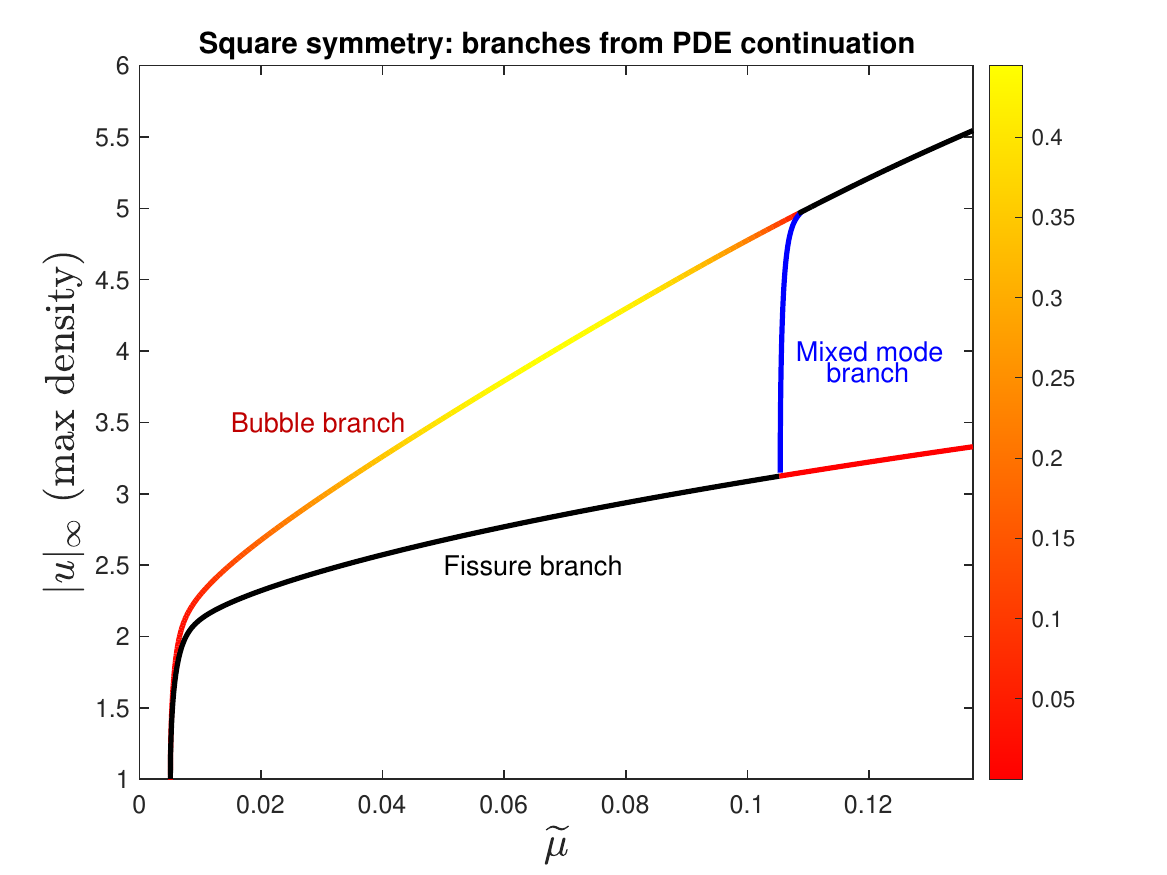}
    \includegraphics[width=0.49\textwidth]{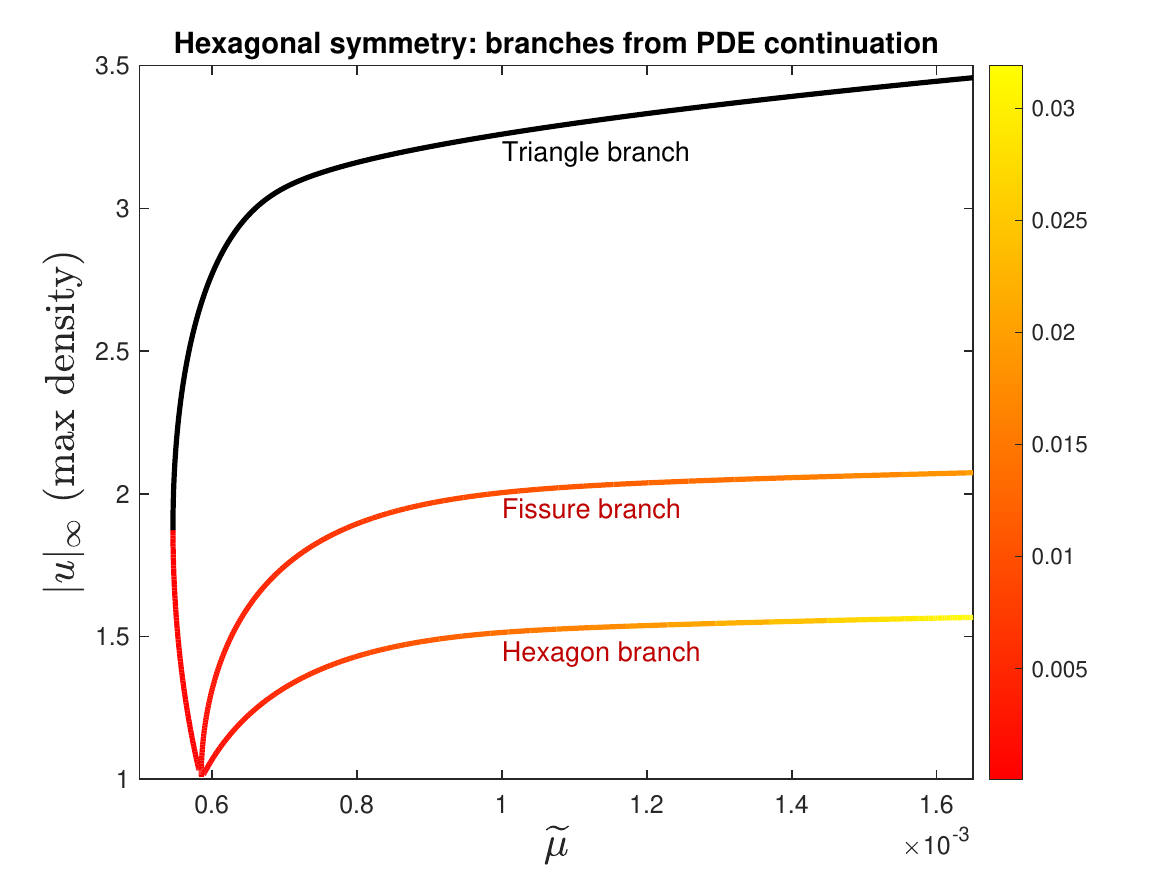}
    \caption{ Bifurcation diagrams from secant continuation for the full equation with diffusion, $\ep = 0.05$ (square symmetry, left), $\ep = 0.01$ (hexagonal symmetry, right).
      The color along the fissure and bubble branches represents the size of the largest numerically computed eigenvalue of the Jacobian at that equilibrium, so that lighter branches are more unstable.  Black represents a stable branch, where the largest computed eigenvalue is 0 due to translational symmetry.}
    \label{fig:bifdia_pde_cont_stab}
\end{figure}

In addition to being curiously vertical, this secondary bifurcation also occurs precisely at the parameter value where the bubble branch changes topology, that is,  where vacuum bubbles merge and confine the support of the solution $\Omega_0$ to disconnected clusters. In the finite-rank case, this change of topology happens precisely when $A_0 = 0$. At this secondary bifurcation, stability at both fissures and bubbles changes from what emerged from the local bifurcation and the primary vertical branch. Prior to the transition, fissure solutions are stable; afterwards, clusters (the continuation of bubbles) are stable. For square symmetry, then, the only stable solutions are fissures, first,  and clusters, later; solutions with true vacuum  bubbles are unstable; see Fig.~\ref{fig:bifdia_pde_cont_stab} for the bifurcation diagram with numerically computed eigenvalues. 

We did not find  such secondary bifurcation for hexagonal symmetry. In this case, the only stable solutions are triangles both near the bifurcation point and for larger values of $\mnew$; fissure and hexagon solutions are always unstable. This was confirmed both in the numerical continuation of the PDE solution with small diffusion and in the finite-rank continuation: the computed Morse index remains constant along all three solution branches for ${\mnew}>0$.

\begin{figure}
    \centering
    \includegraphics[width=0.52\textwidth]{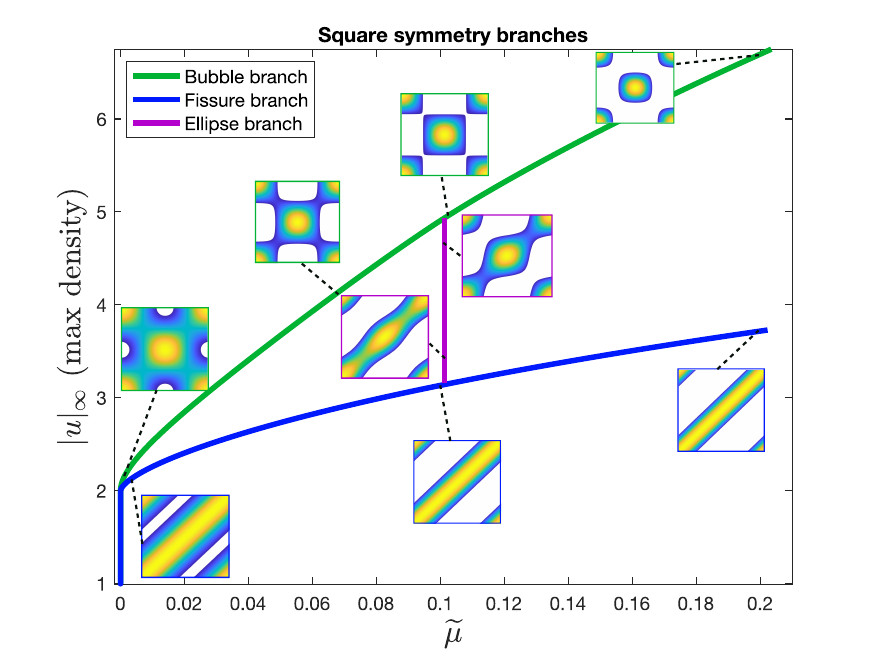}
    \includegraphics[width=0.47\textwidth]{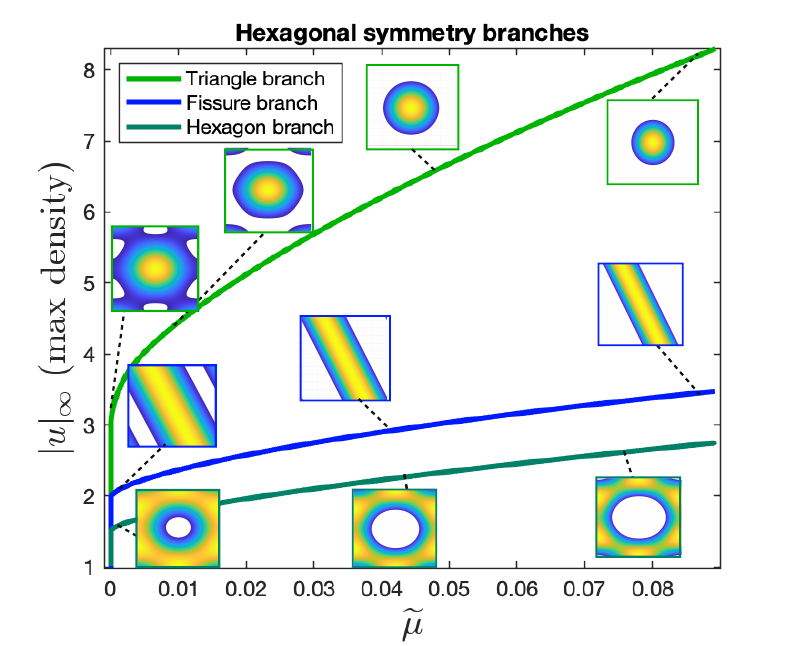}
    \caption{ Bifurcation diagrams from finite rank continuation, with corresponding states along the branches, showing  square symmetry (left) and  hexagonal symmetry (right). In the left panel, one sees in purple the interpolating branch of mixed-mode solutions. For hexagonal symmetry, we find no such branch. The computed Morse index remains constant along all 3 solution branches, with no observed crossing which would indicate a bifurcation. }
    \label{fig:rank-3-sq}
\end{figure}



\subsection{Vertical Branch II: The Connecting Branch}\label{ss:2nd_vert_branch}

For \eqref{e:MV_timedep_nodiff}, without diffusion and with square symmetry, we find an explicit vertical mixed-mode branch at $\morig = 2\mnew_*$. These solutions all have $A_0 = 0$. 

\begin{Proposition} Let $\mnew = \frac{2}{\pi^2}, V (x, y) = 2\cos(x) \cos(y)$. For each $\tau \in [0,1]$, there exists a solution $u$
to equations \eqref{e:MV_gen}-\eqref{e:MV_boundary_gen} of the form
\[
u(x,y) = \left(A_1 \cos (x) \cos (y) + \tau A_1\sin (x)\sin (y) \right)_+,
\]
for a unique $A_1=A_1(\tau) > 0$. 
    
\end{Proposition}

\begin{Proof} Let $\Omega_0' = \left\{ (x,y) \in [-\pi,\pi^2] \ \big| \ A_1 \cos(x) \cos(y)+ B_1\sin(x)\sin(y) > 0\  \right\}$. 
Substituting the ansatz $\mnew = \frac{2}{\pi^2},\ \  u = \left(A_1 \cos (x) \cos (y) + \tau A_1\sin (x)\sin (y) \right)_+$ into \eqref{e:MV_gen}-\eqref{e:MV_boundary_gen}, we obtain the system
\begin{align}
 \label{e:dens2} 0& = 4\pi^2 -  \iint_{\Omega_0'} \left( A_1 \cos(x) \cos(y)+ \tau A_1\sin(x)\sin(y) \right) d x d y \\
   \label{e:a1}    0& = A_1 -   \frac{2}{\pi^2} \iint_{\Omega_0'}  A_1 \cos^2(x) \cos^2(y) d x d y\\
        \label{e:b1}    0& = \tau A_1 -   \frac{2}{\pi^2} \iint_{\Omega_0'} \tau A_1\sin^2(x) \sin^2(y) d x dy.
\end{align}
We note two key facts: 
\begin{enumerate}
    \item the complement $(\Omega_0')^c$ is exactly equal to $\Omega_0'$, shifted by $\pi$ in the $x$-direction (or, equivalently, by $\pi$ in the $y$-direction). This follows immediately from the symmetries $\cos(x + \pi) = -\cos(x),$ and $ \sin(x+\pi) = -\sin(x)$;
    \item the functions $\cos^2(x) \cos^2(y)$ and $\sin^2(x) \sin^2(y)$ are both $\pi$-periodic in $x$. 
\end{enumerate}
These facts imply that 
\begin{align*}
 \iint_{\Omega_0'}  \cos^2(x) \cos^2(y) d x d y &= \frac{1}{2}\int_{-\pi}^\pi \int_{-\pi}^\pi \cos^2(x) \cos^2(y) d x d y, \\
 \iint_{\Omega_0'}  \sin^2(x) \sin^2(y) d x d y &= \frac{1}{2}\int_{-\pi}^\pi \int_{-\pi}^\pi \sin^2(x) \sin^2(y) d x d y.
 \end{align*}
Then \eqref{e:a1} and \eqref{e:b1} are automatically satisfied no matter the value of $A_1, \tau$. 
Considering \eqref{e:dens2}, we note that for fixed $\tau$, the regions $\Omega_0, \Omega_0'$ do not depend on $A_1$. Then the quantity 
\[\iint_{\Omega_0'} \left( \cos(x) \cos(y)+ \tau\sin(x)\sin(y) \right) d x d y \] is simply a number, independent of $A_1$, and we can solve as
\[
A_1(\tau) = \frac{4\pi^2}{\iint_{\Omega_0'} \left( \cos(x) \cos(y)+ \tau\sin(x)\sin(y) \right) d x d y }.
\] \end{Proof}

Along the bubble branch, we note that such a solution with $A_0 = 0$ corresponds exactly to the transition between vacuum bubbles and isolated clusters. 

\begin{Remark}\label{r:triangle}
We emphasize that the bifurcation diagram in the case of the hexagonal lattice, shown in Fig.~\ref{fig:rank-3-sq}, does not include a vertical branch. Bifurcation diagrams with the symmetry of the hexagonal lattice computed in \cite{buzgol} sometimes exhibit such secondary vertical branches, consisting of solutions with $D_3$-symmetry. These do not appear to exist in this setting. This is corroborated by numerical calculations of stability, implied in Fig.~\ref{fig:bifdia_pde_cont_stab} which does not appear to change supercritically. 
\end{Remark}

\section{Comparison with evolution of finitely many particles}\label{s:discrete}

We also compared predictions from the continuum limit with direct simulations and numerical continuations of finite-size particle systems. We find that finite particle dynamics largely reflect predictions from the continuum model, with some hysteretic effects due to the finite size of the system. Notably, on the square lattice, continuum model dynamics predict patterns well despite the fact that particles tend to prefer a hexagonal microstructure. 

{}\smallskip\textbf{Finite-Particle ODEs.} In two dimensions, we consider interactions between $N^2$ agents driven by  the following gradient flow,
    \begin{align}\label{e:discrete_main}
        \dot{x}_{\underline{\ell}} &= -\frac{1}{N^2} \sum_{\underline{j} \neq \underline{\ell}}^{N} \partial_{1} W(x_{\underline{\ell}}-x_{\underline{j}}, y_{\underline{\ell}}-y_{\underline{j}} ) \\
        \dot{y}_{\underline{\ell}} &= -\frac{1}{N^2} \sum_{\underline{j} \neq \underline{\ell}}^{N} \partial_{2} W(x_{\underline{\ell}}-x_{\underline{j}}, y_{\underline{\ell}}-y_{\underline{j}} )
\end{align}
where $\underline{\ell} = (\ell_1, \ell_2)$ is the label for a particle at some position $(x,y)$ in the plane. Modeling repulsion by a Dirac-$\delta$ is no longer meaningful for the ODE and we therefore model the repulsive interaction by a potential  $K^\zeta(x,y) = (2\pi\zeta)^{-2}K_0(\frac{\sqrt{x^2+y^2}}{\zeta})= \mathcal{F}^{-1}(\frac{1}{4\pi^2(1+\zeta^2(k^2+l^2))}),$ for $\zeta$ fixed, small. Here $K_0$ represents the modified Bessel function of the second kind. For the attractive part of the potential, we again take $V = 2\cos(x)\cos(y)$ for the square lattice, and $V = \cos(x+\frac{y}{\sqrt{3}})+\cos(x-\frac{y}{\sqrt{3}})+\cos(\frac{2y}{\sqrt{3}})$ for the hexagonal lattice.

The counterpart here to the uniform density state in the continuum model is the crystal solution, where particles are equally spaced in their respective lattice. Such a solution $(\bar{x}_{\underline{\ell}}, \bar{y}_{\underline{\ell}})$,  is given  for the square lattice by
\begin{align*}
    \bar{x}_{\underline{\ell}} &= (\ell_1+\ell_2) \cdot \frac{\pi}{N}, \qquad
    \bar{y}_{\underline{\ell}} = (\ell_1-\ell_2) \cdot \frac{\pi}{N},
\end{align*}
and for the hexagonal lattice by
\begin{align*}
    \bar{x}_{\underline{\ell}} &= \frac{2\pi \ell_1}{N}-\frac{\pi \ell_2}{N},
   \qquad \bar{y}_{\underline{\ell}}= \frac{\sqrt{3}\pi \ell_2}{N}. 
\end{align*}
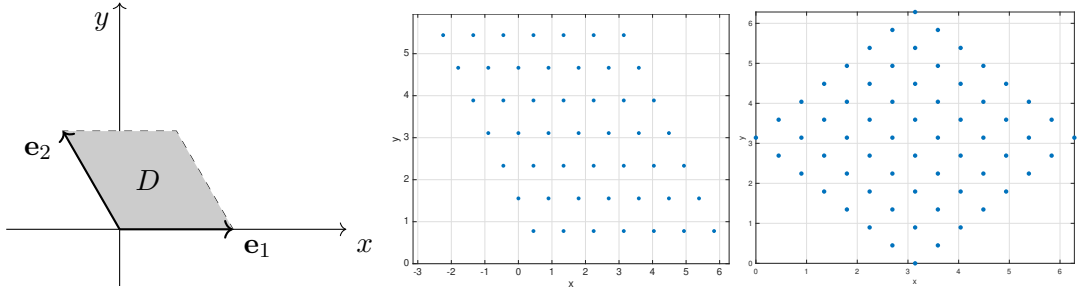
\begin{figure}
    \centering
    \begin{tikzpicture}[scale = 1.5]
      \draw [->](-1,0) -- (2,0) node [below right] {$x$}; 

      \draw [->](0,-0.5) -- (0,2) node [below left] {$y$}; 
      
      \draw[dashed] (0,0) -- (1,0) -- (1/2,{sqrt(3)/2}) -- (-1/2,{sqrt(3)/2}) -- (0,0); 
      \path[fill=black!20] (0,0) -- (1,0) -- (1/2,{sqrt(3)/2}) -- (-1/2,{sqrt(3)/2}) -- (0,0);
      
      \node at (0.25, {sqrt(3)/4}) {$D$};   
      \draw [thick, ->](0,0) -- (1,0) node [below right] {$\mathbf{e}_1$};
      \draw [thick, ->](0,0) -- (-1/2,{sqrt(3)/2}) node [below left] {$\mathbf{e}_2$};
    \end{tikzpicture}
     \includegraphics[width=.27\linewidth]{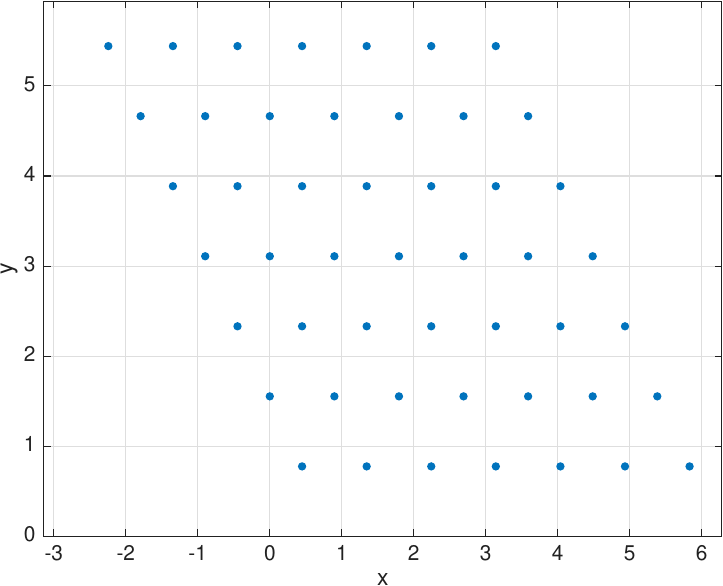}
     \includegraphics[width=.27\linewidth,trim={0.7in 0.2in 0.7in .36in},clip]{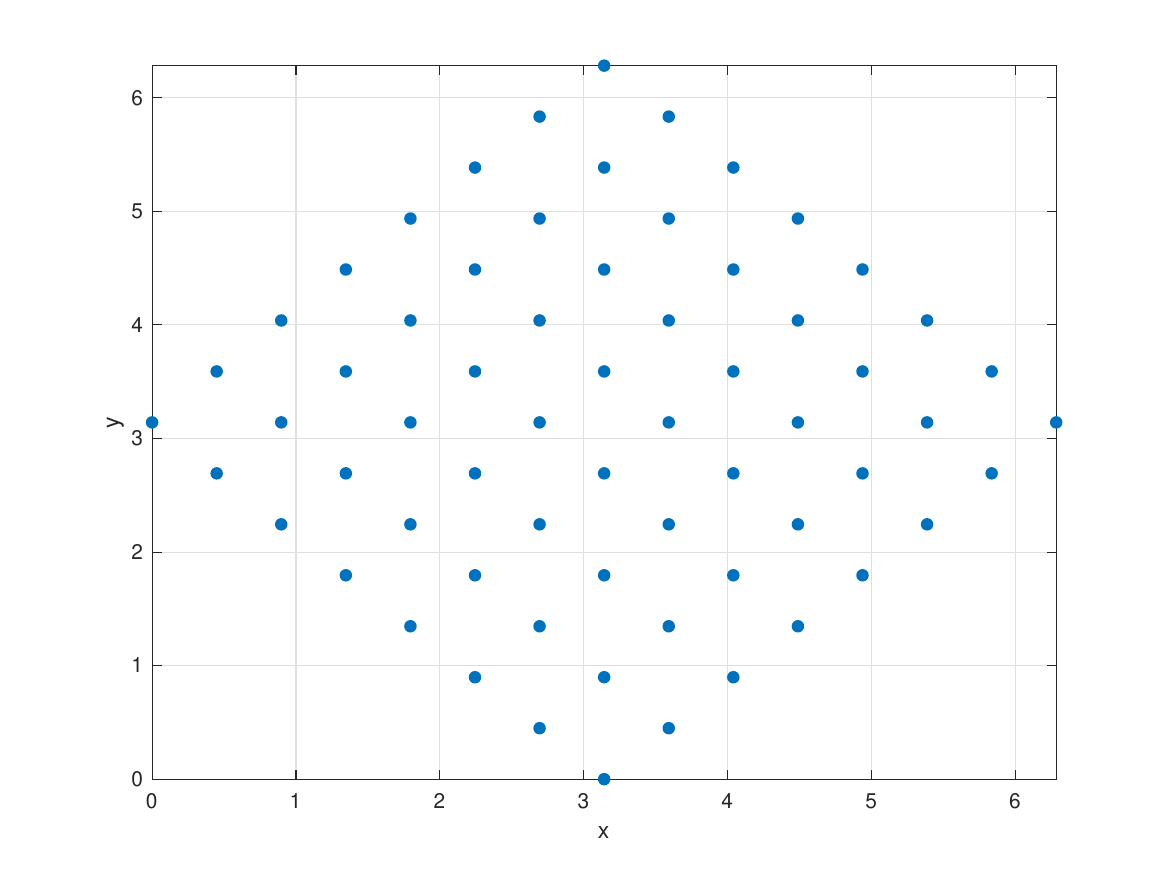}
    \caption{Periodic domain for hexagonal lattice (left) and  crystal states for the hexagonal and square lattices, respectively (right).}
    \label{fig:hexdom}
\end{figure}

{}\smallskip\textbf{Onset of Bifurcation: Effect of Repulsive Potential.}
In addition to finite size corrections, the bifurcation point also changes due to the change in repulsive potential. Replacing the Dirac-$\delta$ by $K^\zeta$, and linearizing at the constant state, we find a linearized operator
$
\mathcal{L}v = \Delta((K^\zeta + V)*v).
$
The continuum bifurcation points with this potential are thus $\morig_* = \frac{1}{\pi^2(1+2\zeta^2)}$ (square),  and $\morig_* = \frac{1}{\sqrt{3}\pi^2(1+2\zeta^2)}$ (hexagon), and we do find these in large-$N$ limits in the discrete model below.  
\subsection{Numerical Continuation}
We implement secant continuation to find parameter values where non-trivial branches bifurcate from the crystalline state in \eqref{e:discrete_main} and track branches of steady states. The Bessel potential is of course not periodic and needs to be replaced by its periodization, that is, the sum over all lattice translates. Since this sum is not explicit, we approximate the periodized potential by the sum of nine translated copies of $K^\zeta$. The translates are centered at all combinations of the respective lattice's basis vectors around and including the origin (i.e. $(0,0), \pm\mathbf{e}_1, \pm \mathbf{e}_2$, etc.).

Adding phase conditions in both  $x$- and $y$-directions, similarly to \S\ref{s:num_pde}, we solve 
\begin{align*}\label{e:secant_discrete}
        0 &= -\frac{1}{N^2} \sum_{\underline{j} \neq \underline{\ell}}^{N} \partial_{1} W(x_{\underline{\ell}}-x_{\underline{j}}, y_{\underline{\ell}}-y_{\underline{j}} ) + s^x\mathbf{1},\\
     0&= -\frac{1}{N^2} \sum_{\underline{j} \neq \underline{\ell}}^{N} \partial_{2} W(x_{\underline{\ell}}-x_{\underline{j}}, y_{\underline{\ell}}-y_{\underline{j}} )+ s^y\mathbf{1},&&\textrm{(evolution equations)} \\
       0 &= \frac{1}{N^2}\sum_{\underline{j}} x_{\underline{j}}-\overline{x}_{\underline{j}},  \\
       0 &= \frac{1}{N^2}\sum_{\underline{j}} y_{\underline{j}}-\overline{y}_{\underline{j}}, &&\textrm{(phase conditions)}
\end{align*}
where $\mathbf{1} = (1, 1, ... 1)$, and $s^x,s^y$ are dummy variables associated with the phase conditions. There is clearly no mass constraint necessary, here.  We solve the system in $2N^2+3$ variables $(x,y,,s^x,s^y,\mnew)$ using Newton's method and a secant normalization.
All results described in the following  were obtained with a fixed $\zeta = 0.3$. 

\begin{figure}[h!]
    \centering
    \includegraphics[width=0.49\textwidth]{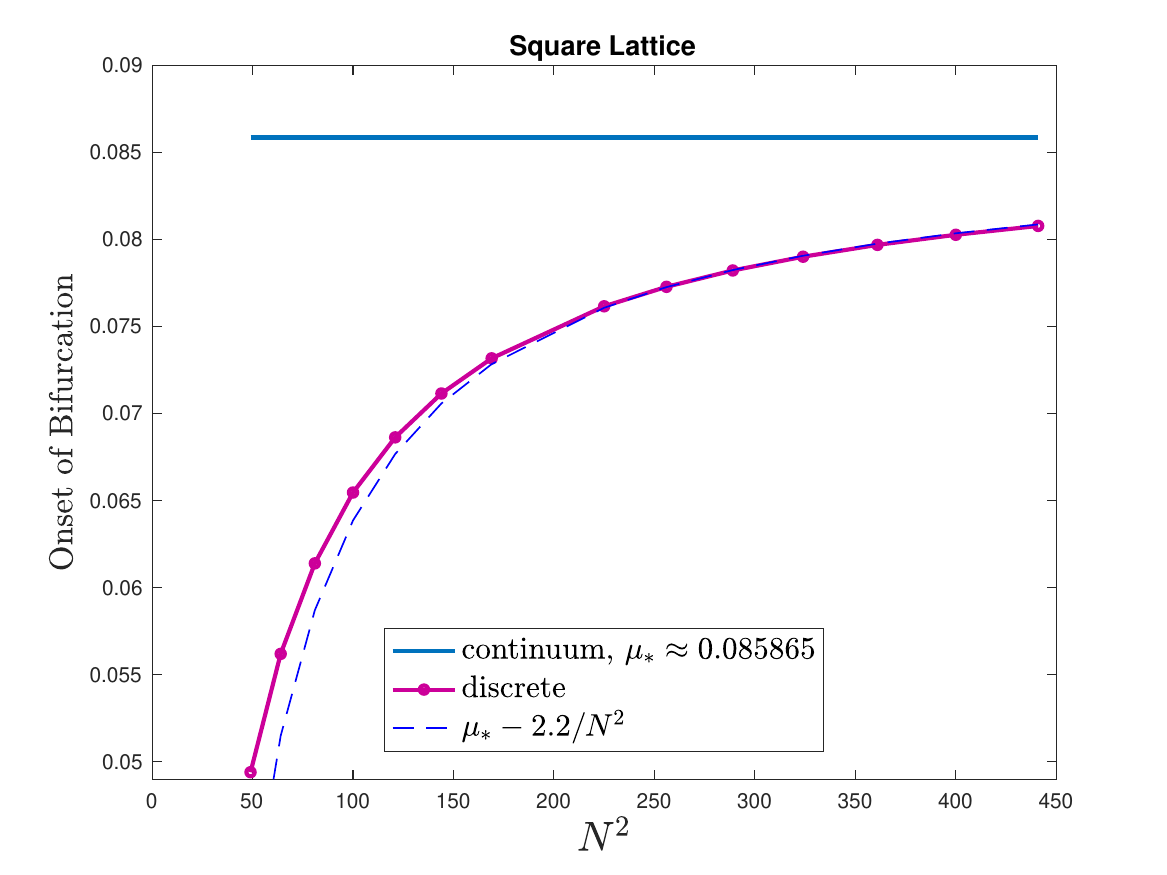}
    \includegraphics[width=0.49\textwidth]{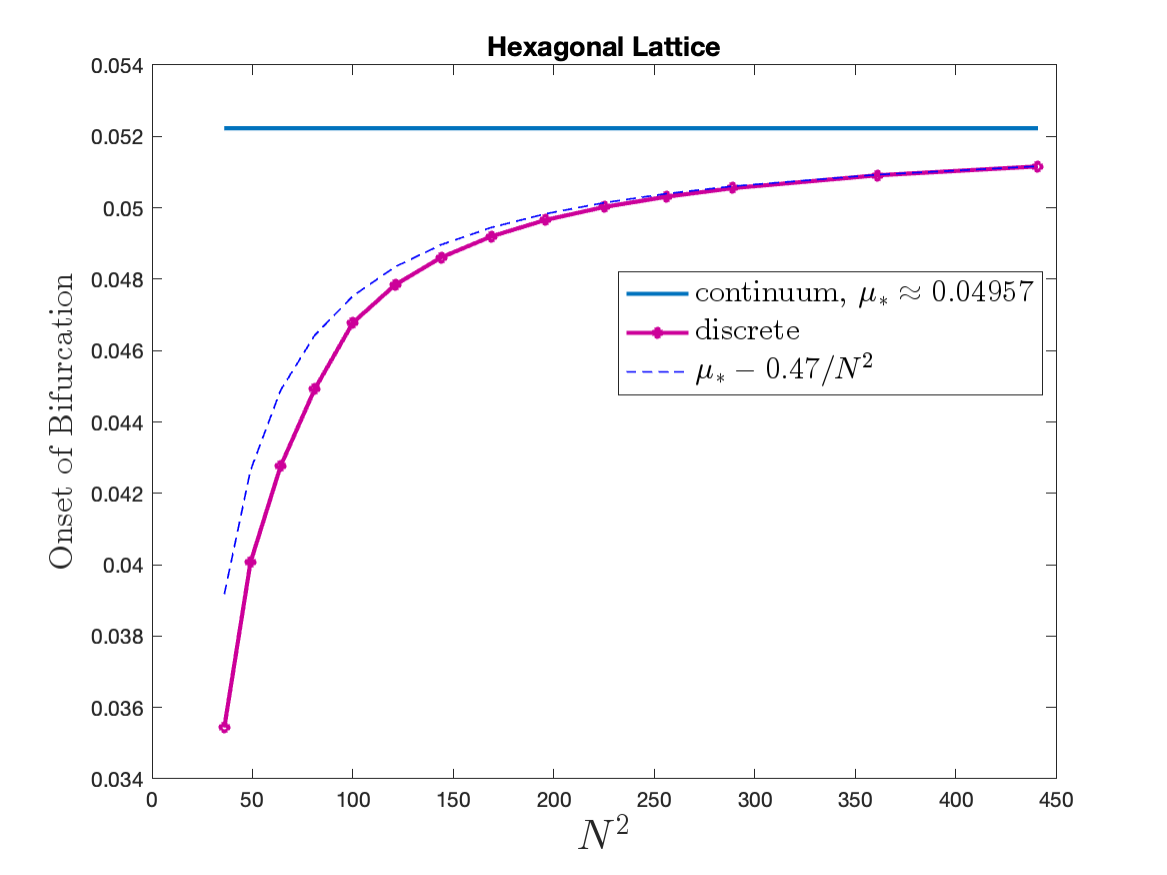}
    \caption{Onset of instability of the pure crystalline state found using secant continuation for fixed $\zeta=0.3$ and varying total numbers of particles $N^2$ on the square lattice (left) and the hexagonal lattice (right). The finite-$N$ corrections to the bifurcation point appears to be $\mathcal{O}(N^{-2})$.
    }
    \label{fig:bif_onset_discrete}
\end{figure}

\begin{figure}[h!]
    \centering
    \includegraphics[width=0.5\linewidth]{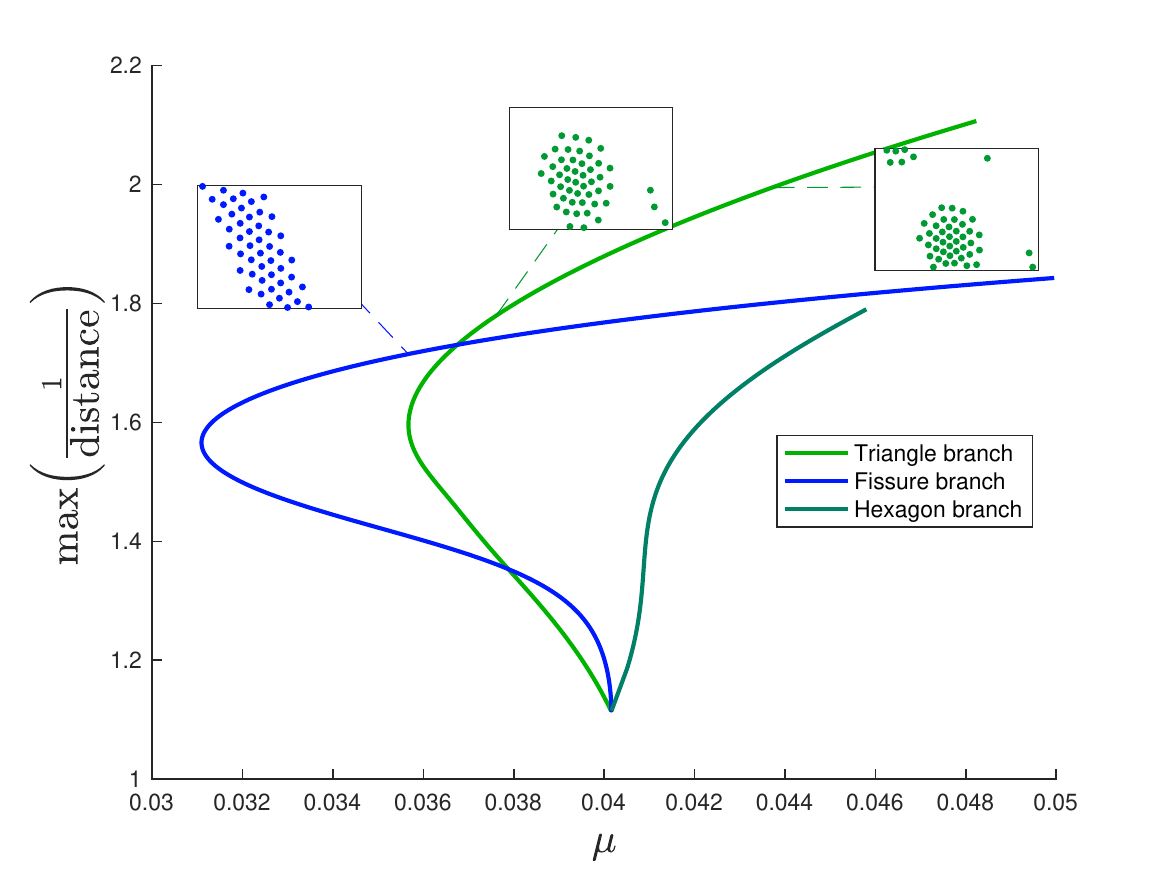}
    \caption{ Bifurcation diagram obtained using secant continuation on the hexagonal lattice for $N^2= 49$ particles, with sample particle configurations observed in direct simulations. Triangle branch and hexagon branch are in green and dark green, respectively, fissure branch in blue. Hexagons are not observed in direct simulations. On the vertical axis, we show the inverse of minimal distances between particles as a proxy for the maximal density. }
    \label{fig:bif_inset_hex}
\end{figure}

Fig.~\ref{fig:bif_onset_discrete} shows the numerically calculated onset of instability of the crystalline state depending on the particle number, compared to the prediction from the continuum limit. We find convergence proportional to the inverse of the total particle number in both square and hexagonal lattice configuration. 
Fig.~\ref{fig:bif_inset_hex} shows a numerically computed bifurcation diagram in the case of the hexagonal lattice. We find triangle, hexagonal, and fissure branches. As predicted in the continuum limit, triangle branches are stable after an initial saddle-node bifurcation, while hexagon branches are unstable. Fissure  branches are stable only for subcritical parameter values. The discreteness notably causes the bifurcation of fissures  to be weakly subcritical, as opposed to the weak supercriticality caused by small noise $\eps$ in \S\ref{s:diffusion}, a phenomenon also observed in the one-dimensional case \cite{stevensscheel}.
We noted that during continuation there are several secondary bifurcations, and we do not claim at all that this bifurcation diagram is complete. An intuitive reason for secondary bifurcations are possible rearrangements and emergence of defects in the particle configuration, an effect which is notably absent in the one-dimensional case \cite{stevensscheel}. 
In fact, such rearrangements and secondary bifurcations prevented us from creating the analogous figure in the case of a square lattice using numerical continuation. We therefore now turn to direct simulations, which do, in the case of square lattice symmetry, show a striking similarity to our results in the continuum limit. 

\subsection{Direct Simulations and Observed States}
We collect here some observations from direct simulations. 

{}\smallskip\textbf{Square lattice.}
On the square lattice, as in the continuum, we find that for a range of $\mnew$ values above onset, perturbations from the crystal state lead to particles arranging to form fissures. Past a secondary bifurcation point, particles organize only transiently into fissures, which give way ito single clusters. 
In all cases observed, although the crystal state has square particle microstructure, the vacuum states exhibit hexagonal lattice microstructure (with occasional defects as expected in finite-particle simulations). In other words, in the supercritical pattern-forming regime, the predictions from the macrostructure of the problem appear to hold independently of particle microstructure. 


\begin{figure}
    \centering
    \includegraphics[width=0.16\linewidth]{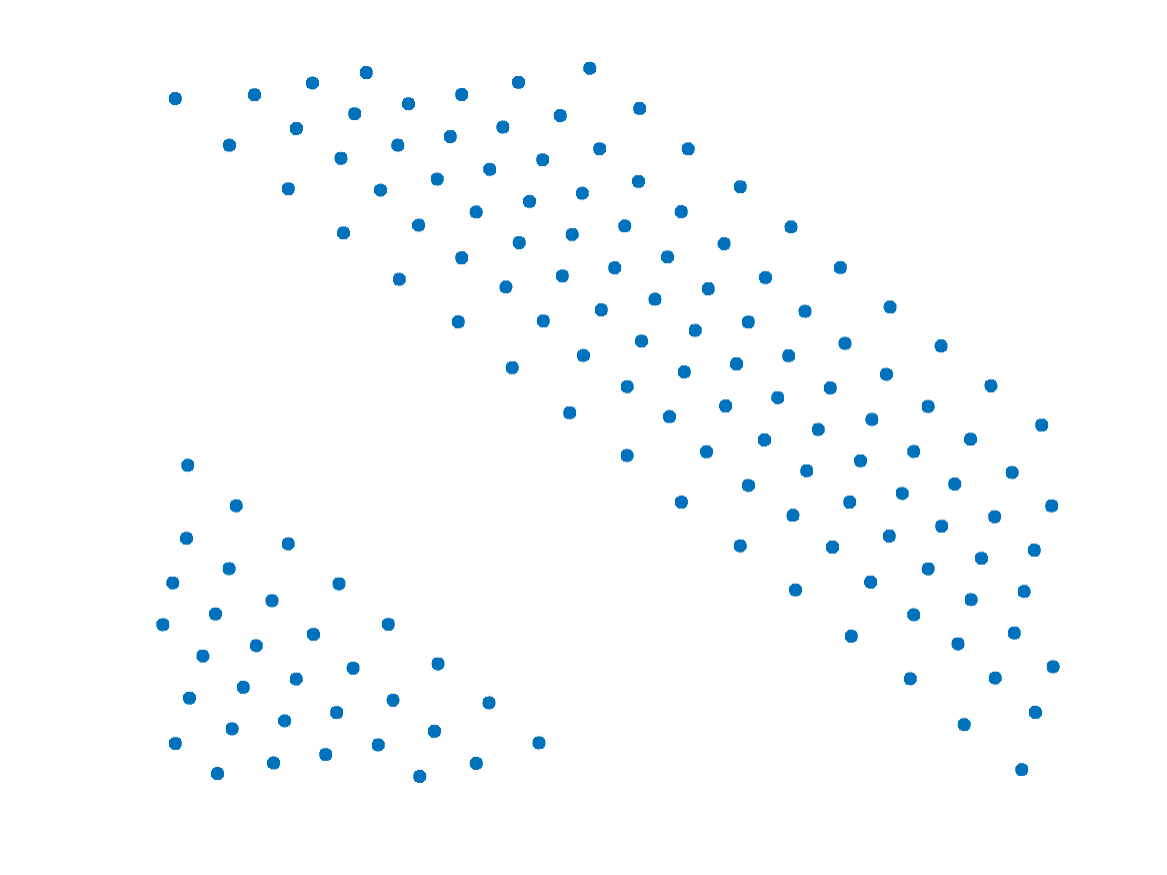}
    \includegraphics[width=0.16\linewidth]{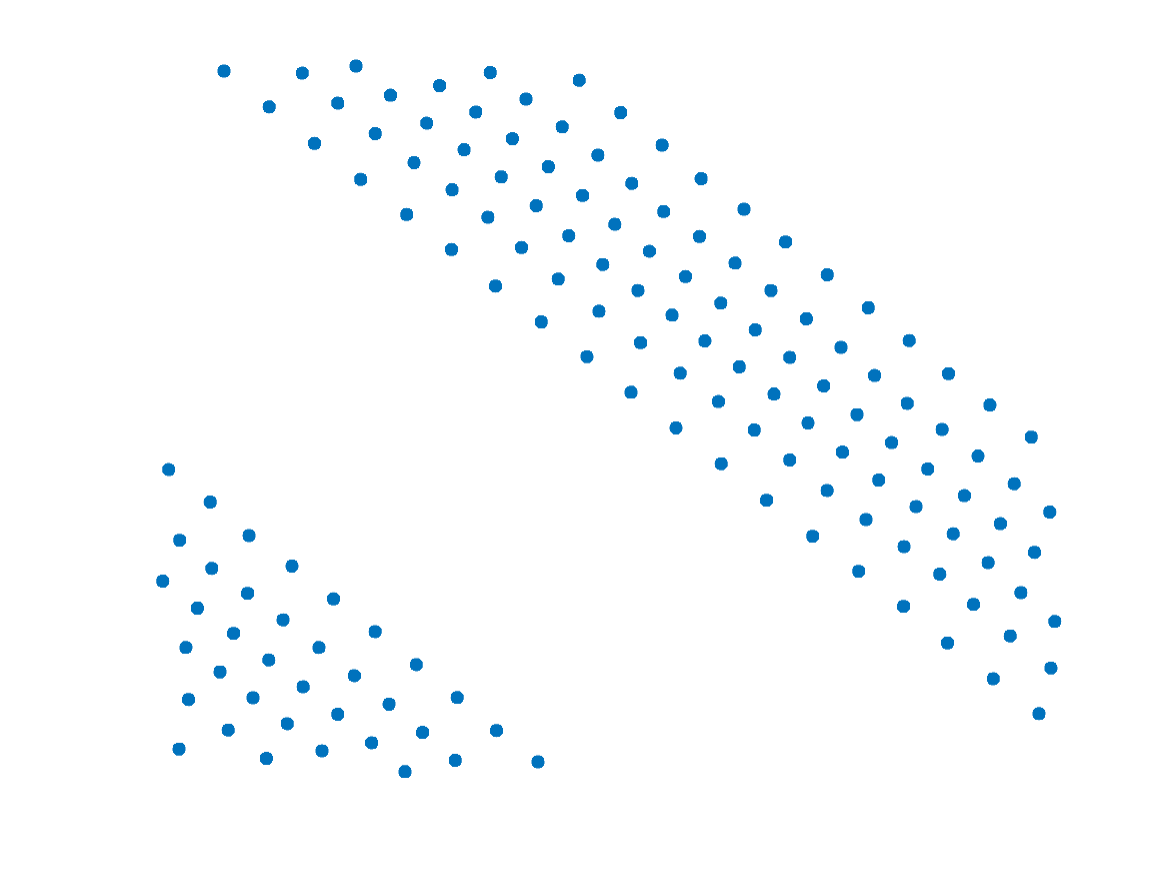}
    \includegraphics[width=0.16\linewidth]{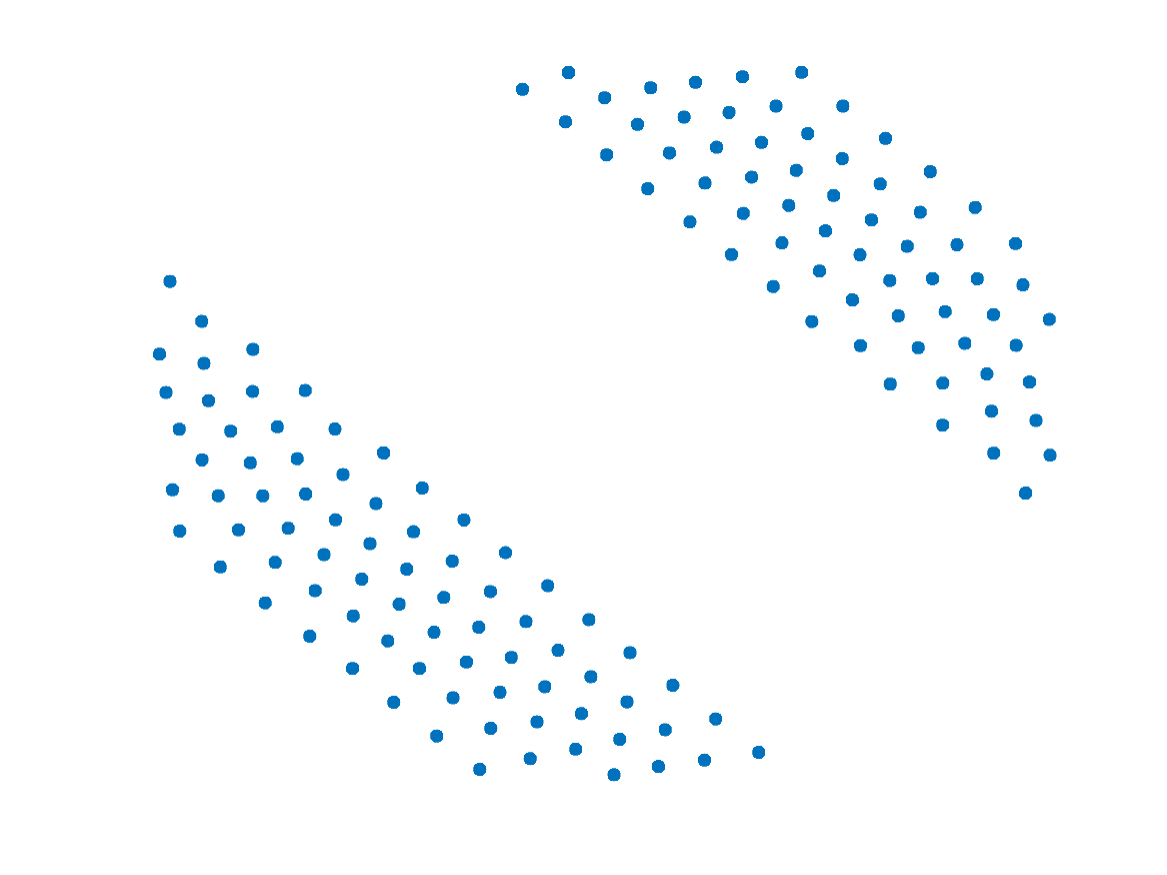}
    \includegraphics[width=0.16\linewidth]{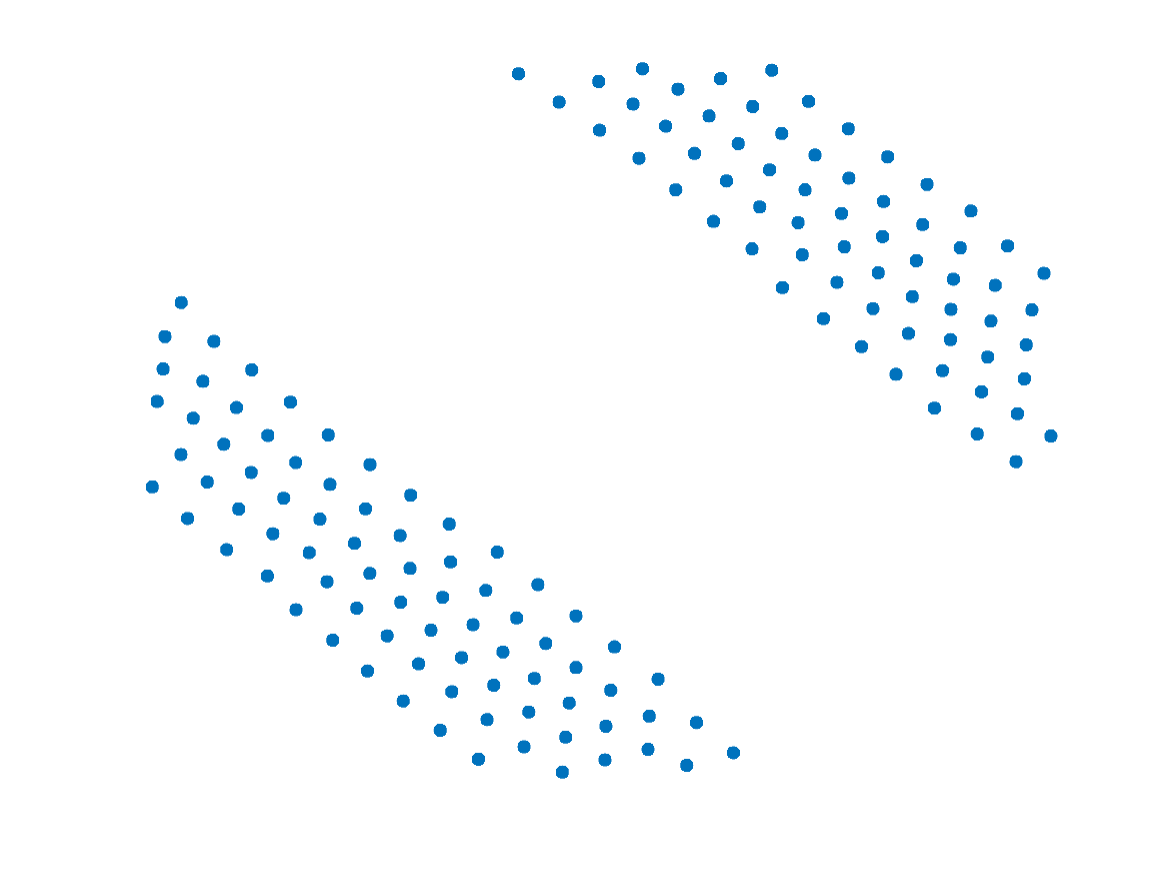}
     \includegraphics[width=0.16\linewidth]{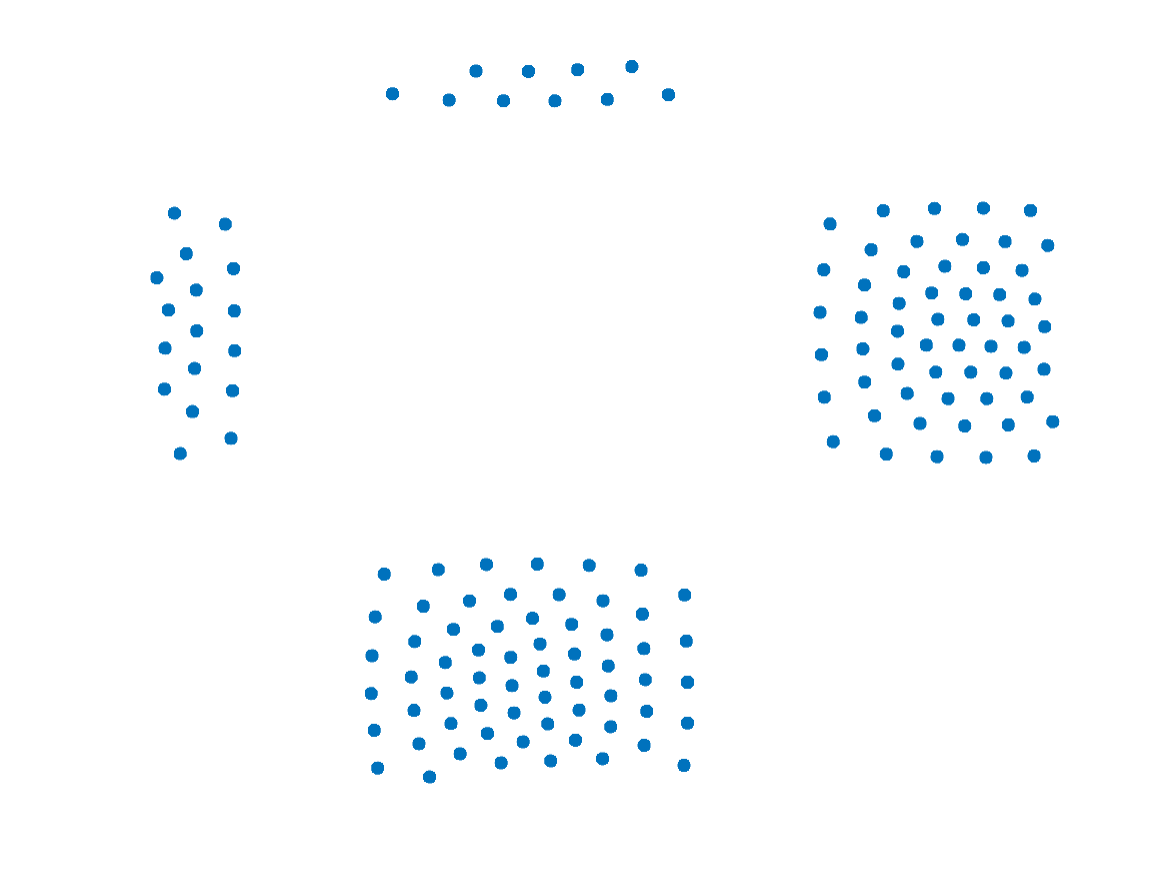}
     \includegraphics[width=0.16\linewidth]{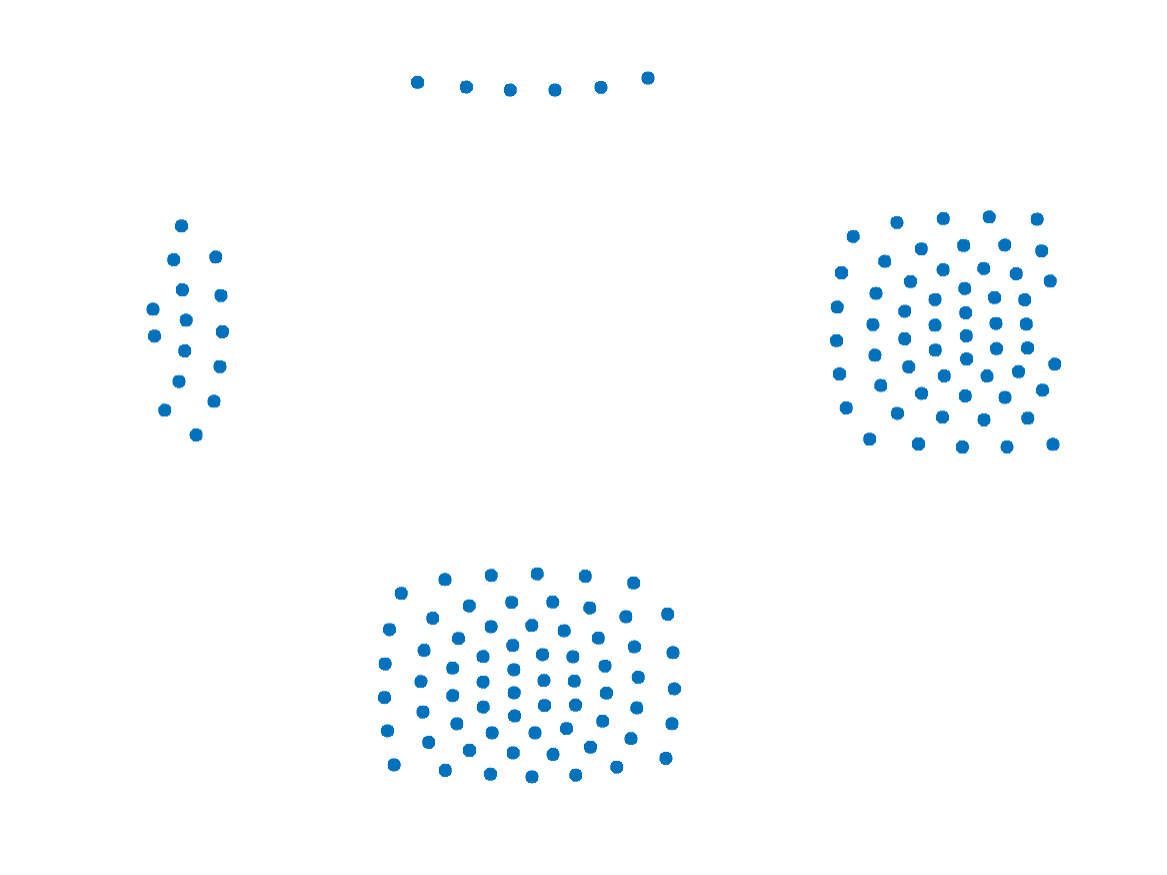}


    \caption{ Bifurcation on the square lattice, $N^2 = 144$ particles. States observed at $t = 10000$ in direct simulations for various values of $\morig$, 
    from left to right: $\morig$ = 0.075, 0.1, 0.125, 0.15, 0.16, 0.2. Initial conditions were given by perturbing the crystal state with random noise at $t=0$, in all cases except $\morig = 0.075$. In that case, the initial perturbation was in the direction of the unstable eigenvector $\cos(\overline{x})\cos(\overline{y})$, due to the fact that random perturbations favored the instability of the square lattice toward a uniform state with hexagonal lattice microstructure.  }
    \label{fig:discrete_selected_states_n_11}
\end{figure}

{}\smallskip\textbf{Square lattice: hysteresis near bifurcation points.}
Near both the initial and secondary bifurcation points, the discrete structure induces hysteresis, more pronounced for small numbers of particles. 
In the subcritical region where the branches bend backwards, both fissures and bubbles can be seen at the same values of $\morig$, depending on whether the system evolves from a fissure or cluster state. In each case, the selected state appears stable to small perturbations in the direction of the other. 
We note that for square symmetry of the potential, neither of the continuum models (the original system \eqref{e:MV_gen}-\eqref{e:MV_boundary_gen} or the corresponding system with diffusion) has a backward-bending branch -- this form of hysteresis is an intrinsically discrete phenomenon. 

Similarly, for a range of $\morig$ near the secondary bifurcation point where fissures and bubbles exchange stability, both fissures and clusters can be seen at the same values of $\morig$, depending on initial conditions. 
\begin{figure}
        \includegraphics[width=.21\textwidth, trim={0.08in 0.05in 0.05in .2in},clip]{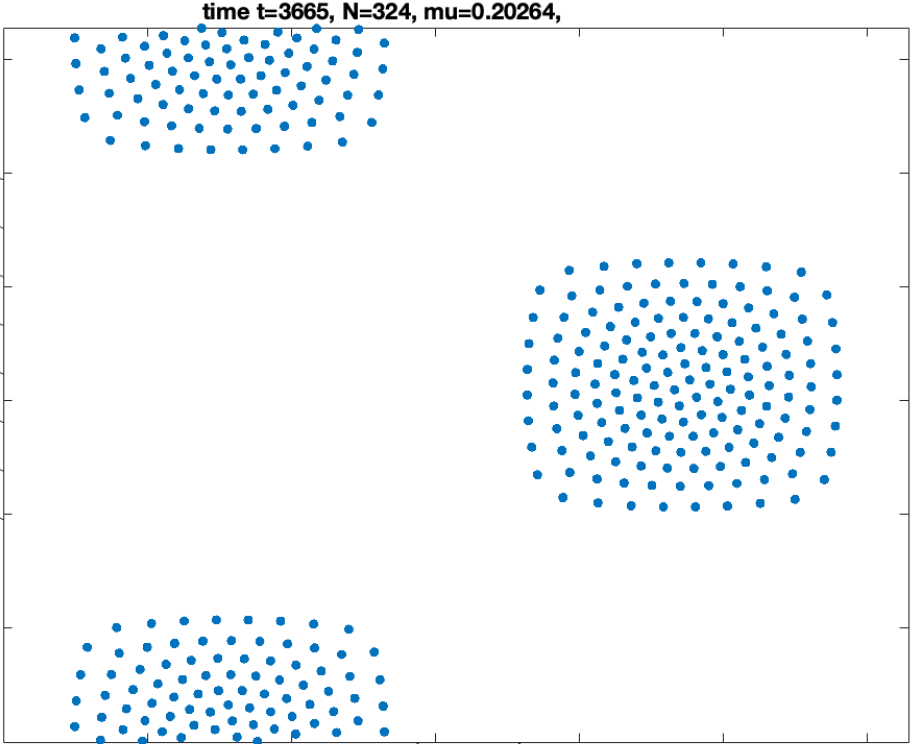}\hfill
        \includegraphics[width=.21\textwidth, trim={0.05in 0.05in 0.05in .27in},clip]{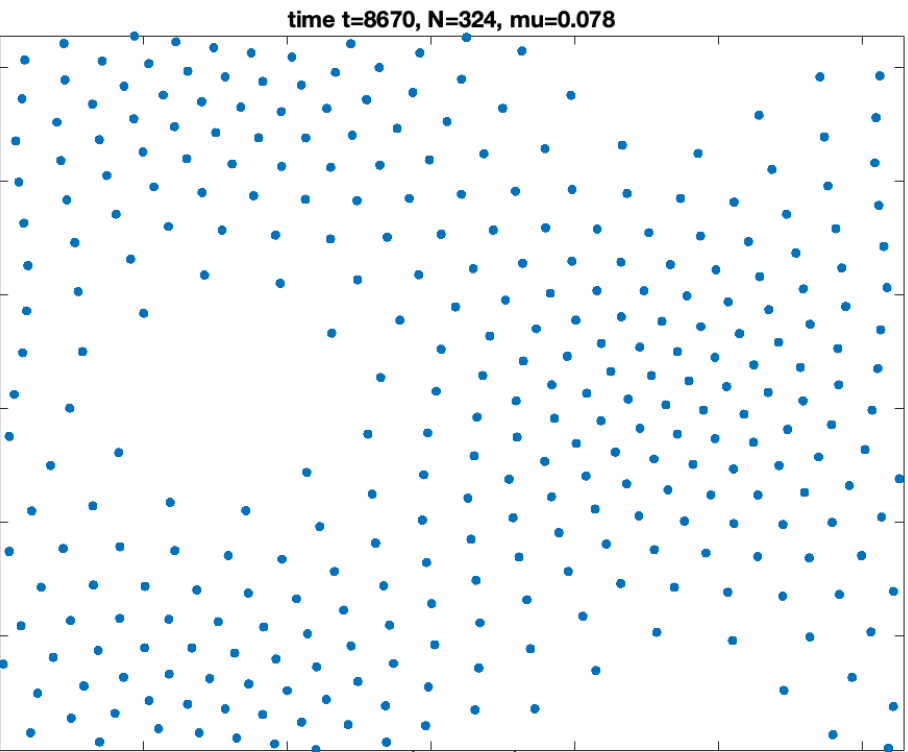}\hfill
          \includegraphics[width=.21\textwidth, trim={0.05in 0.05in 0.05in .25in},clip]{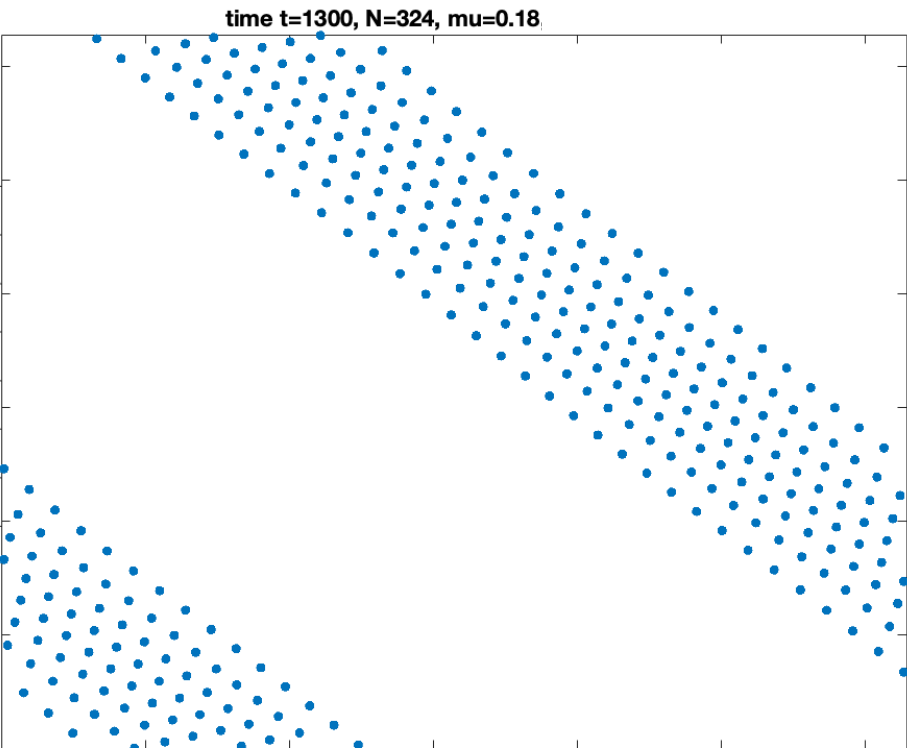}\hfill
        \includegraphics[width=.21\textwidth, trim={0.1in 0.05in 0.05in .2in},clip]{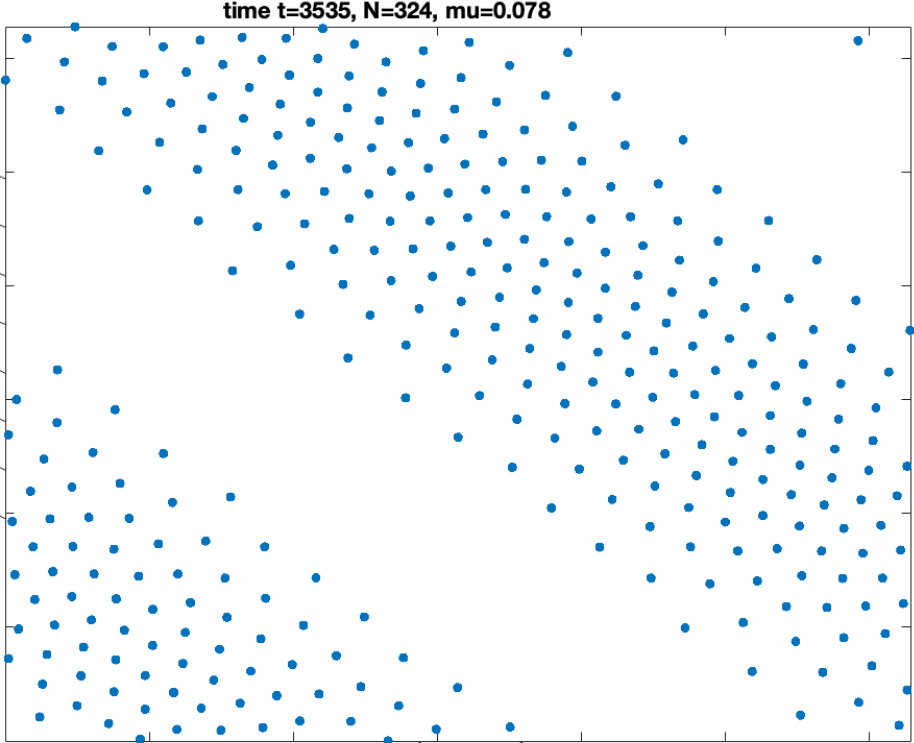}
        \caption{ Hysteresis near initial bifurcation in discrete model (square lattice), $N^2 = 324$ particles. Bubble and fissure states can be observed at the same $\morig$ values in the subcritical region where the branches bend backwards (2nd and 4th panels). 
        The states in the 2nd and 4th panels were obtained by first increasing $\morig$ to produce the states seen in the 1st and 3rd panels, and then decreasing $\morig$ to 0.078, which is in the subcritical region for $N^2 = 324$.}
        \label{fig:sq_vac_bistability}
\end{figure}

\begin{figure}
        \includegraphics[width=.21\textwidth]{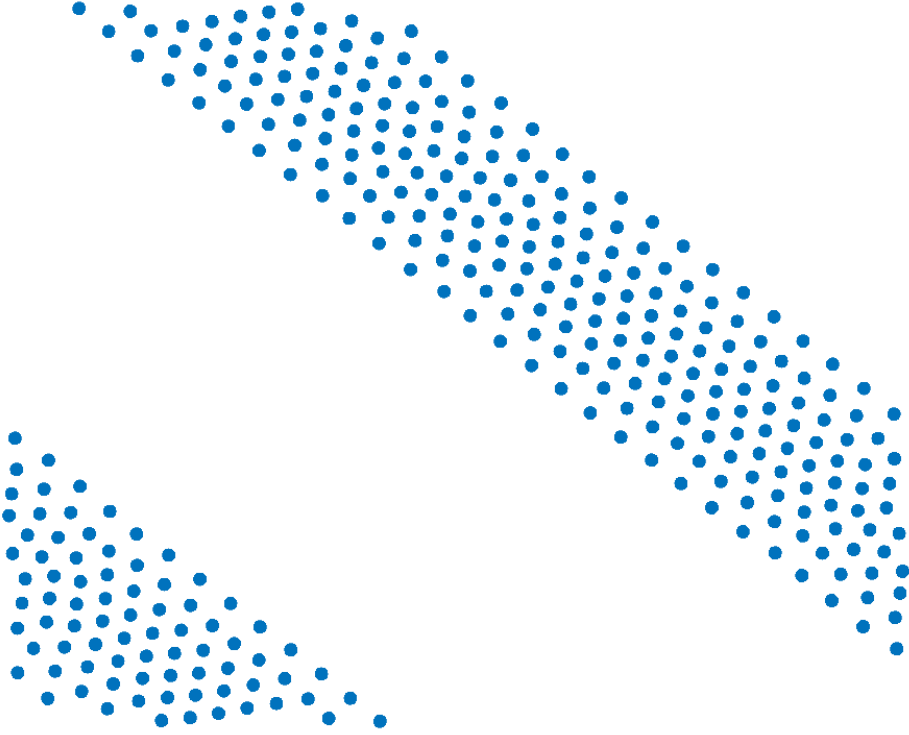}\hfill
        \includegraphics[width=.21\textwidth]{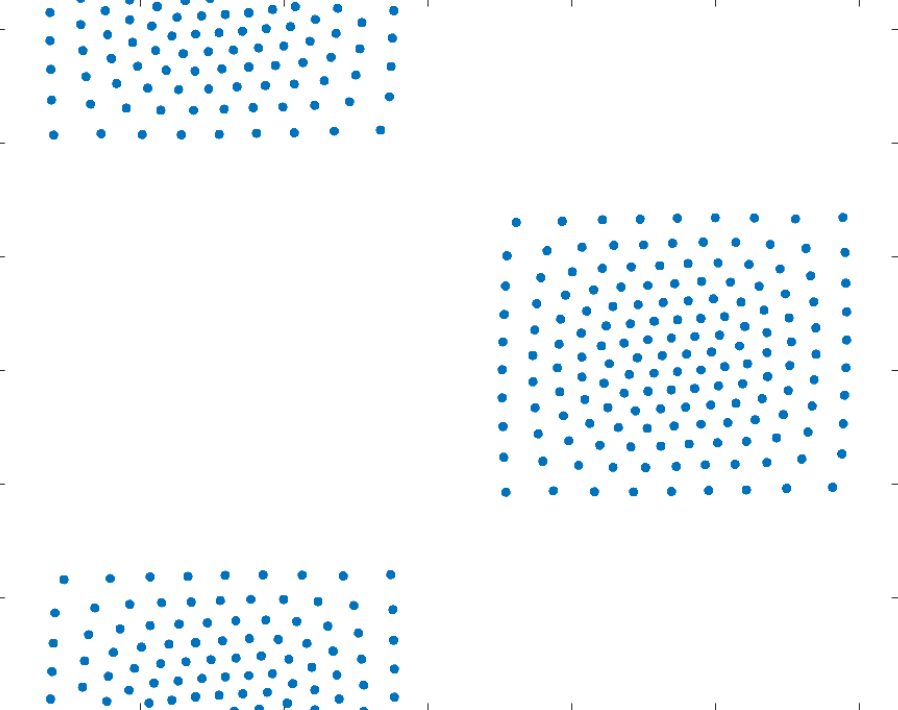}\hfill
          \includegraphics[width=.21\textwidth]{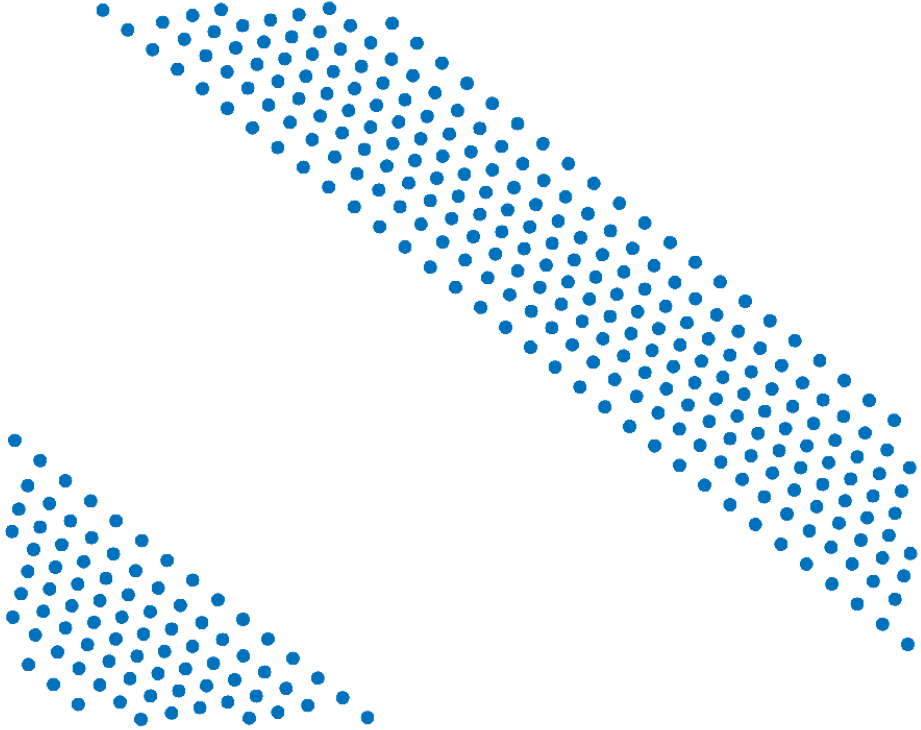}\hfill
        \includegraphics[width=.21\textwidth]{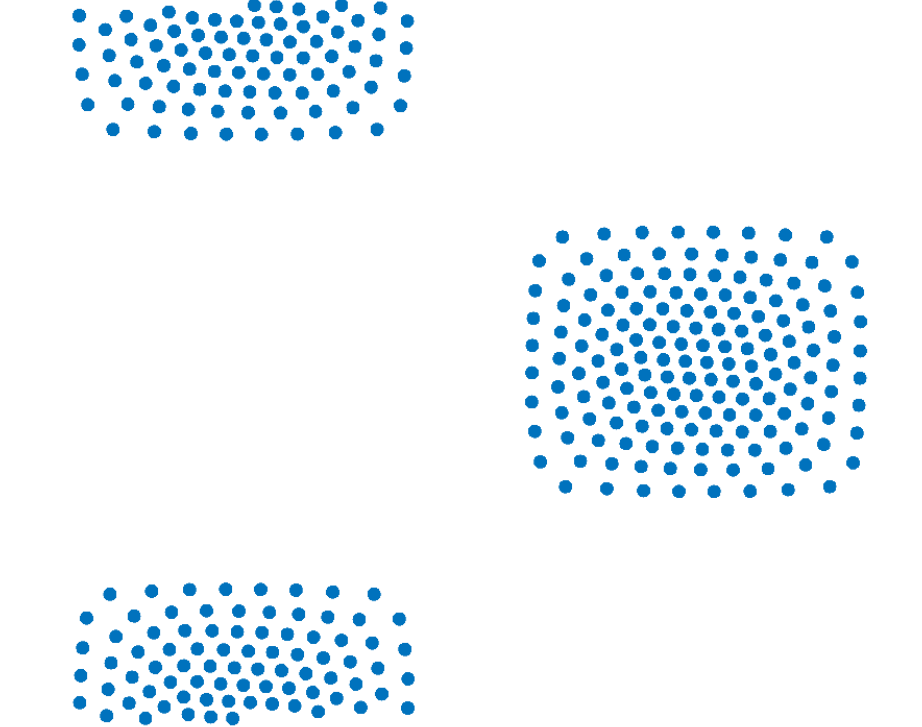}
        \caption{ Hysteresis near secondary bifurcation in discrete model (square lattice), $N^2 = 324$ particles. Both bubble and fissure states can be observed after long times for a range of $\morig$ values near the secondary bifurcation, depending on the initial condition. Left two panels: $\morig = 0.155$. Right two panels: $\morig = 0.175$. For $N^2 = 324$, the interval $(0.155,0.177)$ appears to be the approximate range of bistability.}
        \label{fig:sq_vac_bistability_2ndbif}
\end{figure}

{}\smallskip\textbf{Hexagonal lattice.}
On the hexagonal lattice, in the supercritical range of $\morig$, only clusters are observed. Fissures and hexagons are not seen in direct simulations. 
This corroborates findings from \S\ref{s:diffusion} and \S\ref{s:num_pde} concerning stability. 

All states observed in the supercritical regime were single connected clusters; true bubble solutions were seen only in the subcritical backward-bending part of the branch. It is worthwhile to note, however, that in the continuum model, the topology change from bubbles to clusters happens at quite a small value $\mnew \approx  0.003$. It is therefore likely that a very high number of particles, larger than the maximal number $N^2\sim 400$ considered here,  would be needed to observe true bubble solutions supercritically.
In the subcritical region, similarly to the square lattice, both fissures and bubbles can be observed, depending on initial conditions. 

\begin{figure}
    \centering
    \includegraphics[width=0.5\linewidth]{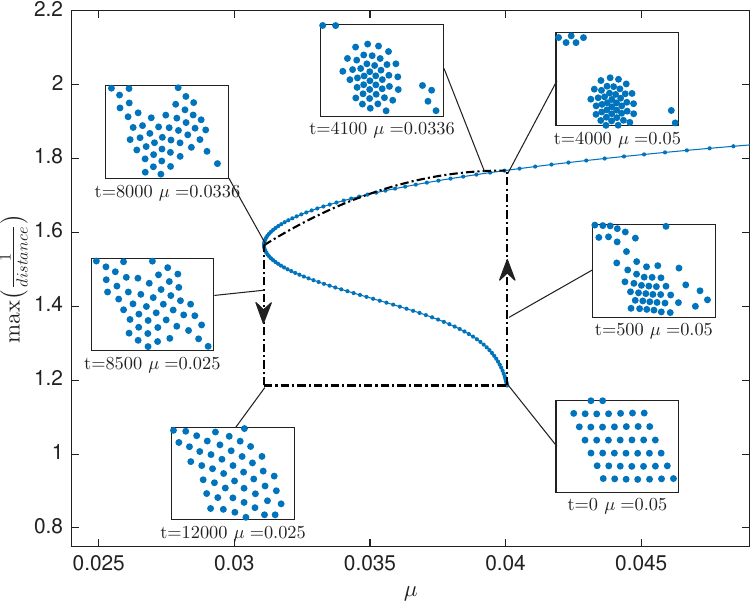}
     \caption{ Hysteresis loop on the hexagonal lattice, $N^2 = 49$ particles.}
     \label{fig:hysteresis_loop}
\end{figure}

Fig.~\ref{fig:hysteresis_loop} shows a hysteresis loop on the hexagonal lattice,
where $\morig$ is initially increased past the bifurcation point, then decreased, so that the solution travels back along the upper bifurcation branch. Once the saddle-node is reached, when $\morig$ is decreased just a bit more,
the solution``falls off'' this branch and returns to a crystal-like formation. 

\section{Discussion}\label{s:dis}

We studied clustering in interacting-particle systems in the presence of competing short-range repulsion and long-range attraction. Most of our results are concerned with a continuum limit that we study in two-dimensional, periodic geometry. We focus on a transition from uniform distribution of particles to states with vacuum regions which is curiously reversible, that is, it does not display hysteresis in the parameter governing the strength of long-range attraction. Our results give expansions for sizes of vacuum regions, shapes of vacuum regions, and even changes in topology. Sizes of vacuum regions differ significantly depending on their shape, as our asymptotics demonstrate. We did not find an apparent simple connection between, say, comparative  sizes of vacuum regions, maximal densities, and stability of states. Analyzing the effect of diffusion, we found weakly bent branches. Notably, the strength and even direction of bending depend on the type of branch. Bubbles with hexagonal symmetry bifurcate subcritically, which contradicts a naive intuition that diffusion tends to destroy clusters and therefore necessitates stronger attractive potentials for the transition to clustered states. We also demonstrate relevance of our observations for dynamics of finitely many interacting particles. 

We left open some questions concerning the completeness of the bifurcation diagrams we construct. There is strong evidence that the scenario here is rigid in the sense that the diagram does not depend on the specific shape of the attractive potential and diagrams do not exhibit  branches with submaximal isotropy, although we do not prove this rigorously. Similarly, noise appears to destroy the presence of solutions with submaximal isotropy even without vacuum regions. 

There are clearly numerous generalizations one might wish to consider. Generalizing to potentials that are not of finite (minimal) rank would be a first step, which was carried out for the one-dimensional case in \cite{stevensscheel}. While the vertical branch still persists and the perturbation argument in \S\ref{s:diffusion} should go through, finding solutions with vacuum regions now cannot be reduced to a finite-dimensional problem through an explicit ansatz of the form \eqref{e:rk3}, say, and requires parameterization of the free boundary of the support $\partial\Omega_0$ in \eqref{e:MV_gen}--\eqref{e:MV_boundary_gen} by an auxiliary unknown function. One may also wish to consider the effect of smoothed repulsive potentials, as used for instance in the discrete setting and analyzed in \cite{stevensscheel} in the one-dimensional setting.

Beyond these more immediate questions, one may wish to consider more general lattices, rhombic in 2-dimensional configuration space, or 3-dimensional configurations. A rough calculation suggests that the radius of spherical bubbles grows as $\mnew^{\frac{1}{n+2}}$ in $n$-dimensional space. As analyzed in the one-dimensional setting \cite{stevensscheel}, the arguments here should also give valid predictions for sorting in multiple-species particle mixtures. 

Lastly, it would be interesting to study the dynamics beyond the somewhat artificially imposed periodic geometry. In more generic pattern-forming systems, such as for instance the Swift-Hohenberg equation \cite{crosshohenberg}, periodic boundary conditions give good predictions for the basic crystallographic structures, while an analysis of $x\in\R^n$ can capture information on defects, such as grain boundaries. 
In other settings, for instance phase separation processes modeled with the Cahn-Hilliard equation \cite{fife}, periodic boundary conditions fail to capture instabilities of initial patterns formed in spinodal decomposition and ensuing coarsening processes. In our setting, one might expect that periodic boundary conditions give good predictions when the interaction potential inherently selects a finite wavenumber, as does the linearization in the Swift-Hohenberg equation. To make this more precise, one might choose periods $L$ different from $2\pi$ and  assume that the now $L$-dependent bifurcation point $\mu_*(L)$ has a minimum for some finite $L$, which one would then choose as a system-selected period in the periodized analysis pursued here. There may of course also be very interesting situations where $\mu_*(L)$ has a minimum at $L=\infty$, as one finds for instance in the Cahn-Hilliard equation, in which case we expect that periodic boundary conditions have limited predicitve value. 

%
%

\end{document}